\documentclass[fleqn,10pt]{olplainarticle}

\usepackage{subcaption}
\usepackage{multirow}
\usepackage{url}
\usepackage{tabularx}
\usepackage{booktabs}
\usepackage{xcolor}
\usepackage{hyperref}
\usepackage{cleveref}
\usepackage{soul}
\usepackage{verbatim}
\definecolor{commone}{RGB}{76,120,168}
\definecolor{commtwo}{RGB}{245,133,24}
\definecolor{commthree}{RGB}{84,162,75}
\definecolor{commfour}{RGB}{228,87,86}
\definecolor{commfive}{RGB}{114,183,178}
\definecolor{commsix}{RGB}{178,121,162}
\definecolor{commseven}{RGB}{255,157,166}
\definecolor{commeight}{RGB}{157,117,93}
\definecolor{commnine}{RGB}{138,138,138}
\definecolor{commten}{RGB}{47,75,124}
\definecolor{commother}{RGB}{181,166,66}
\newcommand{\communityswatch}[1]{\raisebox{0.65ex}{\fcolorbox{black}{#1}{\rule{0.0ex}{0.0ex}}}}

\hypersetup{colorlinks=true, citecolor=teal, urlcolor=blue, linkcolor=purple}

\newcommand{\edo}[1]{\textcolor{orange}{#1}}
\newcommand{\fab}[1]{{\color{magenta}#1}}

\title{Beyond Direct Retweets: Multi-Step Pathways in Italian COVID-19 Twitter}

\author[1,2,3,$\dagger$]{Edoardo Maggioni}
\author[2,$\dagger$]{Ren Manfredi}
\author[4]{Fabio Saracco}
\author[5]{Rossana Mastrandrea}

\affil[1]{Department of Computer Science, University of Pisa, Pisa, Italy}
\affil[2]{IMT School for Advanced Studies, Lucca, Italy}
\affil[3]{Institute of Informatics and Telematics (IIT), National Research Council (CNR), Pisa, Italy}
\affil[4]{``Enrico Fermi'' Research Center, Rome, Italy}
\affil[5]{School of Economics and Management, University of Turin, Turin, Italy}
\affil[$\dagger$]{These authors contributed equally to this work.}


\begin{abstract}
We study how retweet interactions in large-scale Twitter debates are organized beyond direct links alone. 
Focusing on Twitter debate in Italy during the first phase of the COVID-19 pandemic, we combine a validated community-reconstruction pipeline with a higher-order random-walk framework to examine how short multi-step pathways redistribute attention across discursive communities. 
Rather than reconstructing observed cascades of individual tweets, we use motif-based random-walk paths as a structural device to compare direct community-to-community connectivity with the distribution of multi-step endpoints. We find that attention is initially concentrated within communities, but that this concentration weakens as path length increases. At the same time, the resulting cross-community redistribution is not uniform: some communities become increasingly relevant as endpoints of longer pathways, while others lose relative prominence. These differences are not fully captured by community size or by first-order retweet connectivity alone, and they also display important directional asymmetries when the network is analyzed under the reversed orientation. In summary, the results show that moving beyond direct retweets changes the community-level representation of online debate and reveals higher-order structural patterns that remain invisible in first-order analyses.
\end{abstract}

\begin{document}

\maketitle

% ===================== INTRODUCTION =====================

\section{Introduction}

Social media platforms have deeply transformed how information is produced, circulated, and amplified in public discourse. Therefore, understanding how communication fluxes organize on social platforms and what are the patterns that information flows describe is instrumental to shed light on how public debate unfolds. A substantial body of research has shown that online discussions tend to organize into relatively distinct communities, often associated with shared political alignments, topics of interest, or common interpretive frames~\citep{watts2007influentials, gomezrodriguez2010inferring, wu2011who, conover2011political,del2016spreading,goel2016structural,becatti2019extracting, cinelli2021echo,flamino2023political, serafino2024analysis,caldarelli2025physics}. This line of work has highlighted the emergence of communities, polarization, and echo chambers as central features of online debates. In parallel, another line of research has examined how information propagates through social networks, for example, through diffusion models or the analysis of retweet cascades~\citep{banos2013role,osullivan2015mathematical,sobkowicz2018opinion,chang2021analytical}. These two perspectives have both proved useful, but they are often developed separately: community-based approaches describe the modular structure of interaction networks, while diffusion-oriented approaches focus on propagation processes.

This work is situated at the intersection of these perspectives, but adopts a different standpoint. Rather than attempting to reconstruct actual information cascades or the temporal diffusion of specific pieces of content, we focus on the structural organization of retweet interactions and on how they constrain possible flows of attention. In particular, we ask what changes when interactions are observed not only through direct links between users or communities, but also through short sequential paths, that is, sequences of successive retweet interactions on the network. 

To address this question, we combine two elements. First, we identify discursive communities through the procedure defined in~\citep{becatti2019extracting,guarino2024verified} (see Sections~\ref{ssec:lit_rev} and~\ref{sec:methods} for more details). Second, we analyze random-walk paths on the weighted directed network and map these paths from the user level to the community level. This allows us to compare direct community-to-community connectivity with the distribution of endpoints generated by paths of different lengths, providing a higher-order description of how attention is structured across communities.

The purpose of this analysis is not to claim that these paths correspond to observed trajectories of single tweets, nor that they provide a direct measure of causal influence. Instead, they are used as a descriptive device to characterize how the aggregated network structure channels possible interactions between communities. From this point of view, the work is concerned less with diffusion in a strict temporal sense and more with the structural organization of attention and amplification patterns in retweet networks.

This perspective is useful for at least three reasons. First, links alone may provide only a partial view of how different parts of the network are structurally connected. Second, comparing short and longer paths makes it possible to assess whether some communities are disproportionately represented as destinations of multi-step transitions, or whether attention remains largely confined within the same community structure already visible in direct interactions. Third, the presence of longer path motifs indicates a non-trivial organization of the information flow, suggesting the existence of mediators of the information generated by the main sources, see~\cite{wu2011who,hilbert2017onestep,bracciale2018fromsuper,dubois2020who,serafino2024analysis}. The analysis is therefore intended as a way to characterize how higher-order pathways reshape the distribution of attention relative to first-order connectivity.

Empirically, the paper focuses on Italian Twitter discussions related to the first phase of the COVID-19 pandemic. This context is particularly suitable because it combines intense political conflict, institutional communication, media activity, and issue-specific public debate within the same retweet environment. The resulting network provides a useful setting to examine how different discursive communities are positioned not only in terms of direct retweet exchanges, but also in terms of their role in multi-step pathways.

The analysis proceeds as follows. We first describe the construction of the dataset and the retweet network, together with the procedure used to identify and interpret the main communities. We then introduce the random-walk framework used to generate path motifs on the weighted directed network, and define the community-level quantities used in the comparison between direct interactions and multi-step pathways. Finally, we examine how within and between community interactions vary across higher-order structures, highlighting which patterns appear robust and which should be interpreted more cautiously.
In light of these goals, the paper addresses the following research questions:

\paragraph{RQ1:} How is retweet activity distributed across discursive communities, and how does this distribution change when moving from direct interactions to higher-order pathways?

\paragraph{RQ2:} Do communities occupy different structural positions when pathways are analyzed in the original and reversed network orientations?\\
\\
Overall, the aim of the paper is not to offer a definitive account of information diffusion on Twitter, but to develop and apply a structural, higher-order framework for studying how community-level patterns of attention and amplification change when retweet networks are observed beyond direct links alone.

% ===================== LITERATURE REVIEW =====================
\subsection{Literature Review}\label{ssec:lit_rev}

\paragraph{Online debate, polarization, and discursive communities.} 
A large body of work has shown that online political debate tends to organize into relatively distinct communities, often associated with echo chambers, shared political alignments, selective exposure, and varying degrees of polarization and informational segregation \citep{garrett2009echo,cinelli2021echo,del2016spreading}.
Within this broader discussion, Twitter has been studied extensively as a platform in which retweet interactions provide a useful trace of endorsement, amplification, and audience alignment within online debate, since they do not simply connect users but also reveal patterned forms of visibility and support \citep{conover2011political,becatti2019extracting}.
Early studies showed that retweet networks display a clear political structure, with users clustering into segregated communities and with important asymmetries across partisan groups \citep{conover2011political,conover2012partisan}. These findings established retweet networks as a meaningful object for studying the structure of online discourse, especially in politically salient contexts.

More recent work has moved beyond describing polarization alone and has developed more rigorous methods to identify nontrivial communities in noisy online environments. In particular, entropy-based and maximum-entropy null models have provided a general statistical framework for distinguishing meaningful structural patterns from random co-occurrences in complex networks \citep{cimini2019statistical,caldarelli2026thephysics}. Building on this logic, recent studies show that discursive communities, i.e. communities of users contributing to the the formation of a common discourse, can be inferred by exploiting the overlap in the audiences of prominent content producers, especially verified users, and then extending these community labels to the broader retweet network \citep{becatti2019extracting,caldarelli2021flow,guarino2024verified}. This line of research is especially relevant for the present paper because it offers one suitable and reproducible way to reconstruct the structure of public debate from retweet behavior rather than from manually assigned labels or a priori political classifications.

Within this methodological framework, discursive communities have increasingly been treated not merely as descriptive clusters, but as structured mesoscopic formations whose internal organization can itself be studied. For instance, recent work has analyzed the structural organization of Twitter discursive communities and shown that they exhibit nontrivial internal patterns \citep{mattei2022bow}, while related contributions have connected these communities to the semantic organization of online debate, showing how structural and content-related dimensions can be studied jointly \citep{radicioni2021analysing}. In the Italian COVID-19 context, this perspective has also proved useful for reconstructing how public attention, misinformation, and politically differentiated communication unfolded on online social networks during the pandemic, and has also been extended to the identification of echo chambers through the overlap between discursive communities and shared news engagement patterns \citep{caldarelli2021flow,guarino2021information,pratelli2024entropy}. %Additional work on information disorder and on leader-audience asymmetries further supports the relevance of bipartite and audience-based approaches for identifying coherent communities in large-scale online debates \citep{guarino2021information}. 

\paragraph{Diffusion, pathways, and higher-order structure.}
Alongside research on polarization and discursive communities, a large literature has examined how information propagates through online networks, often through diffusion models, cascade analyses, and other process-oriented approaches \citep{singh2025information}. In the context of social media, these perspectives have shown that different spreading mechanisms can shape public discourse in different ways \citep{gong2023broadcast}. At the same time, diffusion and cascade studies typically address questions that are partly different from the one addressed in this section, which is less concerned with the temporal propagation of specific items and more with how the aggregated retweet network organizes possible multi-step pathways across communities.

This distinction becomes relevant when the relational patterns of interest are not well described by pairwise interactions alone, but involve dependencies across paths, memory, or sequential constraints. 
%\fab{[Distinzione un po' di lana caprina, ma che ci può stare, vedi~\cite{peixoto2026graphs}: loro sono un po' stronzi (il modo di presentare l'argomento è pessimo, come spesso succede ai tre caballeros. In questa stagione anche tirato dentro anche un quarto che non conosco), ma il punto lo tengono bene]}. 
A growing body of work in network science has argued that first-order representations may be inadequate in such cases, and that higher-order representations can provide a more appropriate description \citep{fischer2015sampling,lambiotte2019networks,scholtes2017network}. Related contributions have shown that higher-order organization can affect not only local path statistics but also broader structural properties, including centrality, community detection, and flow patterns \citep{benson2016higher,salnikov2016using,scholtes2016higher}. In parallel, work on sequential motifs has further emphasized the analytical relevance of recurring path configurations, showing that motifs can be defined not only over static graph neighborhoods but also over observed walks and other sequence-based structures \citep{larock2022sequential}.

Importantly, in this work we use the term higher-order in a path-based sense, rather than to imply a departure from graph-based representations. The analysis remains defined on a standard weighted directed graph, while the higher-order component refers to the sequential organization of random-walk paths on that graph. This distinction is consistent with recent arguments that graph-based models can represent rich forms of higher-order or multivariate structure without necessarily requiring hypergraph-based representations \citep{peixoto2026graphs}. %\edo{L'ho citato così, va bene?} FS: ottimo!

Within this broader landscape, random-walk frameworks offer one useful bridge between structural and process-based perspectives, since they make it possible to characterize how weighted networks channel possible trajectories without requiring the reconstruction of actual cascades of individual items. Random walks on weighted networks constitute a well-established family of analytical tools \citep{riascos2021random}, and recent work has extended this logic to the study of weighted network motifs interpreted as random-walk patterns \citep{picciolo2022weighted}. This is especially relevant for the present study, because it provides a principled way to move beyond direct retweet links and describe multi-step pathways as structural configurations of attention, where attention is understood in a network sense as the distribution of visibility, reachability, and amplification induced by retweet connectivity rather than as a psychological property of individual users. In this sense, the contribution of this chapter lies in combining a community-based reading of online debate with a higher-order, path-based framework for analyzing how attention is redistributed across communities beyond first-order connectivity. The relevance of work on sequential motifs lies mainly in showing that recurring path configurations can be analytically meaningful, even though the paths studied here are analytically generated random-walk paths on an aggregated network rather than observed temporal sequences of interactions.

\paragraph{Bridging communities and higher-order pathways.}
%Taken together
In summary, these literatures suggest that the main components of the problem addressed here have already been studied, but mostly in separate ways. On the one hand, research on retweet networks and discursive communities has shown how online debate tends to organize into relatively coherent mesoscopic structures, which can also be linked to differentiated patterns of information exposure and echo chambers \citep{becatti2019extracting,pratelli2024entropy}. On the other hand, a large literature on information diffusion has examined cascades, retweet propagation, and the structural features of spreading processes on online platforms \citep{goel2016structural,vosoughi2018spread}. In parallel, network science has developed higher-order and sequence-based frameworks showing that paths, memory, and flow constraints may require representations that go beyond first-order connectivity \citep{lambiotte2019networks,salnikov2016using,peixoto2017modelling,picciolo2022weighted,larock2022sequential}. These contributions are relevant to the present section, but they are not identical in their questions or objects of analysis. What matters here more specifically is whether moving beyond direct retweet links changes how cross-community attention is distributed when short multi-step pathways are taken into account. This question appears comparatively less explored, especially in empirical settings where discursive communities are first reconstructed from retweet networks and are then used to study structural attention patterns beyond direct community-to-community connectivity. Against this background, the present chapter selectively combines insights from these partly distinct traditions to address that question from a community-based and higher-order perspective.

% ===================== METHODS =====================
\section{Methods} \label{sec:methods}

The analysis is structured as a sequential workflow composed of two main components.

First, we reconstruct discursive communities from the retweet network through a validated bipartite projection of verified and unverified users, followed by community detection and label propagation. This step provides a community-level representation of the debate, which serves as the structural basis for the subsequent analysis.

Second, we analyze higher-order pathways on the resulting weighted directed network using a random-walk framework. This allows us to move beyond direct retweet links and to characterize how attention is structurally redistributed across communities through short multi-step transitions.

The two components are conceptually distinct but methodologically connected: the higher-order analysis is defined on the network representation obtained from the community-reconstruction pipeline, and all community-level quantities introduced below rely on this mapping.

\subsection{Data description}

The dataset consists of approximately 4.5 million tweets in Italian language, collected between February 21 and April 20, 2020. The original data was extracted by \cite{caldarelli2021flow} via the Twitter Streaming API by employing a series of keywords and hashtags related to the COVID-19 pandemic. Table \ref{hashtag_twitterapi} lists a subset of hashtags employed by \cite{caldarelli2021flow}. 
Tweets were collected as Twitter objects, structured JSON units characterized by a number of different attributes, such as unique identifier, body of the message, and creation date. Twitter objects also include nested child objects, dictionaries containing user and contextual metadata (attachments, hashtags, links, and mentions), as well as geographic information.\\
Among the information available in the Twitter object, we are interested in two specific attributes. The first concerns the user status, which is differentiated into verified and unverified users. Verified users are accounts whose identity and authenticity have been verified by the platform\footnote{During the period under consideration, Twitter conducted the user verification process manually, according to internal criteria that were not made public (\cite{8750923}).}. These are mainly users of public interest across multiple sectors, from institutions to public figures such as politicians, entertainers, journalists, etc. Beyond providing a guarantee of authenticity in a predominantly anonymous digital landscape, verified users possess distinctive characteristics that differentiate them from the unverified population. A primary distinction lies in their role as producer of original content, while unverified users, which represent the vast majority of accounts, function mainly as ``spreaders'' who amplify information through retweeting activity \citep{becatti2019extracting}. This structural role positions verified users as main drivers and anchors of online public discourse. Furthermore, because these accounts are held by prominent public figures, their political, ideological, or professional affiliations are often well-documented or easily discernible \citep{becatti2019extracting,guarino2024verified}. This facilitates the identification and classification of the broader discursive communities and ideological clusters that might form around these users.\\
%Another important attribute provided by the Twitter object is the ``retweeted status'', which allowed us to identify retweets from original tweets. 
In the following, we will focus on the retweet network. Retweets consist of reposts of existing tweets by users other than the author: the original tweet is then shared with the followers of the user who retweeted it, promoting its diffusion. Therefore, a retweet serves as an indicator of approval or endorsement of content created by other users, expressed through a specific action \citep{conover2011political}. This behavioral feedback provides us with a more reliable and objective tool for assessing the opinions of retweeters, usually unverified users.\\
\noindent For the present analysis, we work on the time-aggregated retweet data derived from this collection. The resulting retweet-event edgelist contains 2,996,072 rows and involves 334,633 unique users overall. The verified--unverified bipartite construction used in the community-reconstruction pipeline includes 2,581 verified users and 149,141 unverified users. 

\begin{table}
\centering
\begin{tabular}{l} \toprule
    \textbf{Hashtags} \\ \midrule
    \#coronavirus\\
    \#coronaviruschina\\
    \#coronaviruses\\
    \#coronaviruswuhan\\
    \#covid2019\\
    \#covid-19\\
    \#nCoV\\
    \#nCov2019\\
    \#ChinaCoronaVirus\\
    \#ChinaWuHan\\
    \#CoronavirusOutbreak\\
    \#WuhanCoronavirus\\
    \#COVID19\\
    \#SARSCoV2\\
    \midrule
\end{tabular}
\caption{Subset of hashtags used to collect tweets. Adapted from \cite{caldarelli2021flow}.}
\label{hashtag_twitterapi}
\end{table}

\subsection{Bipartite-network-induced Discursive Communities} 

To identify the main discursive communities, we follow the procedure originally presented by \citet{becatti2019extracting} and further reviewed by \citet{guarino2024verified}.
As defined in \citet{guarino2024verified}, discursive communities are groups of users that aggregate “organically around shared narratives, values, and ideological affinities”. In the previous paragraph, we saw that the main drivers of public discourse and producers of the majority of original content are verified users, while unverified users are primarily responsible for the dissemination and amplification of such content. For this reason, we will focus on the former by constructing a network of verified users, where each link between two nodes represents the similarity between the corresponding users. 
To determine the existence of a genuine similarity, we will employ an indirect measure of similarity based on the shared audience of two verified accounts. Conceptually, the method is based on the idea that if the content of two verified users is shared by the same public, this is a sign that unverified users perceive them as similar. As we have already seen, this behavioral indicator can be fairly represented by the retweet action.
Nevertheless, such a measure is particularly noisy, as retweets may be due to genuine interest, as well as to the virality of a post or to accidents. Therefore, to focus on the genuine signal of similarity, we employ a maximum entropy bipartite configuration model (BICM, \cite{saracco2015randomizing}) as a statistical benchmark. The result of this statistical validation is a monopartite binary network among verified users, in which links indicate non-trivial similarities among their audiences. Communities in this network capture groups of verified users sharing the same audience. The information about the communities of verified users can be further propagated to unverified users using a label propagation algorithm on the retweet network. The procedure described above is called ``Bipartite-network-induced Discursive Communities'', or BiDC~\cite{guarino2024verified}. The entire procedure is described in detail in Appendix~\ref{appendix:biDC}.

\subsection{Random-walk network representation}

We consider a directed, weighted retweet network represented as a graph $G=(V,E)$, where $V$ is the set of nodes (users), $|V|=N$, and $E$ is the set of directed edges, $|E|=L$. An edge $(i,j)\in E$ indicates that user $i$ retweeted user $j$, and it is associated with a positive weight $w_{ij}$ equal to the total number of retweets from $i$ to $j$ over the observation period.

The network can be represented by a weighted adjacency matrix $W=(w_{ij})$, where $w_{ij}>0$ if and only if $(i,j)\in E$. 
Each node is also associated with a community label $c(i)$, obtained through the pipeline described above. These labels allow us to move from a node-level description of the network to a community-level representation used in the subsequent analysis.

In what follows, we distinguish clearly between three levels of representation: (i) the user-level retweet network, (ii) the aggregated community--community network of direct retweet links, and (iii) the community-level representation of random-walk paths. This distinction is crucial to avoid ambiguity when comparing direct and higher-order structures.

\subsection{Random-walk paths}

To characterize how attention is structurally redistributed beyond direct retweet interactions, we adopt a random-walk framework that generates short sequences of retweet links on the aggregated network.
A key element of this construction is the introduction of an absorbing sink node, which allows each realized path to terminate in a well-defined way. This is not only a technical device, but a crucial modeling choice: it ensures that paths of different lengths can be treated as distinct and independent events, which can therefore be counted separately.
As a result, the frequencies of different path configurations can be compared and interpreted as alternative structural patterns of attention flow across the network, where transitions are weighted by the number of retweet interactions, without implying the reconstruction of observed cascades or temporal sequences of individual tweets.

We formalize this construction by representing the weighted retweet network as a random-walk process on an augmented directed graph, following \cite{picciolo2022weighted}. Starting from the weighted adjacency matrix $W=(w_{ij})$, we define node out-strength and in-strength for each user $i$ as
\begin{equation}
 s_i^{\mathrm{out}} = \sum_j w_{ij}, \qquad s_i^{\mathrm{in}} = \sum_j w_{ji}.
\end{equation}

In general, $s_i^{\mathrm{out}} \neq s_i^{\mathrm{in}}$, and this imbalance affects the set of feasible random-walk paths. Since the random walk describes forward transitions from a node to its neighbors along outgoing retweet links, transition probabilities are naturally defined in terms of outgoing strength.
To preserve this forward exploration while ensuring that paths can terminate, we introduce an absorbing sink node. The choice of balancing excess in-strength through the sink is therefore not arbitrary: it allows nodes with higher incoming than outgoing strength to act as structural endpoints of the process, reflecting their role as attention sinks in the network.

Formally, we augment the network with an absorbing sink node labeled $N{+}1$. For each node $i$ such that $s_i^{\mathrm{out}} < s_i^{\mathrm{in}}$, we introduce a directed edge from $i$ to the sink with weight
\begin{equation}
 w_{i,N+1} = s_i^{\mathrm{in}} - s_i^{\mathrm{out}}.
\end{equation}
Nodes with $s_i^{\mathrm{out}} \geq s_i^{\mathrm{in}}$ have no outgoing edge to the sink. The sink node $N{+}1$ is absorbing and has no outgoing edges.

Following \cite{picciolo2022weighted}, we then define the \emph{corrected out-strength} $ss_i^{\mathrm{out}}$ of each node as
\begin{equation}
 ss_i^{\mathrm{out}} =
 \begin{cases}
 s_i^{\mathrm{out}}, & \text{if } s_i^{\mathrm{out}} \geq s_i^{\mathrm{in}},\\
 s_i^{\mathrm{out}} + w_{i,N+1}, & \text{if } s_i^{\mathrm{out}} < s_i^{\mathrm{in}}.
 \end{cases}
\end{equation}

The random walk is then defined as a Markov process on this augmented graph. Let $M=(m_{ij})$ denote the transition matrix. The transition probabilities from node $i$ to any neighbor $j$ in the original network are given by
\begin{equation}
 m_{ij} = \frac{w_{ij}}{ss_i^{\mathrm{out}}},
\end{equation}
while the transition probability from node $i$ to the sink is
\begin{equation}
 m_{i,N+1} = \frac{w_{i,N+1}}{ss_i^{\mathrm{out}}}.
\end{equation}

For nodes with $s_i^{\mathrm{out}} < s_i^{\mathrm{in}}$, this implies
\begin{equation}
 m_{i,N+1} = \frac{s_i^{\mathrm{in}} - s_i^{\mathrm{out}}}{s_i^{\mathrm{in}}}.
\end{equation}
which reflects local imbalances between in- and out-strength and helps characterize how likely a node is to act as a structural endpoint of a random-walk path.
This construction ensures that transition probabilities are properly normalized over outgoing edges while preserving the heterogeneity of weighted flows, as discussed in \cite{picciolo2022weighted}.

In this framework, motifs are not interpreted as weighted versions of static subgraphs, but as short random-walk patterns that describe how sequences of retweet interactions are realized on the weighted network. 
The presence of the absorbing sink node is central to this construction: it ensures that different path configurations can be treated as distinct and independent events, so that their relative frequencies provide a meaningful characterization of higher-order structural patterns.
In this sense, the motif distribution captures how the network channels possible multi-step transitions, reflecting the interplay between edge weights and local flow imbalances. In the present study, we use this framework as an analytical device to characterize structural patterns of attention across communities, rather than as a model of observed diffusion processes or a description of individual information cascades.

Each path starts from a node $u_0$, sampled uniformly at random over all nodes. This choice provides a neutral starting rule at the node level, in the sense that all nodes have the same probability of being selected as initial points of the process. As a consequence, larger communities are sampled more frequently in proportion to their size. 
This should therefore be interpreted as a modeling assumption on how paths are initiated, rather than as an empirical description of how attention is distributed or triggered on the platform. 
From this initial node, the random walk is generated by sampling the next node according to the transition probabilities defined above. The walk continues until the sink node $N{+}1$ is reached or until three steps are completed, whichever comes first. 
The choice of limiting paths to three steps follows the original formulation in \cite{picciolo2022weighted}, where this setting allows the motif frequencies to be computed analytically. In our case, paths are generated through simulations, but we retain the same three-step setting so that the simulated motif frequencies remain directly comparable with their theoretical counterparts. This also allows us to verify that the simulation reproduces the expected theoretical frequencies, providing a useful internal consistency check for the random-walk construction adopted here.
In this way, realized paths can have effective length one, two, or three before absorption.

Importantly, these paths should be interpreted as sequences of retweet interactions across users, rather than as traces of a single propagating item. A multi-step path therefore identifies a chain of users linked by successive retweets, which we interpret as a structural pattern of attention and amplification.
In this context, retweets are understood as signals of endorsement or alignment with previously shared content, rather than as direct measures of causal influence between users.

\subsection{Motif selection}
\begin{figure}[h!]
    \centering
    \includegraphics[width=0.6\linewidth]{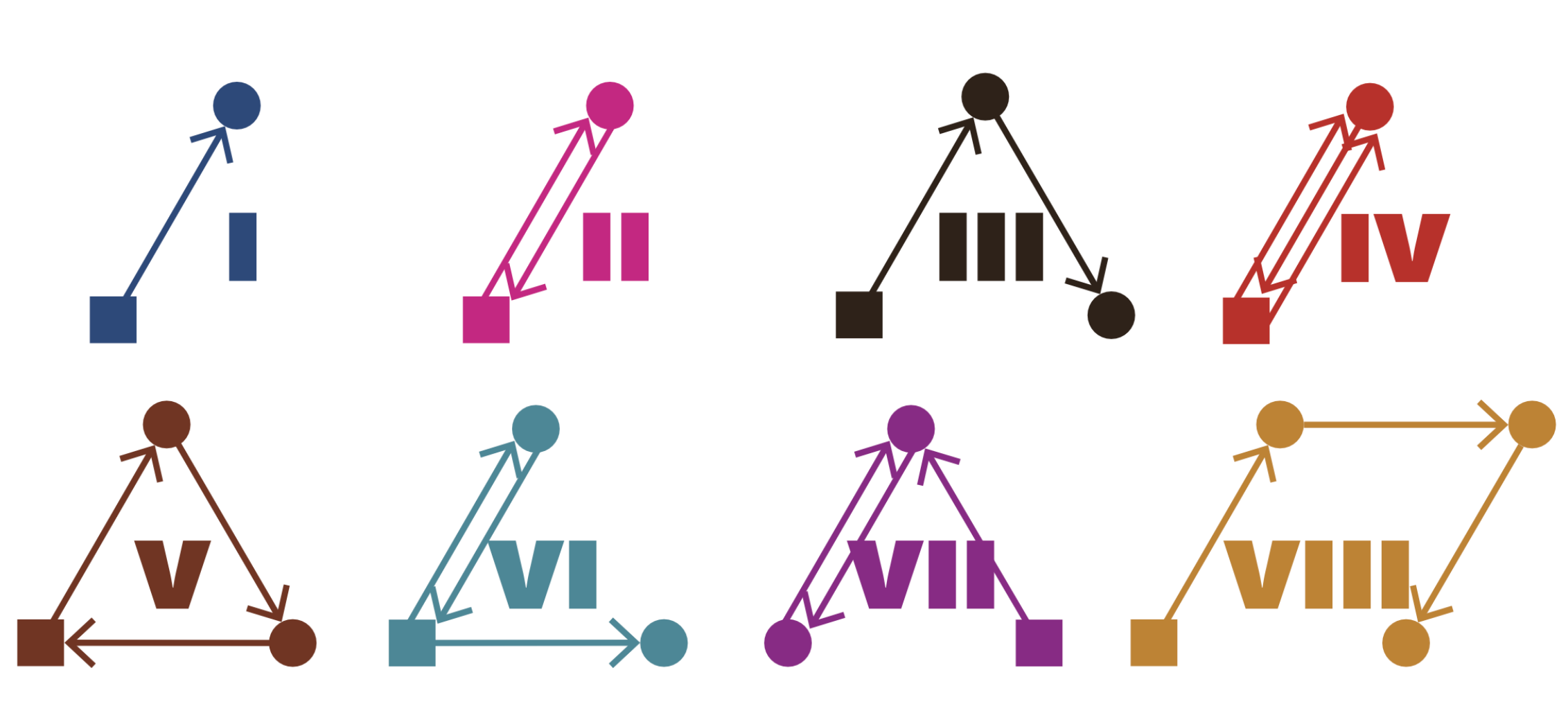}
    \caption{The eight weighted random-walk motifs defined by \cite{picciolo2022weighted}; the square represents the starting node for the random walker. Each motif corresponds to a distinct pattern of node visits and transitions before absorption. The motifs are mutually exclusive and collectively exhaustive for paths of length up to three steps. Figure adapted from \cite{picciolo2022weighted}.}
    \label{fig:Motifs_3}
\end{figure}

Each realized walk is classified into one of the eight mutually exclusive weighted random-walk motifs introduced by \cite{picciolo2022weighted}. 
The classification depends on the number of effective steps completed before absorption and on whether the walk revisits already visited nodes or reaches a previously unvisited node. Paths that terminate immediately in the sink (i.e., after a single transition) are excluded by construction. The possible motifs are illustrated in Fig.\ref{fig:Motifs_3}.

%\ross{Questo lo direi solo dopo aver mostrato i primi risultati.Toglierei tutto da "In this work..."alla fine. Non possiamo parlare di motifs profile senza aver ancora mostrato nulla.} \edo{Ok, tolto. Però in questa maniera nei metodi non c'è traccia di motif profiles, motif profiles per comunità di partenza e Jensen-Shannon distance, che invece sono strumenti che usiamo per analizzare i risultati. } 

%In this work, we focus on Motifs~1, 3, and 8, which represent progressively longer chain-like pathways in the random-walk construction. For ease of reference, we refer to Motif~1 as the \emph{Single-step attention chain}, to Motif~3 as the \emph{Two-step chain}, and to Motif~8 as the \emph{Extended chain}. These motifs will be used to compare short- and longer-range attention structures in the network.For descriptive purposes, we also consider the distribution of motifs generated by paths originating from each community (motif profiles), and compare communities based on the similarity of these distributions using Jensen--Shannon distances.

\subsection{Final-jump transitions}

The main analyses in this section focus on \emph{final-jump} transitions, namely the mapping from the starting node of a realized walk to its terminal node in the original network. 

If a realized walk visits the node sequence
\begin{equation}
\label{def:ell}
 u_0 \to u_1 \to \cdots \to u_\ell,
\end{equation}
with $\ell \in \{1,2,3\}$ before absorption, we summarize it at the community level as
\begin{equation}
 c(u_0) \to c(u_\ell),
\end{equation}
where $c(\cdot)$ denotes the community label.

This representation isolates the community destination reached after a multi-step retweet pathway and is therefore aligned with the goal of quantifying cross-community attention at different path lengths.

\subsection{From node paths to community pathways}\label{ssec:path2path}

Node-level paths are mapped to community-level transitions by considering only the communities of their endpoints. This yields a community--community representation of higher-order attention pathways.

Formally, for each motif $m$, we define a community--community matrix of path counts.
$\mathbf{W}^{\mathrm{paths},(m)}$, whose entries $w^{\mathrm{paths},(m)}_{c' c}$ counts the number of paths of motif $m$ starting in the community $c$ and ending in the community $c'$.
In the following, we work with these absolute path counts, which quantify the total mass of realized paths connecting pairs of communities for each motif $m$.

To quantify the incoming cross-community attention received by community $c$ through paths of motif $m$, we define the incoming spillover of community $c$ as
\begin{equation}\label{eq:save_maggioni}
    s^{\mathrm{paths},(m)}_{\mathrm{in}, c} = \sum_{c'\neq c} w^{\mathrm{paths},(m)}_{c' c},
\end{equation}
and the corresponding share as
\begin{equation}
    \sigma^{\mathrm{paths},(m)}_{\mathrm{in}, c} = \frac{s^{\mathrm{paths},(m)}_{\mathrm{in}, c}}{\sum_{c'} s^{\mathrm{paths},(m)}_{\mathrm{in}, c'}}.
\end{equation}
This quantity captures the global allocation of incoming attention across communities for motif $m$, normalized by the total mass of paths exchanged between different communities. It therefore allows us to compare how that allocation changes across path lengths.

Similarly, we define the within-community persistence of attention for community $c$ under motif $m$ as
\begin{equation}
    s^{\mathrm{paths},(m)}_{\mathrm{within}, c} = w^{\mathrm{paths},(m)}_{cc},
\end{equation}
and the corresponding share as
\begin{equation}
    \sigma^{\mathrm{paths},(m)}_{\mathrm{within}, c} = \frac{s^{\mathrm{paths},(m)}_{\mathrm{within}, c}}{\sum_{c'} s^{\mathrm{paths},(m)}_{\mathrm{within}, c'} + \sum_{c'} s^{\mathrm{paths},(m)}_{\mathrm{in}, c'}} = \frac{w^{\mathrm{paths},(m)}_{cc}}{\sum_{c,c'} w^{\mathrm{paths},(m)}_{c c'}}.
\end{equation}
This quantity captures the share of attention that remains within community $c$ for motif $m$, relative to the total attention mass generated by that motif.

It should be noticed that the two share, i.e. $\sigma^{\mathrm{paths},(m)}_{\mathrm{in}, c}$ and $\sigma^{\mathrm{paths},(m)}_{\mathrm{within}, c}$, have different normalization, as they capture distinct aspects of attention allocation. The incoming share $\sigma^{\mathrm{paths},(m)}_{\mathrm{in}, c}$ is normalized over cross-community flows only, as it is designed to measure how attention is redistributed across different communities. By contrast, the within-community share $\sigma^{\mathrm{paths},(m)}_{\mathrm{within}, c}$ is normalized over the total volume of paths, including both within- and cross-community contributions, so as to quantify how much of the overall attention remains internal to each community.

\paragraph{Comparison with community size and direct attention.}
Both $\sigma^{\mathrm{paths},(m)}_{\mathrm{in}, c}$ and $\sigma^{\mathrm{paths},(m)}_{\mathrm{within}, c}$ might reflect not only higher-order pathways, but also structural properties of the communities, especially their size and their first-order retweet connectivity. To account for this, we compare the higher-order quantities with two structural representations of the same network: a \emph{size-only} reference and a \emph{first-order} reference. This comparison makes it possible to evaluate how multi-step pathways reweight the distribution of attention relative to simpler structural descriptions, without relying on probabilistic or null-model-based expectations.

%\subsection{Size-only reference}
\paragraph{Size-only reference}
As a first comparison, we introduce a simple size-based reference to account for the effect of community size on attention allocation. Larger communities tend to receive a higher share of attention simply because they contain more nodes and therefore more potential targets of retweet interactions. Let $|c|$ denote the number of nodes in community $c$. We define the size-based reference distribution as
\begin{equation}
    \sigma^{\mathrm{size}}_{c} = \frac{|c|}{\sum_{c'} |c'|}= \frac{|c|}{N}.
\end{equation}
%The denominator $\sum_{c'} |c'| = N$ corresponds to the total number of nodes in the network.
This quantity represents the share of attention that community $c$ would receive under a purely size-driven allocation, in which attention is distributed uniformly across nodes irrespective of network structure.

The same size-based quantity is used as a descriptive reference for both incoming attention and within-community persistence. 
Comparing $\sigma^{\mathrm{paths},(m)}_{\mathrm{in}, c}$ and $\sigma^{\mathrm{paths},(m)}_{\mathrm{within}, c}$ with $\sigma^{\mathrm{size}}_{c}$ therefore provides a simple diagnostic of how much higher-order attention patterns may reflect community size alone. %This reference is purely descriptive and does not rely on probabilistic or null-model assumptions.
%\edo{Ho tolto (commentando) la parte che faceva riferimento ai null-models.}

\subsection{First-order reference}

We next consider a first-order reference obtained by aggregating direct retweet links between communities. Formally, we define a community--community matrix $\mathbf{W}^{\mathrm{links}}$, whose entry $w^{\mathrm{links}}_{c c'}$ accounts for the sum of direct weighted links from users in community $c$ to users in community $c'$.
%\begin{equation}
%W^{\mathrm{links}} = \bigl(w^{\mathrm{links}}_{c_0 c_1}\bigr),
%\end{equation}
As for the paths, we work with absolute retweet counts.

We define the incoming spillover of community $c$ under direct links as
\begin{equation}
    s^{\mathrm{links}}_{\mathrm{in}, c} = \sum_{c'\neq c} w^{\mathrm{links}}_{c' c},
\end{equation}
and the corresponding share as
\begin{equation}
    \sigma^{\mathrm{links}}_{\mathrm{in}, c} = \frac{s^{\mathrm{links}}_{\mathrm{in}, c}}{\sum_{c'} s^{\mathrm{links}}_{\mathrm{in}, c'}}.
\end{equation}
This quantity captures the global allocation of incoming attention across communities under the first-order representation.

Similarly, we define within-community persistence under direct retweet interactions as
\begin{equation}
    s^{\mathrm{links}}_{\mathrm{within}, c} = w^{\mathrm{links}}_{cc},
\end{equation}
and the corresponding share as
\begin{equation}
    \sigma^{\mathrm{links}}_{\mathrm{within}, c} = \frac{s^{\mathrm{links}}_{\mathrm{within}, c}}{\sum_{c'} s^{\mathrm{links}}_{\mathrm{within}, c'} + \sum_{c'} s^{\mathrm{links}}_{\mathrm{in}, c'}} = \frac{w^{\mathrm{links}}_{cc}}{\sum_{c,c'} w^{\mathrm{links}}_{c c'}}.
\end{equation}
This quantity captures the share of attention that remains within community $c$ relative to the total attention mass under direct retweet interactions. Please note that, following the same rationale as in Subsection~\ref{ssec:path2path}, the normalization for $\sigma^{\mathrm{links}}_{\mathrm{in}, c}$ and $\sigma^{\mathrm{links}}_{\mathrm{within}, c}$ are different.

In summary, we place higher-order pathways and simpler structural references on a common footing by expressing them as destination-based shares. In practice, we compare $\sigma^{\mathrm{paths},(m)}_{\mathrm{in}, c}$ with the corresponding reference shares $\sigma^{\mathrm{links}}_{\mathrm{in}, c}$ and $\sigma^{\mathrm{size}}_{c}$, and likewise compare $\sigma^{\mathrm{paths},(m)}_{\mathrm{within}, c}$ with their first-order and size-based counterparts. The goal is not to match path lengths across representations, but to assess how multi-step pathways reweight the distribution of attention relative to simpler structural descriptions. These comparisons are descriptive and do not rely on probabilistic or null-model assumptions.

\subsection{Comparison with references}

To quantify how higher-order pathways reweight attention relative to the reference structures introduced above, we compare both the incoming-attention distributions $\sigma^{\mathrm{paths}, (m)}_{\mathrm{in}, c}$ and the within-community persistence measures $\sigma^{\mathrm{paths},(m)}_{\mathrm{within}, c}$ with their corresponding reference values.

\paragraph{Log-ratio comparison.}
For a generic reference $b \in \{\mathrm{size}, \mathrm{links}\}$, we define
\begin{equation}
\mathrm{log\_ratio}^{b, (m)}_{c} = \log \left( \frac{\sigma^{\mathrm{paths},(m)}_{\mathrm{in}, c}}{\sigma^{b}_{\mathrm{in}, c}} \right).
\end{equation}
$\mathrm{log\_ratio}$ captures whether a community is over- or under-represented in higher-order pathways relative to size alone or direct retweet connectivity: positive values indicate that community $c$ receives a larger share of attention under motif $m$ than under the chosen reference, while negative values indicate a lower share. We opted for the log to highlight even minimal differences between $\sigma^{\mathrm{paths},(m)}_{\mathrm{in}, c}$ and  $\sigma^{b}_{\mathrm{in}, c}$.\\ 
%\fab{[Anziché chiamare delle variabili come il nome di una variabile di codice, avrei usato un qualcosa di più mathematical friendly (una lettera greca, latina, whatever…).]}

\paragraph{Log-ratio for persistence.}
We define the analogous quantity for within-community persistence as
\begin{equation}
    \mathrm{log\_ratio}^{b, (m)}_{\mathrm{within}, c} = \log \left( \frac{\sigma^{\mathrm{paths},(m)}_{\mathrm{within}, c}}{{\sigma}^{b}_{\mathrm{within}, c}} \right).
\end{equation}
In the case of the size-only reference, this compares the persistence share of a community with its normalized size; in the case of the first-order reference, it compares persistence with the fraction of direct retweet interactions that remain within the same community. Positive values indicate higher-than-reference persistence, while negative values indicate lower-than-reference persistence. 

\paragraph{Differences across path lengths.}
To compare how reference-adjusted attention changes between a couple of different motifs $m$ and $n$, we define
\begin{equation}
\Delta^{(m-n)}_{c}= \log\!\frac{\sigma^{(m)}_{\mathrm{in}, c}}{\sigma^{(n)}_{\mathrm{in}, c}}.
\end{equation}
Positive values indicate that community $c$ becomes relatively more over-represented as an endpoint of motif $m$ than of motif $n$. Conversely, negative values indicate that its prominence is stronger in motif $n$ than in motif $m$. Please note that,  for a given reference $b \in \{\mathrm{size}, \mathrm{links}\}$, $\Delta^{(m-n)}_{c}$ can be expressed as the difference of log\_ratios:
\begin{equation}
    \Delta^{(m-n)}_{c}= \log\!\frac{\sigma^{\mathrm{paths},(m)}_{\mathrm{in},c}}{\sigma^{\mathrm{paths},(n)}_{\mathrm{in},c}}
    = \log\!\frac{\sigma^{\mathrm{paths},(m)}_{\mathrm{in},c}}{\sigma^{b}_{\mathrm{in},c}}
    - \log\!\frac{\sigma^{\mathrm{paths},(n)}_{\mathrm{in},c}}{\sigma^{b}_{\mathrm{in},c}}=\mathrm{log\_ratio}^{b, (m)}_{c} -\mathrm{log\_ratio}^{b, (n)}_{c} 
\end{equation}
%\fab{[Per simmetria, comunque lascerei sia a log\_ratio che a $\Delta$ il pedice \emph{in}, ma up to you.]}
We define the corresponding difference for persistence as
\begin{equation}
    \Delta^{(m-n)}_{\mathrm{within}, c} = \log\!\frac{\sigma^{(m)}_{\mathrm{within}, c}}{\sigma^{(n)}_{\mathrm{within}, c}}.
\end{equation}
This quantity captures how reference-adjusted within-community persistence changes between short and long pathways.\\

%\paragraph{Interpretation.}
Both the log-ratio and $\Delta^{(m-n)}_{c}$, together with their persistence counterparts, are descriptive quantities. The log-ratio measures the deviation of a motif-specific attention share from a chosen reference, whereas $\Delta^{(m-n)}_{c}$ summarizes how that deviation changes between short and long pathways. They do not represent probabilities or causal effects, but provide complementary views on how multi-step pathways reweight attention relative to simpler structural descriptions.

\subsection{Orientation: attention vs information flow}
As different representations of the network have different interpretations of the interactions taking place in the Twitter platform, we analyze the network under two edge orientations. In the \emph{attention} orientation, edges are directed from the retweeter to the original author, so that links represent the direction of attention and amplification and can also be interpreted as signals of endorsement (i.e., $i \to j$ indicates that $i$ amplifies $j$). In the \emph{reversed} orientation, edges are inverted {and correspond} to the conventional information-flow representation in which links point from the source of content to its spreaders. Since the sink is introduced to balance the excess of incoming strength relative to outgoing strength, reversing the edges changes how each node is connected to the sink. Consequently, the two orientations define distinct random-walk processes and do not necessarily produce symmetric motif distributions. Considering both orientations allows us to examine how higher-order pathways depend on the directional interpretation of retweet interactions.

The aim is to identify which communities emerge as structurally privileged endpoints of sequential amplification pathways. In the attention orientation, the focus is on the patterns that users describe when retweeting the main sources of content, while in the reversed representation, the focus is on the flow of information from the sources and the leaders of the debate towards the standard users. 

While the difference may seem merely philosophical, it is not. For higher-order pathways, the reversed orientation is not obtained by re-reading the same path matrices with exchanged margins. Instead, it requires constructing a new random-walk process on the edge-reversed network and generating a corresponding set of motif-specific paths and community-level path matrices.

Within this framework, the incoming-attention quantities introduced above (e.g., $s_{\mathrm{in},c}$ and $\sigma_{\mathrm{in},c}$) are complemented by their outgoing counterparts, defined as
\begin{equation}
    s^{(\cdot)}_{\mathrm{out},c} = \sum_{c'\neq c} w^{(\cdot)}_{c c'}, 
    \qquad
    \sigma^{(\cdot)}_{\mathrm{out},c} = \frac{s^{(\cdot)}_{\mathrm{out},c}}{\sum_{c'} s^{(\cdot)}_{\mathrm{out},c'}}
\end{equation}
where $(\cdot)$ denotes any of the matrices introduced above (paths or links). These quantities are defined in direct analogy with the incoming case, simply by replacing ``in'' with ``out''.

%Under the reversed orientation, outgoing attention in the original network corresponds to incoming attention, thereby providing a complementary directional view of the same structural patterns. In the case of the first-order community matrix of direct retweet links, this correspondence amounts to exchanging rows and columns of the matrix, since reversing link directions swaps sources and destinations. 

%The reversed representation is therefore treated as a directional lens rather than a robustness check: it highlights how sequential pathways are structured when retweet relations are traced in the opposite direction, providing a complementary perspective on how attention pathways are organized. 

% ===================== RESULTS =====================
\section{Results}

\subsection{Global Network}

The original network is a directed, weighted retweet graph and is not fully connected. Its largest weakly connected component (LWCC) contains 308,243 nodes (92.11\% of all nodes) and 1,849,151 directed edges, while the remaining nodes are distributed across 9,742 much smaller weakly connected components. The LWCC is extremely sparse (density $1.946\cdot10^{-5}$), as is typical of large-scale online interaction systems. In the following, we focus on the LWCC only, which provides the connected substrate for all subsequent analyses. Table~\ref{tab:global_network_summary} reports a compact summary of the main global network properties discussed below.

\begin{table}[h]
\centering
\small
\renewcommand{\arraystretch}{1.15}
\begin{tabular}{l r}
\hline
\textbf{Property (LWCC)} & \textbf{Value} \\
\hline
Nodes ($N$) & 308,243 \\
Edges ($M$, directed) & 1,849,151 \\
Directed density & $1.946\cdot10^{-5}$ \\
Average in-degree ($\bar{k}^{in}=\bar{k}^{out}$) & 6.00 \\
Average in-strength ($\bar{s}^{in}=\bar{s}^{out}$) & 9.56 \\
Reciprocity (unweighted) & 0.0132 \\
Degree assortativity (out$\to$in) & $-0.1306$ \\
Degree assortativity (out$\to$out) & $0.1329$ \\
Degree assortativity (undirected projection) & $-0.1569$ \\
Average clustering (undirected projection) & 0.0550 \\
\hline
\end{tabular}
\caption{Summary of global network properties for the largest weakly connected component (LWCC).}
\label{tab:global_network_summary}
\end{table}

The average in-degree (which equals the average out-degree in any directed network) is 6.00, while the average weighted in-strength (and out-strength) is 9.56. Degree and strength distributions are highly skewed: most nodes have few interactions, while a small number of hubs concentrate a large fraction of retweets, as shown in Figures \ref{fig:lwcc_degree_distribution} and \ref{fig:lwcc_strength_distribution} (Appendix). %\fab{\st{Such near power-law, heavy-tailed patterns are a standard feature of social networks and online communication platforms \mbox{\citep{barabasi2016network}.}} [È piuttosto noto.]}

In particular, 6.77\% of the nodes have zero out-degree (they do not retweet anyone, although they are retweeted at least once), while 79.57\% have zero in-degree (they are never retweeted, although they retweet at least once). This highlights a strongly asymmetric retweet structure: a large majority of users do not create any relevant content, whereas a much smaller fraction concentrates most incoming attention. In addition, 48.04\% of the nodes have out-degree equal to 1, while only 7.63\% have in-degree equal to 1, indicating that many users retweet only a single account. In contrast, few users are retweeted by just one account.

Reciprocity is fairly low, with only 1.32\% of edges being reciprocated. This is consistent with reading the original orientation as an attention-oriented representation of the retweet network, in which users typically direct attention toward a relatively small set of accounts without receiving reciprocal attention in return. Degree assortativity is negative in the directed configuration out$\to$in ($r=-0.1306$), indicating that highly active retweeters tend to direct attention toward accounts that already receive many retweets. For comparison, the undirected degree assortativity of the symmetrized network is also negative ($r=-0.1569$), while the out$\to$out configuration is moderately assortative ($r=0.1329$). Triadic closure, measured on the undirected projection of the network, is moderate (average clustering $C=0.055$), indicating the presence of local closure beyond purely tree-like amplification patterns.

Overall, the global retweet network exhibits a strongly asymmetric and broadcast-like structure, in which a small number of users receive a large share of attention while most users have limited visibility. This is reflected in heavy-tailed degree distributions, low reciprocity, moderate clustering, and disassortative mixing in the out$\to$in configuration, all of which point to a system where attention is concentrated on a small set of highly visible users and propagated outward through largely non-reciprocal interactions. 

\paragraph{Intermediate Network Objects.}
Before turning to the community-level results, we briefly summarize the intermediate network objects. Starting from the global retweet network, we extract a verified--unverified bipartite graph that retains only cross-status interactions (unique pairs), yielding 432,131 verified--unverified links between 2,581 verified accounts and 149,141 unverified accounts. This bipartite view captures similarity in shared unverified audiences rather than direct verified-to-verified interactions.

\begin{figure}[h!]
    \centering
    \begin{subfigure}[b]{0.64\textwidth}
        \centering
        \includegraphics[height=0.30\textheight,keepaspectratio]{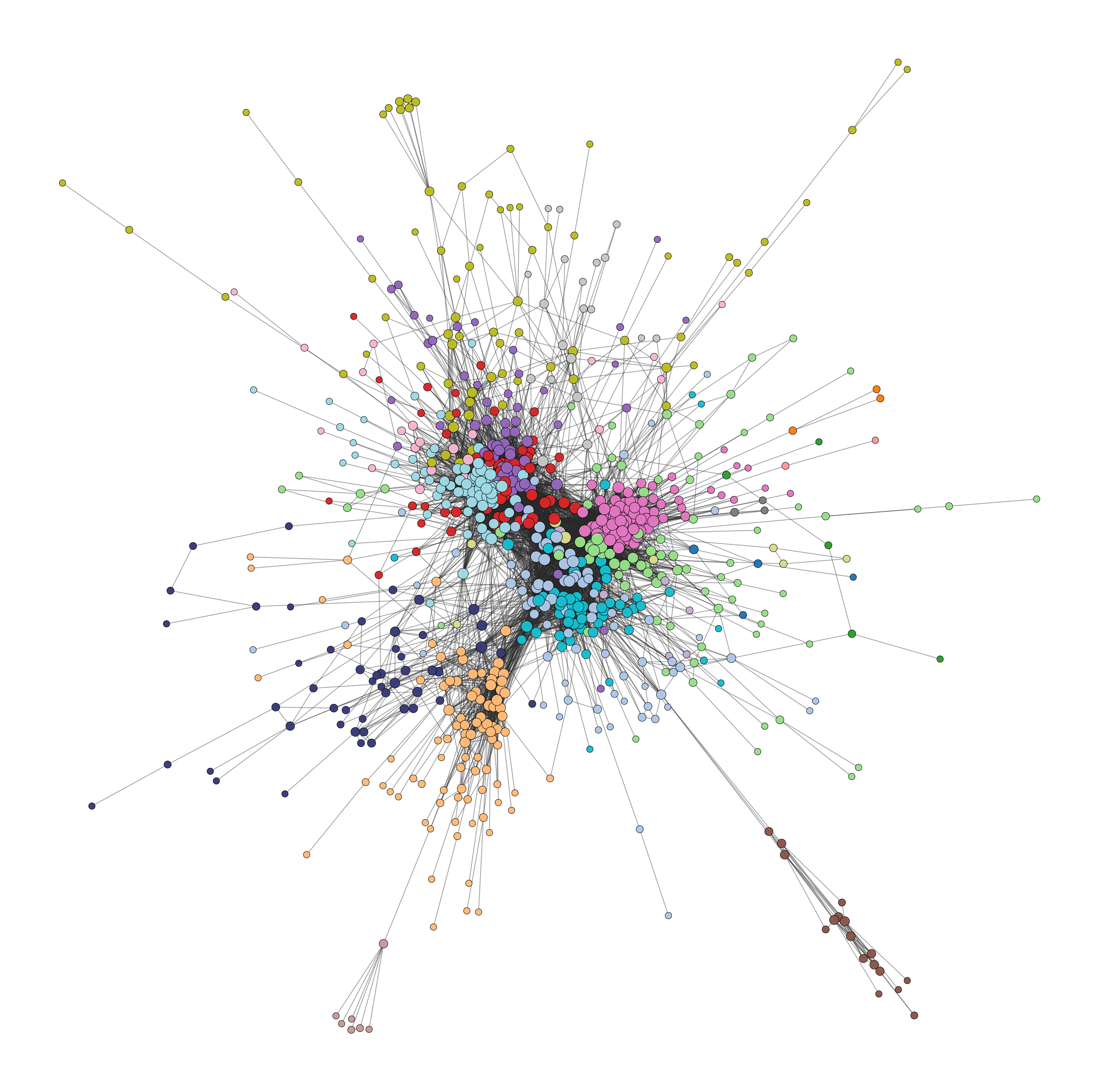}
        \caption{Largest WCC of the validated projection on verified users. 832 nodes and 11,058 edges. 21 communities.}
        \label{fig:ver_lcc}
    \end{subfigure}
    %\hfill
    \begin{subfigure}[b]{0.35\textwidth}
        \centering
        \includegraphics[height=0.26\textheight,keepaspectratio]{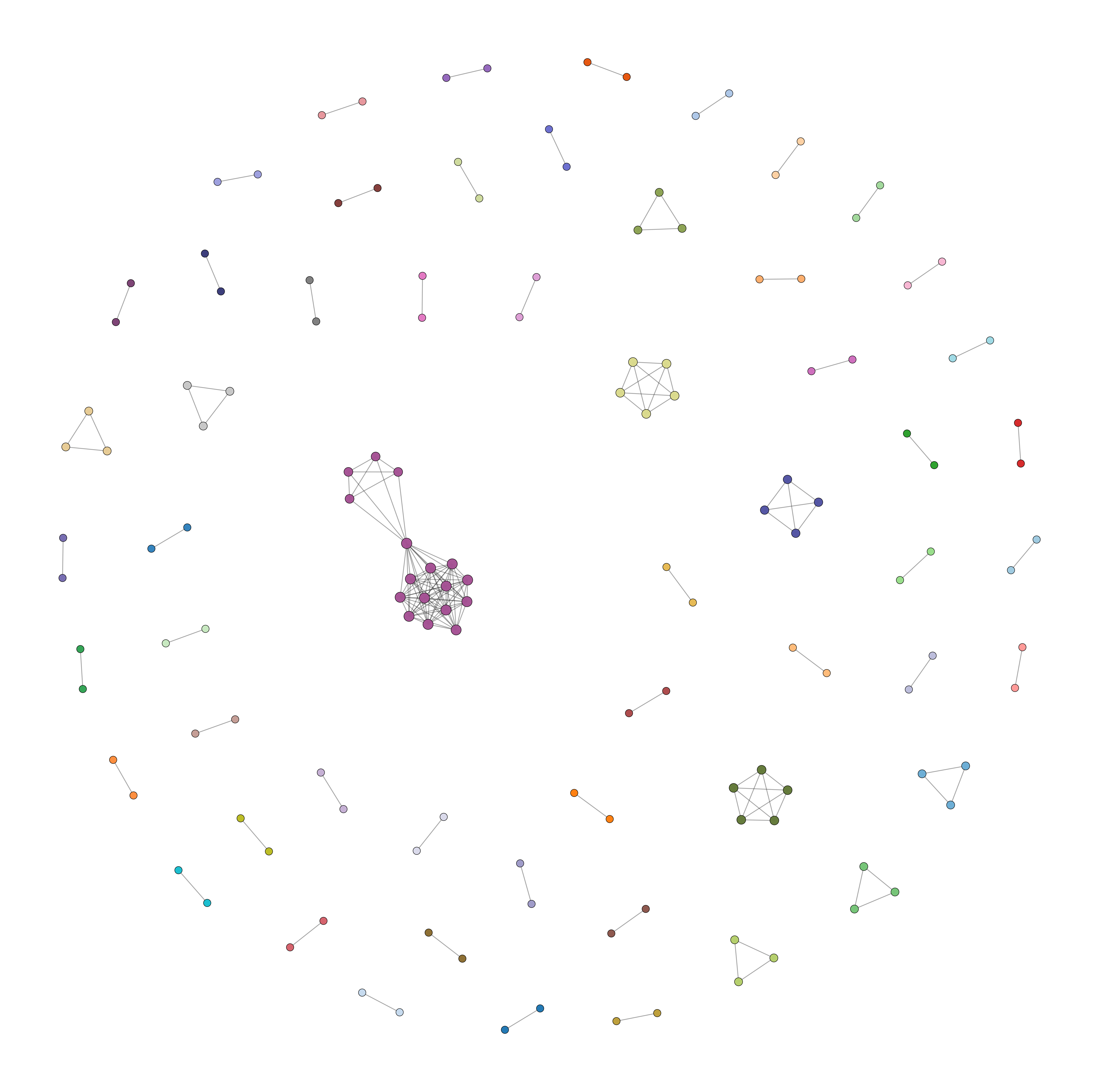}
        \caption{Remaining components. 141 nodes. \\56 communities.}
        \label{fig:ver_remaining}
    \end{subfigure}
    \caption{Connectivity of the validated projection on verified users. Nodes are colored by community. 973 nodes and 77 communities in total. Communities found with Leiden algorithm and resolution parameter $\gamma = 2$. }
    \label{fig:validated_projection_components}
\end{figure}

We then compute a statistically validated projection onto verified users. The validated projection contains 973 verified nodes and 11,235 edges and is not fully connected (its largest component includes 832 nodes and 11,058 edges, with 57 components overall). A graphical representation of the validated projection is shown in Figure \ref{fig:validated_projection_components}, where nodes are colored by community. Community detection is performed on the full validated projection, producing 77 verified communities under our reference specification (Leiden, $\gamma$=2; see Methods for stability checks). These verified-community labels are then propagated to the full network through weighted label propagation (verified seeds kept fixed), and results are subsequently restricted to the LWCC for downstream analyses. For interpretability, we canonicalize labels by descending community size within the LWCC (largest community = 1). In the remainder, we focus on the top 10 propagated communities, which together cover the large majority of LWCC users, plus an aggregated `Other' group 0.

\subsection{Community Landscape}
In Section~\ref{sec:methods}, we described the procedure used to identify the main communities from the retweet network of verified and unverified users. Relative to approaches based on ex ante political labels or manual classification (e.g. \cite{hanna2013partisan,badawy2018analyzing}), this pipeline remains comparatively agnostic, in the sense that it starts from observed retweet behavior and only subsequently assigns an interpretive meaning to the detected groups. The community labels used below should therefore be understood as heuristic and analytically convenient descriptions of the dominant profiles and narratives that emerge from each cluster, rather than as definitive or exhaustive classifications.

To characterize the detected communities, we qualitatively inspected their most central users, using both degree and strength as indicators of centrality, together with the recurring narratives associated with those users. What emerges is a heterogeneous landscape including both politically aligned clusters and less overtly political communities that nevertheless contribute to the broader public debate on the pandemic.

Several of the largest communities are clearly political. Community~1 is centered on a mixed set of medical-scientific figures, economists, and centre-left political actors, and appears to revolve around critiques of the government's handling of the pandemic, especially with regard to competence, transparency, and the management of the transition to the so-called Phase~2. Community~2 is also critical of the government, but is more strongly associated with right-wing and far-right opposition figures and with a more populist, sovereignist, and security-oriented framing of the crisis. By contrast, Communities~8 and~6 are more closely aligned with the governing coalition. Community~8 is dominated by centre-left and pro-European actors, whose discourse emphasizes institutional coordination, European solidarity, and support for government measures. Community~6 is largely associated with the Five Star Movement and allied media, and appears to combine support for government action with criticism of opposition-led regional administrations.

Other communities are organized less around partisan competition than around broader social, informational, or thematic functions. Community~7 brings together journalists, public figures, NGOs, and experts, and appears to focus on the social consequences of the pandemic, including inequality, marginalized groups, and criticism of both populist rhetoric and aspects of government communication. Community~9 is centered on digital-native media and technology-oriented actors, with narratives related to the infodemic, privacy, digital surveillance, and the broader digital transformation accelerated by the pandemic. Community~3 is structured around sports media and sports-related discussion, while Community~10 is more clearly linked to entertainment, lifestyle media, and simplified forms of public communication about everyday coexistence with COVID-19.

Finally, Communities~4 and~5 are the two clusters most clearly associated with the circulation of institutional and news-related information. Community~4 is dominated by mainstream news outlets and press agencies, while Community~5 contains many institutional and public-safety accounts, including ministries, scientific bodies, and emergency services. Both communities appear to function as important channels for the dissemination of updates, official guidance, and information on the evolution of the pandemic.

\begin{table}[!htbp]
    \centering
    \small
    \renewcommand{\arraystretch}{1.15}
    \begin{tabularx}{\textwidth}{c l X}
    \hline
    \textbf{Label} & \textbf{Short name} & \textbf{Brief description} \\
    \hline
    1 & Technocratic Reformists \& Experts & Mixed cluster of scientists (e.g., Burioni), centre-left local administrators, economists (e.g., Cottarelli), and liberal/centrist politicians. Discussion often focuses on critique of pandemic management and economic recovery tools (especially “phase 2” measures). \\
    
    2 & Right-wing Populists & Right-wing party ecosystem (e.g., Salvini, Meloni). Narrative centered on opposition to the incumbent government, populist framing, and recurrent Eurosceptic/anti-EU themes. \\
    
    3 & Sports \& COVID Bulletins & Sports-related discourse about the impact of COVID-19 on leagues, events, and the sport industry. Notable presence of data-driven daily updates (e.g., YouTrend-style bulletins), mixing sports news with neutral statistics. \\
    
    4 & News Media Mainstream & Legacy and online news outlets, newspapers, and press agencies. Predominantly news reporting and dissemination, comparatively less overt partisan framing (relative to explicitly political clusters). \\
    
    5 & Institutions \& Public Safety & Institutional accounts and public safety actors (e.g., Interior Ministry, police, firefighters). Mostly formal communications: guidelines, operational updates, public notices, and service information. \\
    
    6 & Five Star Movement Ecosystem & M5S political actors and allied media (e.g., Il Fatto Quotidiano). Content oriented around party messaging, government/policy disputes, and partisan interpretation of events. \\
    
    7 & Civil Society \& Social Impact & Public figures, NGOs, humanitarian associations, and some scientific experts. Emphasis on the social dimension of the crisis (poverty, migrants, inequalities), with critiques directed at both right-wing narratives (populism/misinformation) and government/centre-left (neglect of vulnerable groups, communication style). \\
    
    8 & Pro-EU Centre-left (PD) & Centre-left and pro-EU actors (PD area). Discourse highlights EU integrity and EU-level financial instruments for crisis management, and often stresses national coordination (e.g., integrated healthcare system), implicitly contrasting Euroscepticism and regional fragmentation. \\
    
    9 & Digital Transformation \& Debunking & Tech and digital-policy community focusing on COVID-driven digital transformations (and needed reforms), often paired with debunking/verification and informational corrections. \\
    
    10 & Entertainment \& Light News & Pop culture and entertainment ecosystem (TV personalities, lifestyle/gossip media, radio). COVID-19 is discussed in a lighter, more emotional or human-interest style rather than policy- or institution-centered framing. \\
    \hline
    \end{tabularx}
    \caption{Interpretive naming of the top-10 LWCC communities (labels follow LWCC size rank). These labels are heuristic summaries of the dominant profiles and narratives associated with each cluster and are used only to facilitate discussion of the structural results.}
    \label{tab:community_naming_top10}
\end{table}

\begin{table}[h]
    \centering
    \begin{tabular}{llrrrrr}
        \hline
        Community & Brief description & Size & Verified & Out-strength & In-strength & Within \\
        \hline
        1 \ \ \communityswatch{commone} & Technocratic Reformists \& Experts & 83,626 & 406 & 239,765 & 146,367 & 551,627 \\
        2 \ \ \communityswatch{commtwo} & Right-wing Populists & 66,453 & 209 & 146,619 & 97,778 & 1,035,269 \\
        3 \ \ \communityswatch{commthree} & Sports \& COVID Bulletins & 57,287 & 222 & 75,323 & 63,987 & 210,035 \\
        4 \ \ \communityswatch{commfour} & News Media & 45,843 & 383 & 135,687 & 155,831 & 218,530 \\
        5 \ \ \communityswatch{commfive} & Institutions \& Public Safety & 16,398 & 223 & 25,865 & 59,881 & 59,737 \\
        6 \ \ \communityswatch{commsix} & Five Star Movement & 10,414 & 195 & 53,754 & 44,838 & 139,065 \\
        7 \ \ \communityswatch{commseven} & Civil Society \& Social Impact & 7,988 & 82 & 7,016 & 53,622 & 11,653 \\
        8 \ \ \communityswatch{commeight} & Pro-EU Centre-left & 4,651 & 127 & 5,463 & 23,965 & 11,672 \\
        9 \ \ \communityswatch{commnine} & Digital \& Debunking & 3,039 & 99 & 3,864 & 13,799 & 5,926 \\
        10  \communityswatch{commten} & Entertainment \& Light News & 2,778 & 67 & 3,115 & 13,276 & 5,861 \\
        0 \ \ \communityswatch{commother} & Other & 9,766 & 340 & 8,757 & 31,884 & 22,286 \\
        \hline
    \end{tabular}
    \caption{Sizes of the 10 major communities retained for the random-walk analysis (LWCC only). Community~0 ("Other") aggregates all remaining communities. Here, out-strength and in-strength include only inter-community links; within reports intra-community links.}
    \label{tab:major_community_sizes_inter_within}
\end{table}

\newpage
\subsection{Global Motif Composition}

\iffalse
\begin{table}[!htbp]
    \centering
    \scriptsize
    \renewcommand{\arraystretch}{1.1}
    \begin{tabular}{lrrrrrrrr}
    \toprule
     & Motif 1 & Motif 2 & Motif 3 & Motif 4 & Motif 5 & Motif 6 & Motif 7 & Motif 8 \\
    \midrule
    Attention-flow (observed RW) & 0.911339 & 0.000085 & 0.074391 & 0.000155 &  0.000015 & 0.000141 & 0.000579 & 0.013295 \\
    %Attention-flow (theoretical) & 0.911285 & 0.000084 & 0.074474 & 0.000153 & 0.000016 & 0.000136 & 0.000574 & 0.013278 \\
    \bottomrule
    \end{tabular}
    \caption{Global motif distribution under the attention flow orientation. RW with $10^7$ tentative paths.}
    \label{tab:global_motif_profile_original}
\end{table}
\fi

Under the random-walk framework, the realized motif distribution is dominated by simple chain structures (Motif 1), with longer-chain motifs (Motifs 3 and 8) accounting for most of the remaining share. Motifs with repeated nodes (2, 4, 5, 6, 7) are negligible. 
%The observed motif distribution closely matches the theoretical expectation based on the transition matrix, confirming the consistency of the sampling procedure.

Unlike longer motifs, Motif 1 mainly reflects immediate absorption after a single retweet transition. Its frequency is therefore strongly influenced by the local imbalance between node in-strength and out-strength, which determines access to the sink in the random-walk construction.

\begin{figure}[!htbp]
    \centering
    \includegraphics[width=0.75\linewidth]{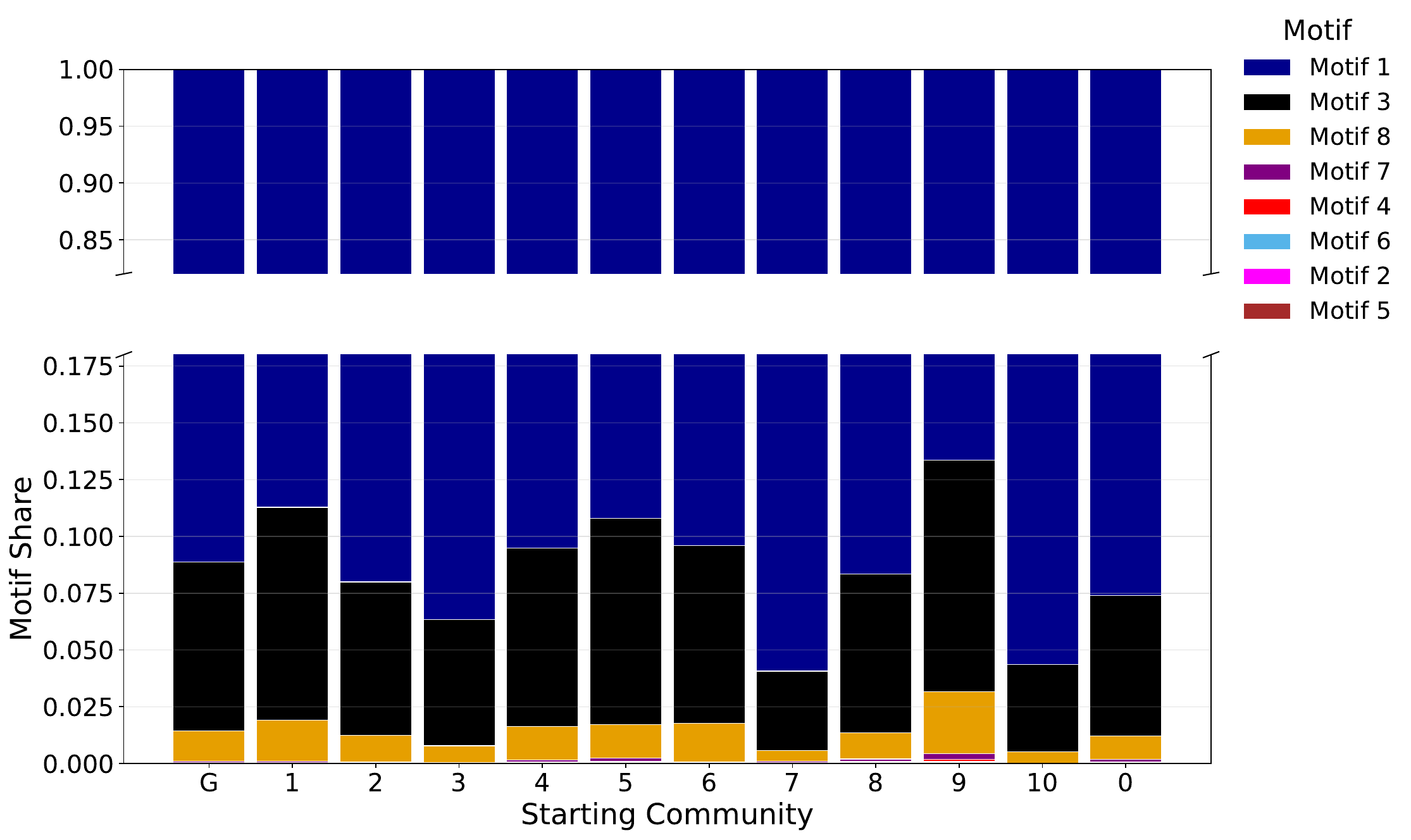}
    \caption{Motif composition by starting community as stacked bars with broken y-axis to expand low-share motifs while retaining the dominant Motif~1 mass. "G" indicates the global motif distribution (all paths). The top 10 communities are ordered by size rank. "0" aggregates all remaining communities.}
    \label{fig:motifs_by_start_ranked_stacked_brokeny}
\end{figure}

To further characterize how motif compositions differ across communities, we consider the motif profile of each starting community, defined as the distribution of random-walk motifs generated by paths originating from that community. The motif profiles of the top 10 communities are visualized as stacked bars in Figure~\ref{fig:motifs_by_start_ranked_stacked_brokeny}, where the y-axis is broken to expand low-share motifs while retaining the dominant Motif~1 mass. The full numerical values of these motif profiles are reported in Table~\ref{tab:motifs_by_start_comm_attention_top10}, in the Appendix~\ref{appendix:global_motifs}, together with alternative visualizations.

\begin{figure}[ht!]
        \centering
        \begin{subfigure}[b]{0.56\textwidth}
            \centering
            \includegraphics[width=\textwidth]{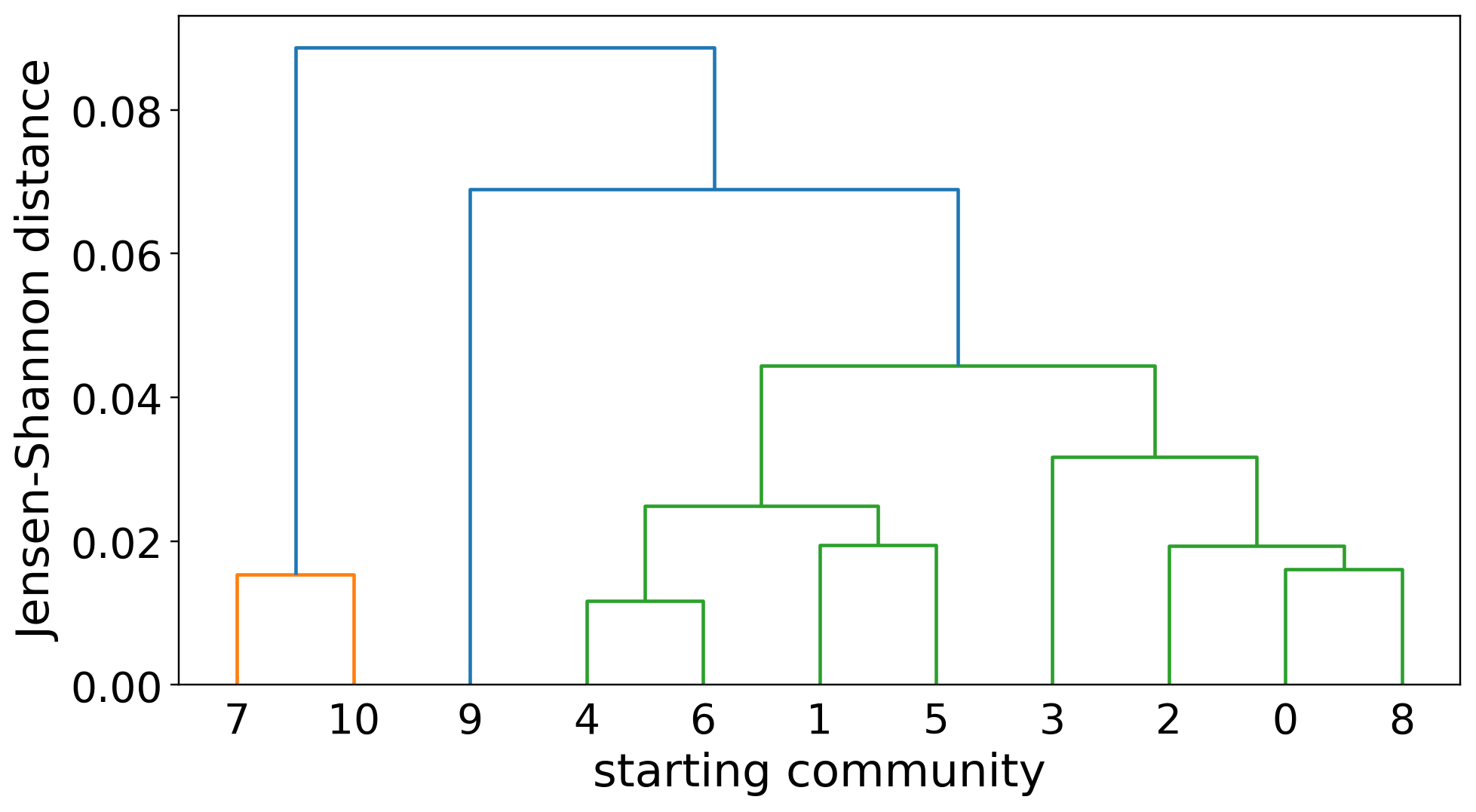}
            \caption{Hierarchical clustering (dendrogram).}
            \label{fig:js_dendrogram_original}
        \end{subfigure}
        \hfill
        \begin{subfigure}[b]{0.38\textwidth}
            \centering
            \includegraphics[width=\textwidth]{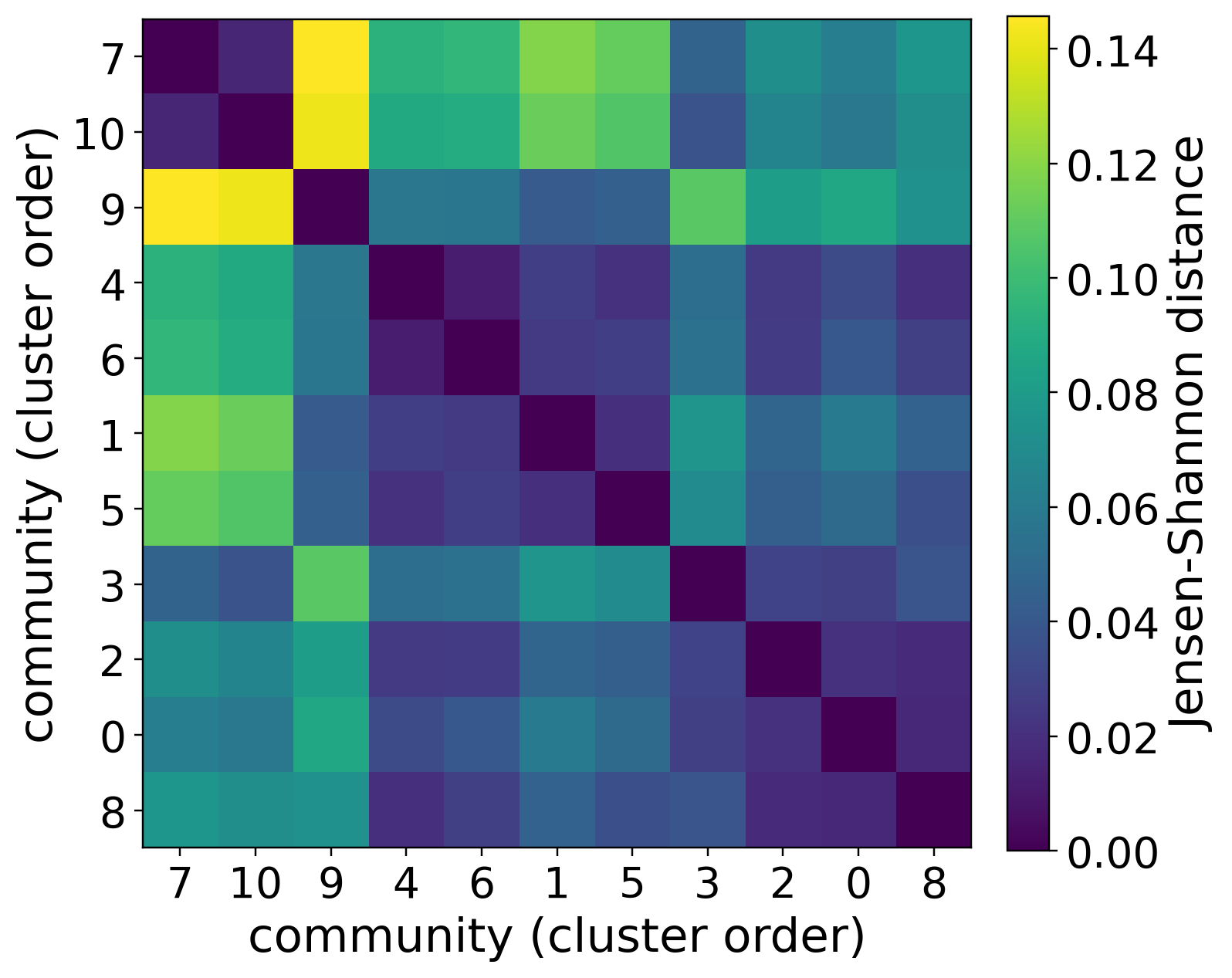}
            \caption{$d_{JS}$ matrix ordered by clustering.}
            \label{fig:js_heatmap_original}
        \end{subfigure}
        \caption{Community similarity in the attention network (original orientation) based on Jensen--Shannon distances between motif-profile distributions (computed by starting community).}
        \label{fig:js_motif_profiles_original}
\end{figure}

To quantify similarity between communities, we compute pairwise Jensen--Shannon distances between these motif profiles. Given two motif-profile distributions $\mathbf{p}$ and $\mathbf{q}$, their Jensen--Shannon distance is defined as $d_{\mathrm{JS}}(\mathbf{p},\mathbf{q})=\sqrt{\frac{1}{2}D_{\mathrm{KL}}(\mathbf{p}\Vert\mathbf{m})+\frac{1}{2}D_{\mathrm{KL}}(\mathbf{q}\Vert\mathbf{m})}$, where $\mathbf{m}=(\mathbf{p}+\mathbf{q})/2$~\citep{lin1991divergence,endres2003new}. $D_{\mathrm{KL}}$ denotes the Kullback--Leibler divergence~\citep{kullback1951information}. This distance is symmetric and remains finite when some components are zero, with lower values indicating more similar motif compositions. The resulting distance matrix is visualized in Figure~\ref{fig:js_motif_profiles_original}, together with a hierarchical clustering (dendrogram).

Overall, communities exhibit heterogeneous motif profiles, with some groups of communities sharing similar motif compositions. In particular, a subset of communities form tightly clustered groups (e.g., communities 7 and 10), indicating highly similar motif compositions, while others (such as community 9) appear more distinct. 

\subsection{Between and Within-Community Direct Connectivity}

We begin by describing the first-order structure of direct retweet interactions at the community level, which serves as a reference point for the higher-order analyses that follow. %Let \fab{$\mathbf{W}^{\mathrm{links}}$} denote the weighted community--community matrix of direct retweet links, and \fab{let $w^{\mathrm{links}}_{c c'}$ be its generic entry accounting for  the total number of retweets from community $c$ to community $c$} within the LWCC. 

\begin{figure}[h!]
    \centering
    \begin{subfigure}[b]{0.55\textwidth}
        \centering
        \includegraphics[width=\textwidth]{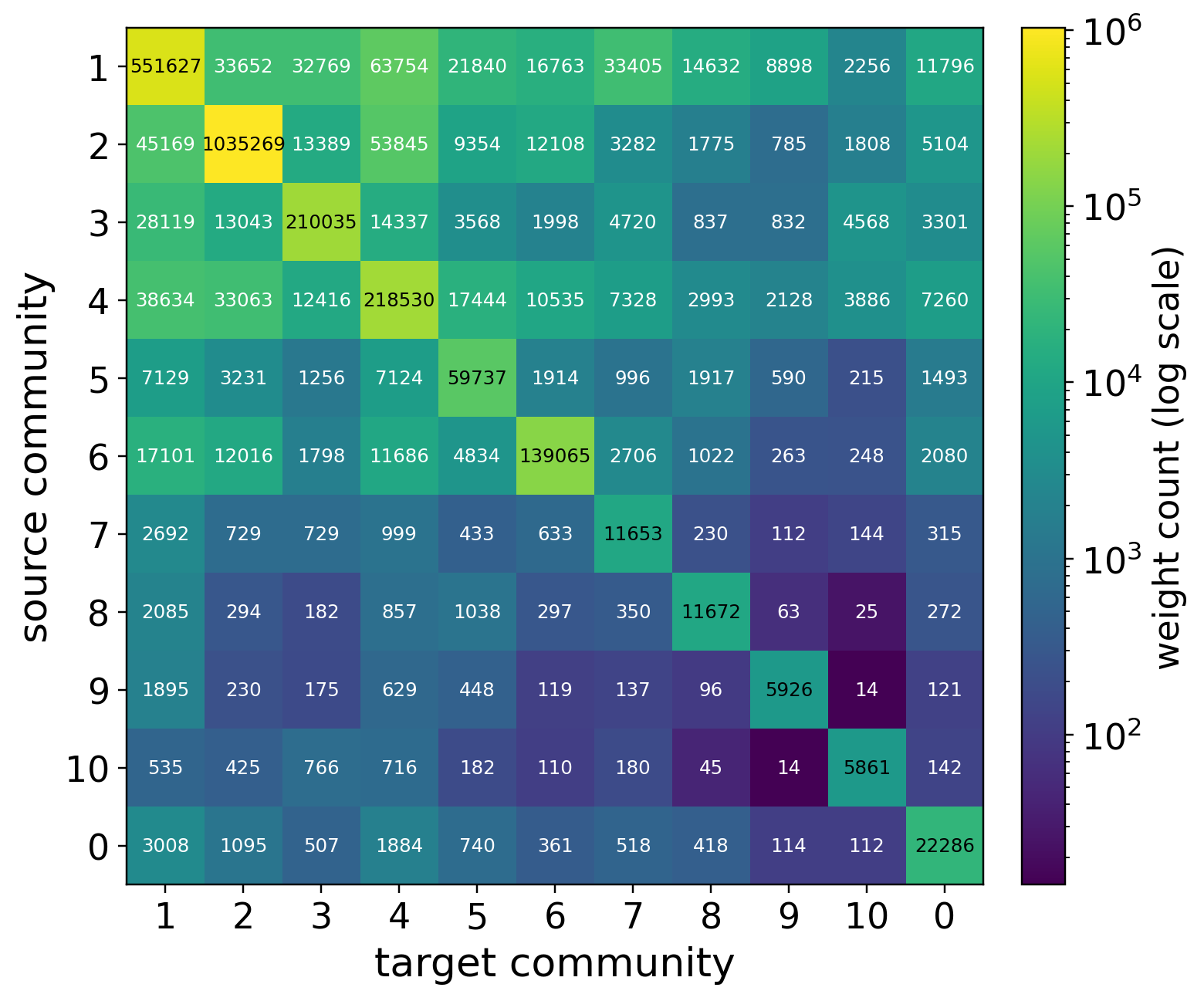}
        \caption{Raw inter-community connectivity (counts).}
        \label{fig:intercommunity_weighted_counts}
    \end{subfigure}
    \hfill
    \begin{subfigure}[b]{0.44\textwidth}
        \centering
        \begin{subfigure}[b]{\textwidth}
            \centering
            \includegraphics[width=0.9\textwidth]{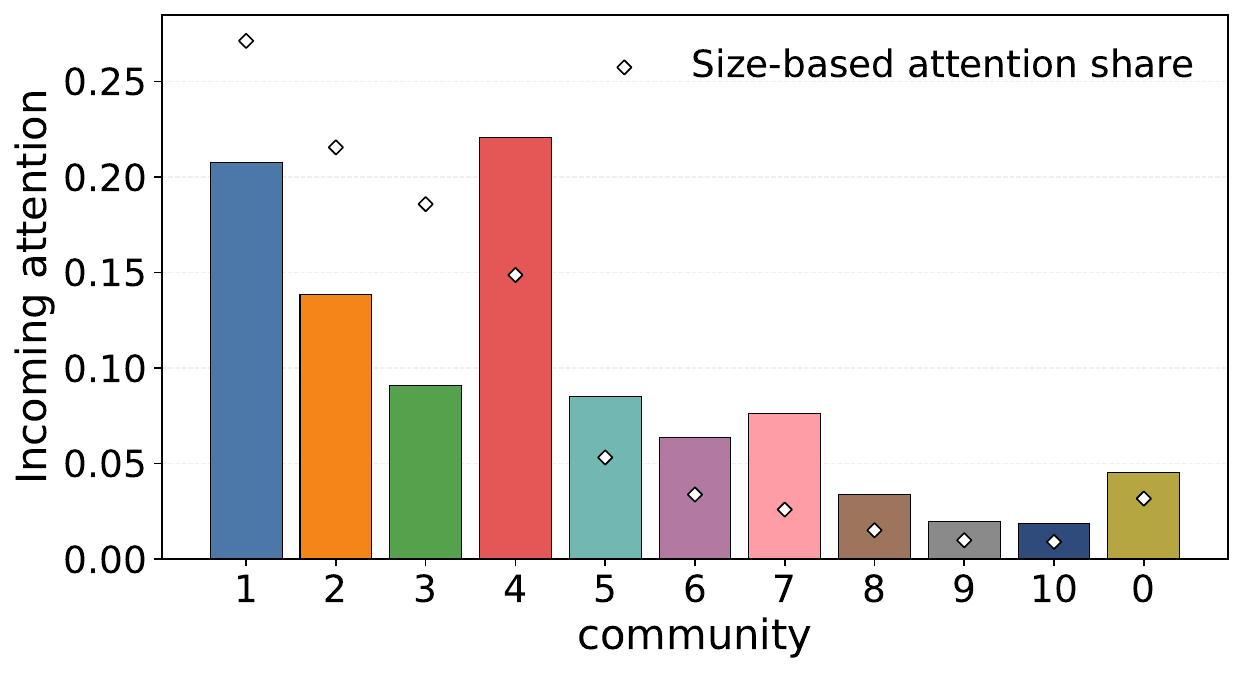}
            \caption{Incoming attention share $\sigma^{\mathrm{links}}_{\mathrm{in},c}$.}
            \label{fig:bar_sigma_links_in}
        \end{subfigure}
        \\
        \begin{subfigure}[b]{\textwidth}
            \centering
            \includegraphics[width=0.9\textwidth]{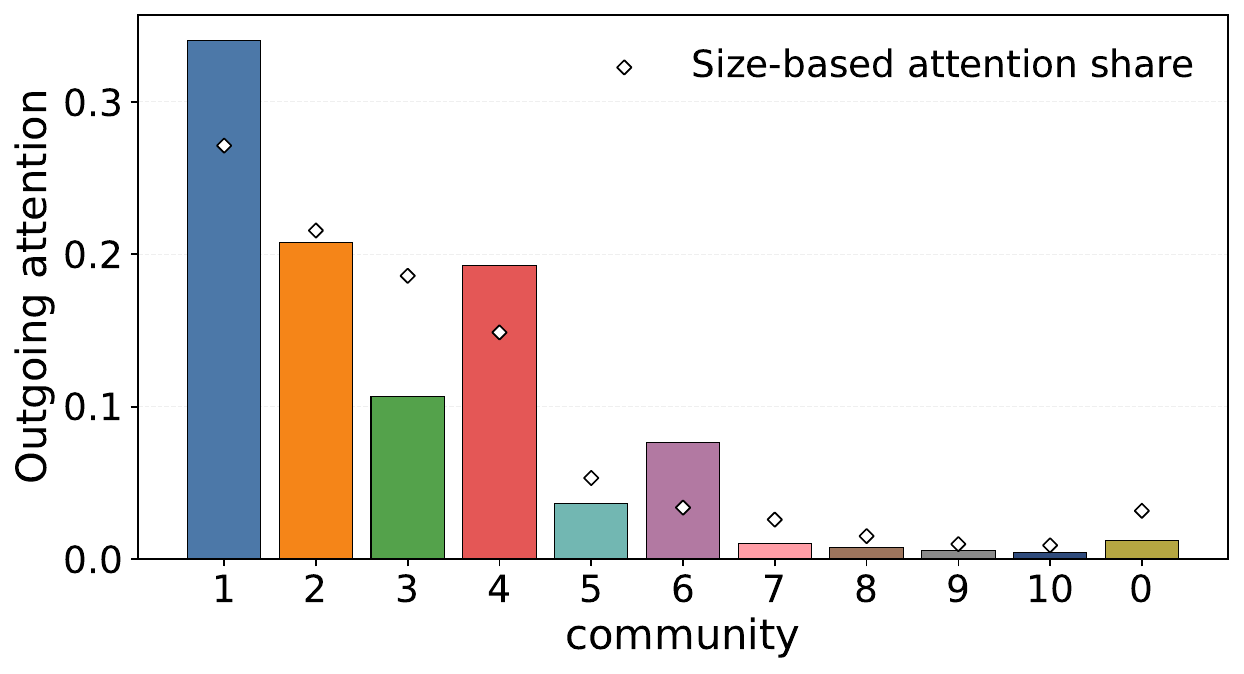}
            \caption{Outgoing attention share $\sigma^{\mathrm{links}}_{\mathrm{out},c}$.}
            \label{fig:bar_sigma_links_out}
        \end{subfigure}
    \end{subfigure}
\caption{First-order inter-community structure. Left: raw connectivity between communities. Right: distribution of incoming and outgoing attention shares across communities. The outgoing quantities are defined analogously to the incoming ones (with in$\rightarrow$out) and anticipate the reversed orientation discussed below.}    \label{fig:intercommunity_connectivity_combined}
\end{figure}

Figure~\ref{fig:intercommunity_connectivity_combined} summarizes the first-order inter-community structure. The heatmap shows the raw matrix of direct retweet counts, while the barplots report the distribution of incoming and outgoing attention shares across communities. 

For completeness, we also report outgoing attention across communities using the quantities introduced in the Methods ($s^{\mathrm{links}}_{\mathrm{out},c}$ and $\sigma^{\mathrm{links}}_{\mathrm{out},c}$). Under the reversed orientation, these correspond to direct incoming attention and therefore provide a complementary directional view of the same first-order structure.\\

We summarize cross-community attention received by each destination community through the incoming spillover quantities $\sigma^{\mathrm{links}}_{\mathrm{in}, c}$ introduced in the Methods. Together, they describe the first-order allocation of incoming attention across communities. As shown in Figures~\ref{fig:bar_sigma_links_in} and \ref{fig:bar_sigma_links_out}, incoming and outgoing attention broadly follow community size, with community~4 standing out as receiving and emitting more attention than would be suggested by size alone.

\begin{figure}[h!]
    \centering
    \begin{subfigure}[b]{0.48\textwidth}
        \centering
        \includegraphics[width=0.9\textwidth]{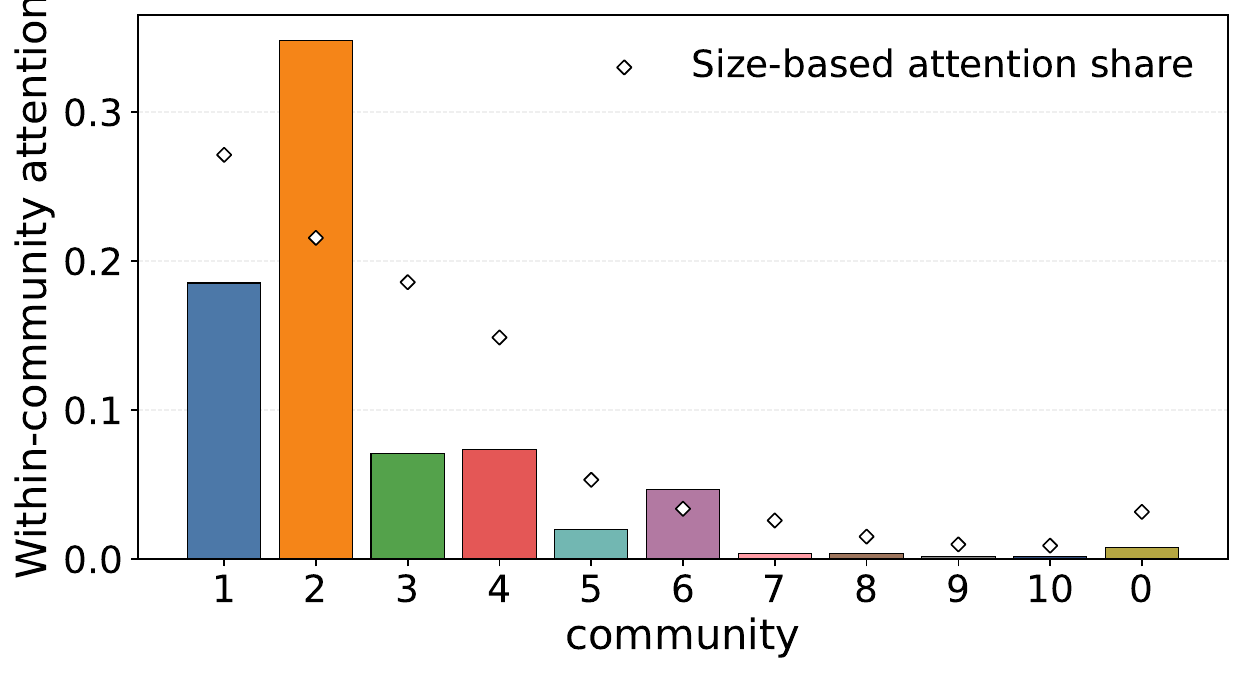}
        \caption{Within-community persistence, $\sigma^{\mathrm{links}}_{\mathrm{within},c}$.}
        \label{fig:bar_sigma_links_within}
    \end{subfigure}
    \hfill
    \begin{subfigure}[b]{0.48\textwidth}
        \centering
        \includegraphics[width=0.9\textwidth]{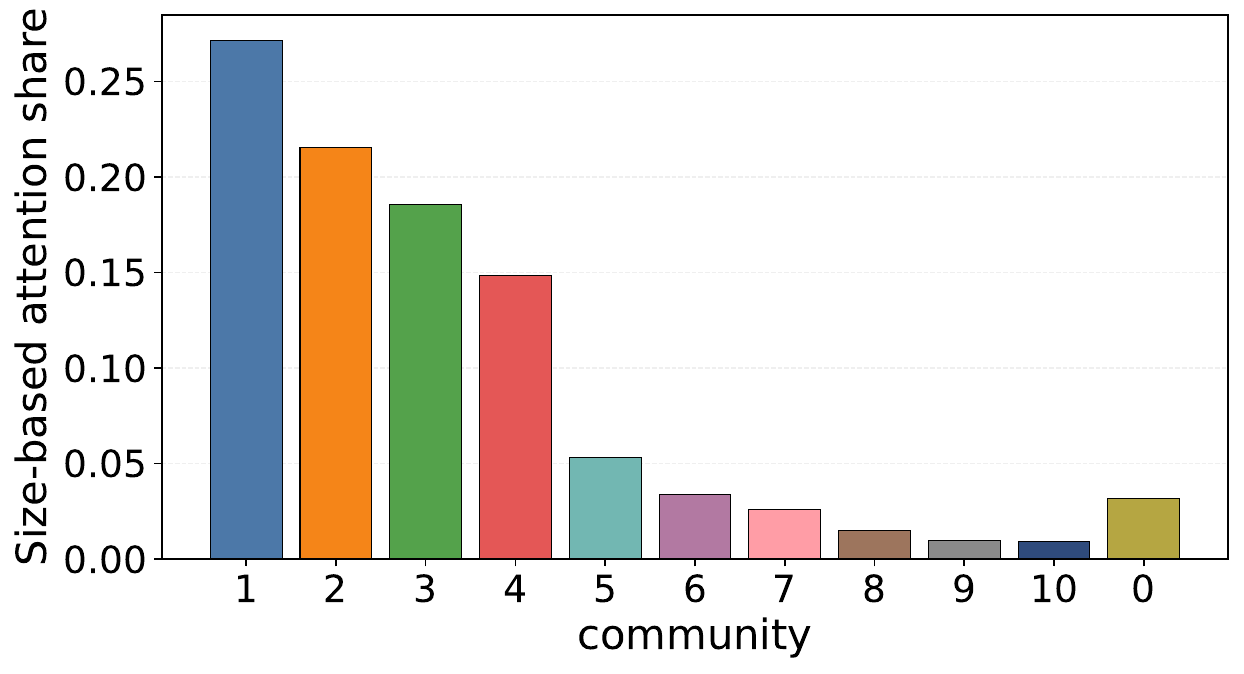}
        \caption{ Size-based reference, $\sigma^{\mathrm{size}}_{\mathrm{within},c}$.}
        \label{fig:bar_sigma_size_within}
    \end{subfigure}
    \caption{First-order within-community persistence across communities. Left: persistence under direct retweet interactions. Right: size-only reference, corresponding to normalized community size.}
    \label{fig:baseline_persistence}
\end{figure}

We also consider the within-community component of attention through the persistence quantities $ \sigma^{\mathrm{links}}_{\mathrm{within}, c} $. These capture the share of total attention that remains within a given community under direct retweet interactions and therefore provide a first-order reference value for within-community retention. Figure~\ref{fig:bar_sigma_links_within} shows that persistence is highly concentrated, with community~2 capturing the largest share of within-community attention.

For comparison, Figure~\ref{fig:bar_sigma_size_within} also reports the size-only reference, corresponding to normalized community size. This provides a coarse diagnostic of how attention and persistence relate to community size alone, as opposed to first-order retweet structure. For brevity, all subsequent comparisons using the size-only reference are reported in Appendix~\ref{appendix:community_diagnostics}, where they serve as a useful diagnostic comparison rather than as the main structural reference used in the main text.

\subsection{Community-level final-jump matrix}

We begin by examining the community-level structure of higher-order pathways through the final-jump matrices. The elements of the matrices $\mathbf{W}^{\mathrm{paths},(m)}$ count the number of random-walk paths of motif $m$ that start from a node in community $c$ and terminate in a node in community $c'$. In this sense, they represent the number of multi-step interaction sequences between communities, where the endpoint corresponds to the last node reached before being absorbed by the sink.
\begin{figure}[h!]
    \centering
    \includegraphics[width=0.99\linewidth]{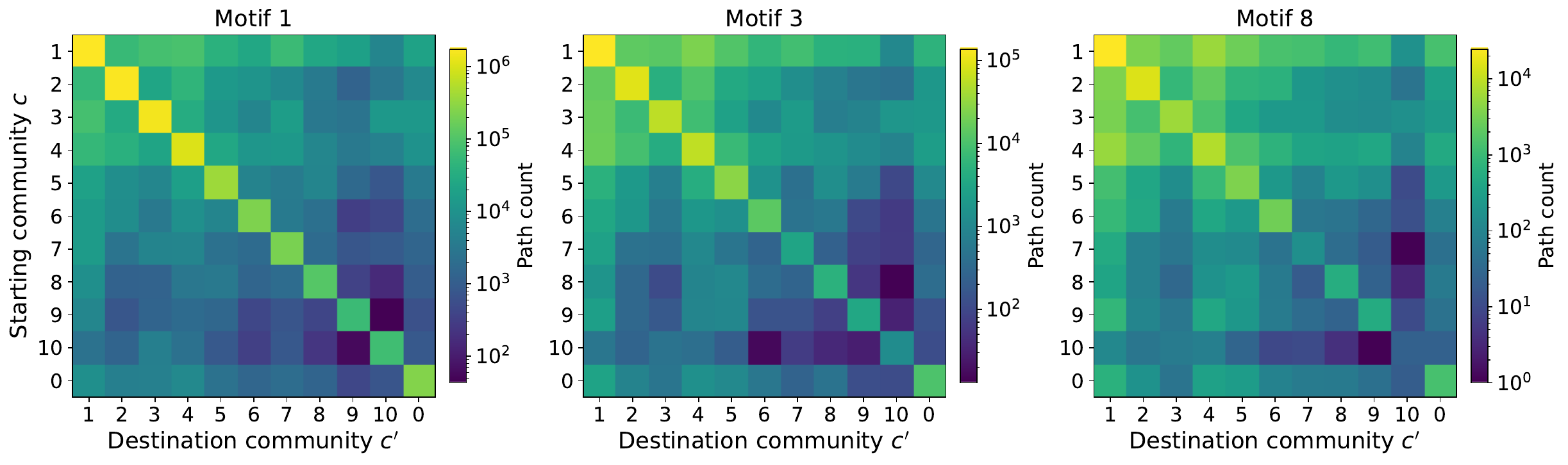}
    \caption{
    Community-level final-jump matrices (raw counts) for Motifs~1, 3, and 8. Each cell $(c,c')$ represents the number of paths starting in community $c$ and ending in community $c'$. Diagonal entries capture within-community persistence, while off-diagonal entries represent cross-community pathways. Color scales are logarithmic and defined separately for each motif.}
    \label{fig:final_jump_heatmaps}
\end{figure}

Across motifs, the matrices show a consistent pattern of strong within-community concentration combined with structured cross-community connectivity. Diagonal entries dominate for short paths (Motif~1), while longer motifs place progressively more mass on off-diagonal entries, see Fig.~\ref{fig:final_jump_heatmaps}. This suggests that, within the random-walk construction, longer motifs are associated with a lower relative concentration on within-community endpoints and a greater cross-community presence.

This observation motivates a more focused analysis of the diagonal component (within-community persistence) and the off-diagonal structure (cross-community spillover) in the following sections.

\subsection{Within-community persistence across motif length}

\begin{figure}[h!]
\centering
    \includegraphics[width=0.85\linewidth,trim={0 0 0 0.6cm},clip]{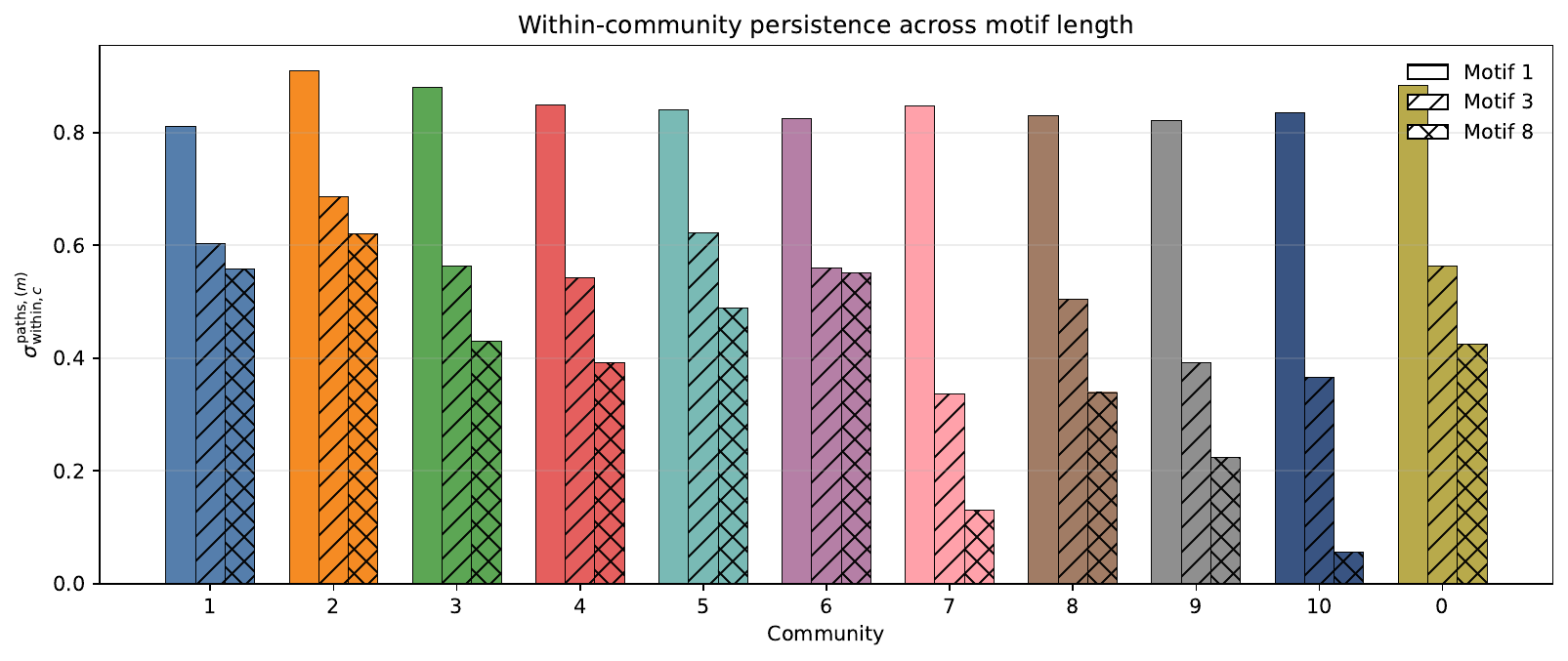}
    \caption{Within-community persistence across motif length. Bars report the share of realized paths that start and end in the same community for Motifs~1, 3, and 8.}
    \label{fig:persistence_raw_by_comm}
\end{figure}

Figure~\ref{fig:persistence_raw_by_comm} reports the motif-specific within-community persistence share, that is, the share of realized paths starting in community $c$ that also end in the same community ($c \to c$), for Motifs~1, 3, and 8.

Across communities, persistence tends to decrease as motif length increases. In descriptive terms, this means that longer realized motifs place relatively less mass on within-community endpoints.

However, this observation alone is not sufficient to characterize how attention is redistributed. Raw persistence levels are strongly influenced by structural factors such as community size and first-order connectivity patterns. We therefore evaluate persistence relative to a reference value in order to identify which communities retain more or less attention than suggested by the corresponding first-order community structure. We next compare observed persistence with the appropriate reference values.

\begin{figure}[h!]
    \centering
    \includegraphics[width=0.85\linewidth]{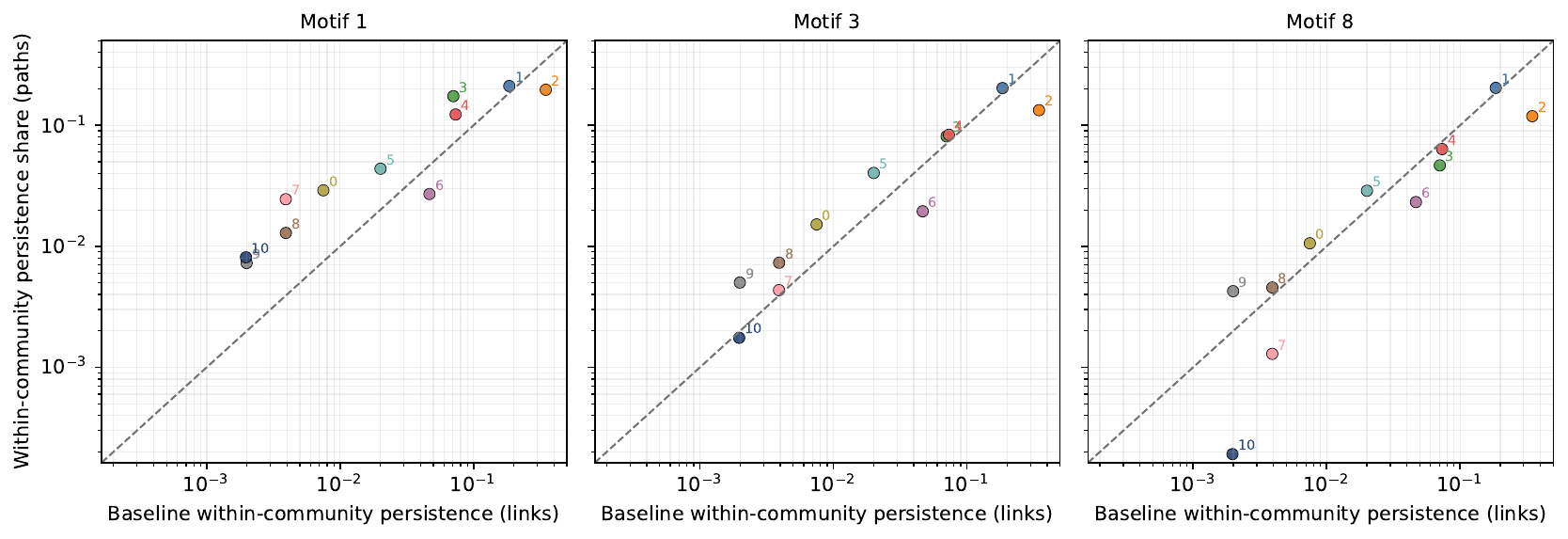}
    \caption{Observed versus first-order reference within-community persistence under $\mathbf{W}^{\mathrm{links}}$. Each point represents a community. The x-axis reports the first-order persistence share $\sigma^{\mathrm{links}}_{\mathrm{within},c}$, while the y-axis reports the corresponding persistence share derived from paths $\sigma^{\mathrm{paths},(m)}_{\mathrm{within},c}$. Both axes are shown on logarithmic scales. The dashed line indicates equality between observed and reference values.}   
    \label{fig:persistence_vs_links}
\end{figure}

To account for these factors, we compare observed persistence with its first-order counterpart. Figure~\ref{fig:persistence_vs_links} reports observed versus reference persistence under $\mathbf{W}^{\mathrm{links}}$, separately for Motifs~1, 3, and 8.

The comparison reveals that observed persistence broadly aligns with the first-order reference, with most communities lying close to the diagonal. At the same time, deviations from that reference are clearly heterogeneous across communities. While some communities remain close to the first-order connectivity pattern, others exhibit higher or lower within-community retention systematically.

This result suggests that the decline in persistence observed in the raw data is not simply a uniform scaling effect, but reflects community-specific patterns of attention retention.

For completeness, the corresponding comparisons using the size-only reference, including both scatterplots and motif-length evolution, are reported in Appendix~\ref{appendix:community_diagnostics} as a coarse diagnostic.

\begin{figure}[h!]
    \centering
    \includegraphics[width=0.85\linewidth]{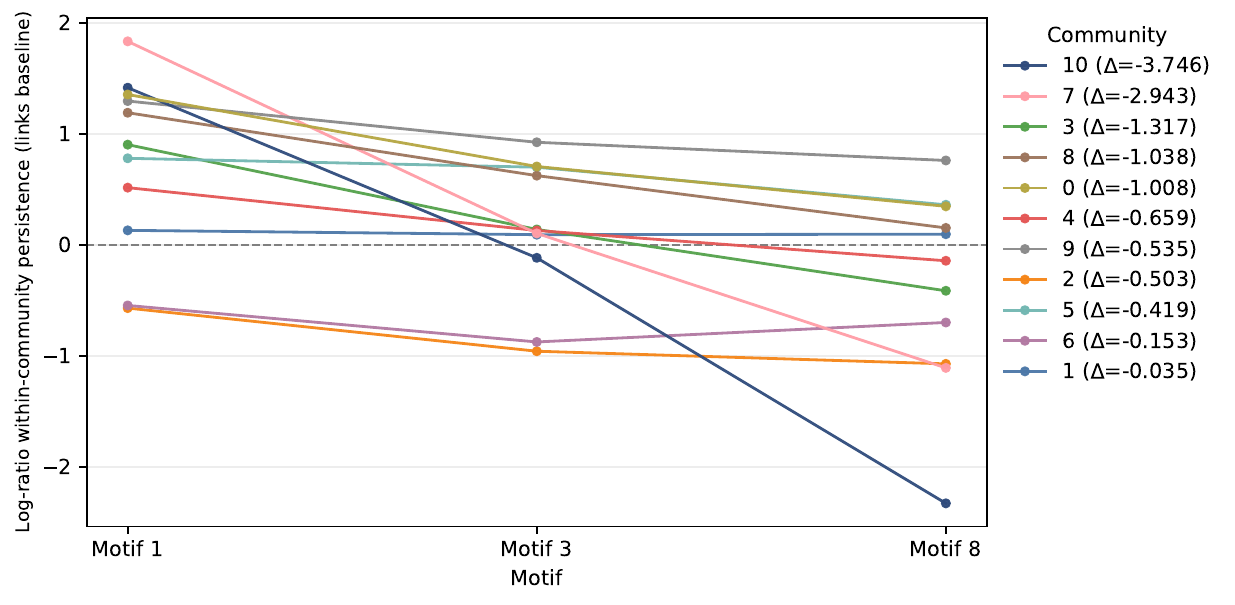}
    \caption{Reference-adjusted within-community persistence across motif lengths. Lines report $\mathrm{log\_ratio}^{\mathrm{links}, (m)}_{\mathrm{within},c}$ for Motifs~1, 3, and 8. Negative values indicate lower persistence relative to the first-order reference. The downward trend highlights the progressive weakening of within-community retention as path length increases.}
    \label{fig:persistence_by_length_links}
\end{figure}

We therefore examine how these deviations evolve across motif length. Figure~\ref{fig:persistence_by_length_links} reports $\text{log\_ratio}^{\mathrm{links}, (m)}_{\mathrm{within},c}$ across Motif~1, Motif~3, and Motif~8. In practice, this compares the observed within-community persistence of each motif with the persistence derived from direct retweet interactions alone.

Across communities, $\text{log\_ratio}^{\mathrm{links}, (m)}_{\mathrm{within},c}$ systematically decreases as motif length increases. This suggests that, relative to the first-order reference, longer motifs place progressively less mass on within-community retention.

This pattern naturally raises the complementary question of how the off-diagonal component changes as motif length increases, and whether this change is distributed uniformly across destination communities. The analyses that follow address this question by examining community-level transitions.

\subsection{Higher-order redistribution of attention}

Building on the previous analysis of within-community persistence, we now examine how the off-diagonal component changes across communities as within-community retention declines. We focus on the share of incoming cross-community attention received by each destination community, $\sigma^{\mathrm{paths},(m)}_{\mathrm{in},c}$, derived from final-jump pathways. %This analysis addresses the complementary question raised in the previous section, namely how cross-community endpoints are reweighted as motif length increases.

\begin{figure}[h!]
    \centering
    \includegraphics[width=\textwidth]{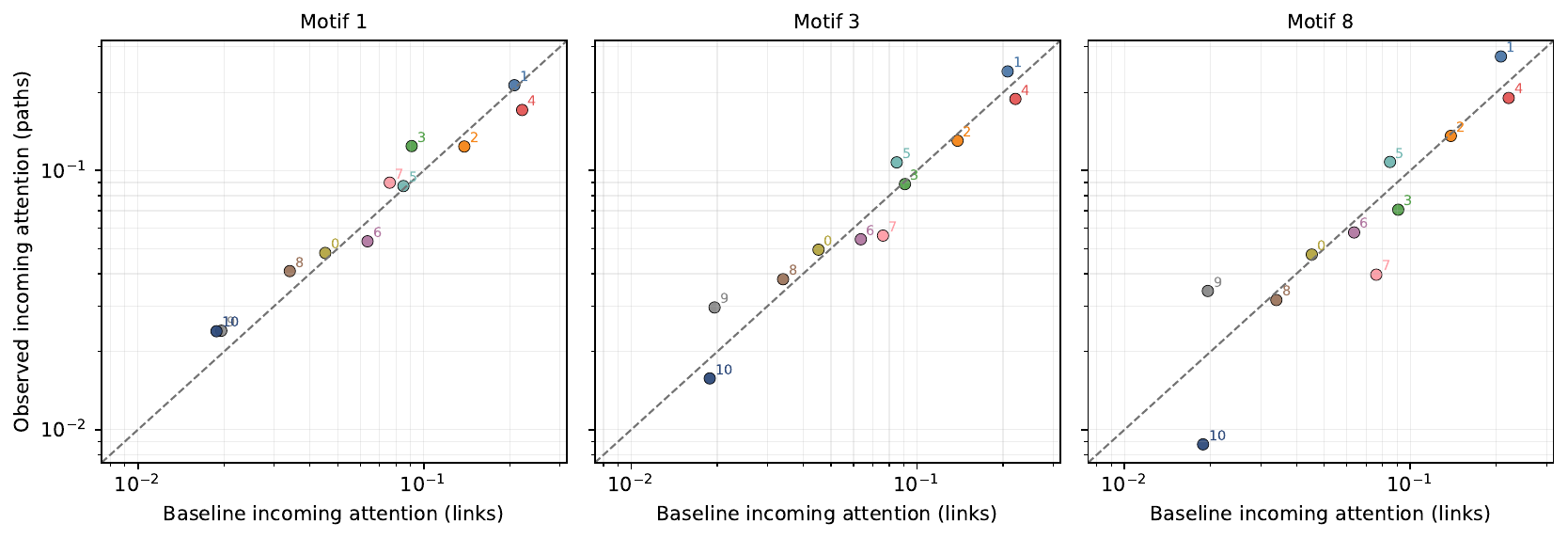}
    \caption{Observed versus first-order reference share of received cross-community attention under $\textbf{W}^{\mathrm{links}}$ (final-jump, attention orientation). Each point is a community; panels correspond to Motif~1, Motif~3, and Motif~8. The dashed line indicates equality between observed and reference shares.}
    \label{fig:obs_vs_links}
\end{figure}

Figure~\ref{fig:obs_vs_links} compares, for the attention orientation and separately for Motifs~1, 3, and 8, the observed share of cross-community attention received by each community with its corresponding first-order reference share under $\textbf{W}^{\mathrm{links}}$. The size-only reference is reported in Appendix~\ref{appendix:community_diagnostics} as a coarse diagnostic. 

Across all three motifs, observed incoming-attention shares remain broadly aligned with the first-order reference. Most communities lie close to the $45^{\circ}$ line, indicating that the higher-order allocation still largely reflects the first-order structure. At the same time, deviations from the diagonal are clearly visible and become more pronounced as motif length increases. Motif~1 exhibits the tightest alignment, Motif~3 shows a modest increase in dispersion, and Motif~8 displays the widest spread. Thus, while the first-order structure remains a strong point of reference, longer motifs introduce systematic over- and under-representation for specific communities.

\subsection{Divergent higher-order trajectories across communities}

\begin{figure}[h!]
    \centering
    \includegraphics[width=0.9\textwidth]{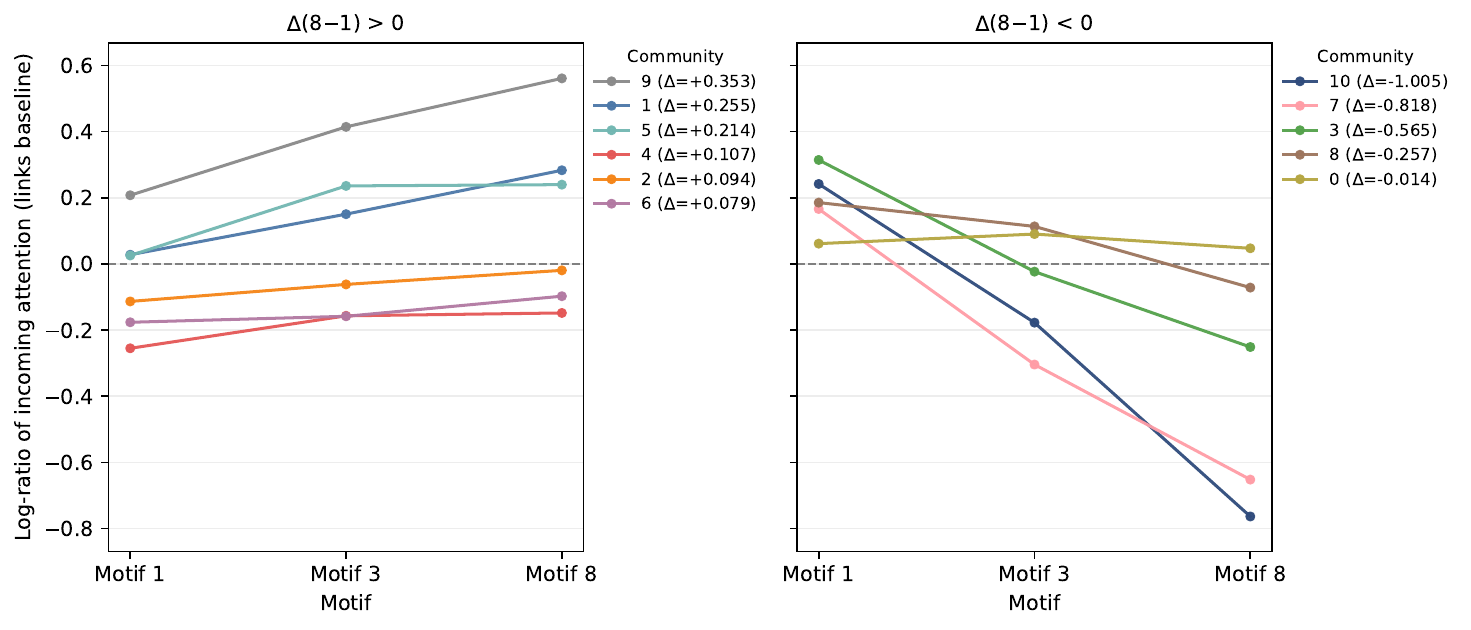}
    \caption{Reference deviations across motif length under $\mathbf{W}^{\mathrm{links}}$ (final-jump, attention orientation). Lines show $\mathrm{log\_ratio}^{\mathrm{links},(m)}_{c}$ from Motif~1 to Motif~8. Left panel: communities with $\Delta^{(8-1)}_{c}>0$ (increasing deviation). Right panel: communities with $\Delta^{(8-1)}_{c}<0$ (decreasing deviation). The dashed horizontal line marks $0$ (no deviation from the reference).}
    \label{fig:split_delta_links}
\end{figure}

To characterize how deviations from the first-order reference evolve with motif length, Figure~\ref{fig:split_delta_links} reports $\mathrm{log\_ratio}^{\mathrm{links},(m)}_{c}$ across Motif~1, Motif~3, and Motif~8, splitting communities by the sign of their motif-length contrast $\Delta^{(8-1)}_{c}$.

The left panel shows communities with positive $\Delta^{(8-1)}_{c}$, namely communities whose relative representation increases as motif length increases. For this group, deviations from the first-order reference remain stable or increase with motif length, indicating that higher-order pathways amplify their attractiveness as destinations of cross-community attention. This pattern is particularly pronounced for a subset of communities with larger positive contrasts, such as community~9 (Digital \& Debunking), community~1 (Technocratic Reformists), and community~5 (Institutions \& Public Safety), with more moderate but still positive effects for communities~4, 2, and 6.

The right panel shows communities with negative $\Delta^{(8-1)}_{c}$, whose relative representation declines as motif length increases. For these communities, $\mathrm{log\_ratio}^{\mathrm{links},(m)}_{c}$ becomes progressively more negative from Motif~1 to Motif~8, indicating that their prominence as endpoints weakens as motif length increases. The strongest declines are observed for community~10 (Entertainment \& Light News), community~7 (Civil Society), and community~3 (Sports \& COVID Bulletins), while communities~8 and~0 exhibit smaller but still negative contrasts.

Importantly, this split is not a minor fluctuation around zero. The magnitude of $\Delta^{(8-1)}_{c}$ is sizable for multiple communities, suggesting that higher-order attention chains selectively reinforce some destinations while weakening others, even though the overall allocation remains broadly aligned with the first-order structure.

This systematic divergence suggests that the observed redistribution is structured rather than uniform across communities. One plausible interpretation of these divergent trajectories lies in the node-level structural properties of the network. In particular, heterogeneity in absorption probabilities ($p_{\mathrm{sink}}$) and in node in-strength jointly shapes where random-walk paths tend to end within this construction. By definition, nodes with higher $p_{\mathrm{sink}}$ are more likely to act as structural endpoints of random-walk trajectories. Among the nodes with high absorption probabilities, a large fraction are also high in-degree nodes, that is, structural hubs of attention. These results indicate that path termination is strongly associated with high sink propensity and that this sink propensity substantially overlaps with hubness in practice. This combination may help explain the observed divergence in community-level trajectories. A more detailed analysis of this relationship is reported in Appendix~\ref{appendix:absorption_hubness}.

\subsection{Reversed directional lens}

We complement the analysis in the attention orientation by repeating the same set of measurements under the reversed orientation, where edges are inverted, and pathways are traced in the opposite direction. As discussed in the Methods, this is not a mere robustness, as it highlights how the structural organization of multi-step pathways changes when  focusing on the flow of information.

We focus on the same quantities introduced above, namely the log-ratio of incoming attention relative to the $\textbf{W}^{\mathrm{links}}$ first-order comparison and the associated motif-length contrast $\Delta^{(8-1)}_{c}$. The full set of figures parallel to the attention orientation (including raw matrices, persistence, and observed-versus-baseline comparisons) is reported in Appendix~\ref{appendix:reversed_orientation}. Here we concentrate on the quantities that provide new information relative to the attention case.

\begin{figure}[h!]
    \centering
    \includegraphics[width=0.9\textwidth]{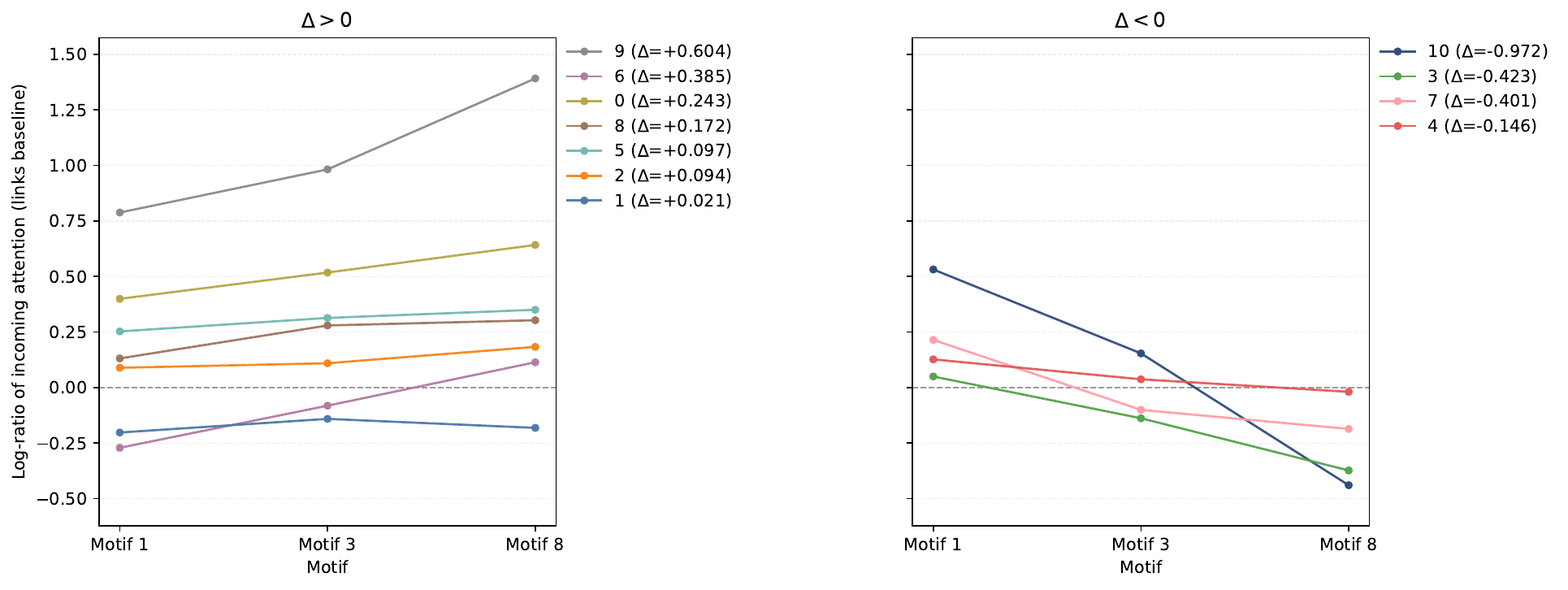}
    \caption{Deviations across motif length relative to the $\mathbf{W}^{\mathrm{links}}$ first-order comparison (final-jump, reversed orientation). Lines show $\mathrm{log\_ratio}^{\mathrm{links},(m)}_{c}$ from Motif~1 to Motif~8. Left panel: communities with $\Delta^{(8-1)}_{c}>0$. Right panel: communities with $\Delta^{(8-1)}_{c}<0$. The dashed horizontal line marks $0$.}
    \label{fig:figR1_reversed_logratio_split_by_delta}
\end{figure}

Figure~\ref{fig:figR1_reversed_logratio_split_by_delta} reports the evolution of $\mathrm{log\_ratio}^{\mathrm{links},(m)}_{c}$ across Motifs~1, 3, and 8 under the reversed orientation, splitting communities according to the sign of $\Delta^{(8-1)}_{c}$. As in the attention case, longer pathways induce systematic divergence, with some communities becoming increasingly over-represented as endpoints of multi-step pathways while others are progressively under-represented.

In this case, the communities with positive $\Delta^{(8-1)}_{c}$ include in particular community~9 (Digital \& Debunking) and community~6 (Five Star Movement), with community~0 (Other) also appearing positive but being less directly interpretable due to its aggregated nature. These communities become more prominent as endpoints of longer pathways under the reversed orientation.

On the other hand, communities with negative $\Delta^{(8-1)}_{c}$ include community~10 (Entertainment \& Light News), community~3 (Sports \& COVID Bulletins), and community~7 (Civil Society). These communities lose prominence as endpoints of longer pathways under the reversed lens, indicating that they are less represented as destinations of extended pathways in this orientation.

Importantly, the magnitude of these deviations remains substantial, indicating that the selective over- and under-representation of communities is not specific to the attention orientation, but a robust structural feature of higher-order pathways. At the same time, the pattern is not identical: the set of communities that gain or lose prominence under longer chains partially differs from the attention case, suggesting directional asymmetries between the two orientations.

To further clarify how these directional asymmetries are distributed across communities, we compare motif-length sensitivity across orientations.

\subsection{Community cartography across orientations}

\begin{figure}[h!]
    \centering
    \includegraphics[width=0.75\linewidth]{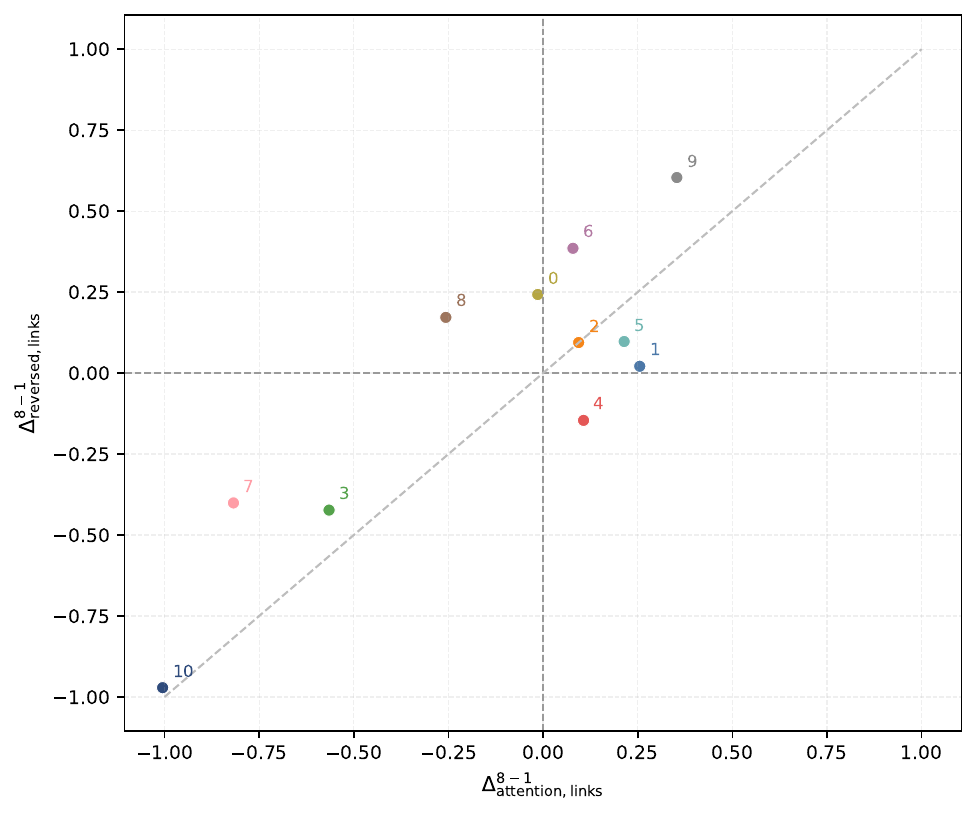}
    \caption{Comparison of motif-length sensitivity across orientations. Each point is a community. The x-axis reports $\Delta^{(8-1)}_{c}$ under the attention orientation and the y-axis reports $\Delta^{(8-1)}_{c}$ under the reversed orientation. Dashed lines mark $0$.}
    \label{fig:figD1_scatter_delta8minus1_att_vs_rev}
\end{figure}

To make this comparison explicit, Figure~\ref{fig:figD1_scatter_delta8minus1_att_vs_rev} reports, for each community, the motif-length contrast under the attention orientation versus the reversed orientation.

Most communities lie in the same-sign quadrants, indicating that the qualitative direction of their higher-order behavior is broadly preserved across orientations. However, the distribution across quadrants highlights important asymmetries.

Communities in the top-right quadrant (positive $\Delta$ in both orientations) are led by community~9 (Digital \& Debunking), with additional contributions from communities~5 (Institutions \& Public Safety), 6 (Five Star Movement), and 2 (Right-wing Populists). Under both orientations, these communities tend to gain prominence in longer pathways.

By contrast, community~8 (Pro-EU Centre-left) lies in the top-left quadrant, indicating positive $\Delta$ under the reversed orientation but negative $\Delta$ under the attention orientation. This suggests that its prominence as an endpoint of extended pathways is stronger under the reversed orientation than under the attention orientation.

Communities~10 (Entertainment \& Light News), 7 (Civil Society \& Social Impact), and 3 (Sports \& COVID Bulletins) fall in the bottom-left quadrant, showing negative $\Delta$ in both orientations. For these communities, longer pathways consistently reduce their prominence as endpoints, suggesting that they remain comparatively peripheral in extended attention chains regardless of the directional lens.

Finally, community~4 (News Media) is located in the bottom-right quadrant, with positive $\Delta$ in the attention orientation but negative $\Delta$ under the reversed lens. This indicates a distinctive asymmetry: it becomes more prominent in extended pathways under the attention orientation than under the reversed one.

Overall, these patterns highlight that higher-order effects are not only heterogeneous across communities, but also direction-dependent, revealing asymmetries that are not captured by first-order connectivity alone.

% ===================== DISCUSSION =====================
\section{Discussion}
The retweet network we examined shows several features commonly associated with large-scale online interaction systems: it is highly asymmetric, strongly broadcast-like, and characterized by a concentration of incoming attention on a relatively small set of nodes. The community structure extracted from the validated projection is also broadly interpretable, with many clusters corresponding to recognizable social and political domains. In parallel, the motif distribution is highly skewed: most realized paths fall into simple chain-like structures (Motifs~1, 3, and 8), while more complex patterns are almost absent.

Against this background, we find that attention is initially concentrated within communities, but that this concentration weakens as path length increases. The share of within-community persistence decreases from Motif~1 to Motif~8, suggesting that, within the random-walk construction, longer pathways are associated with a lower relative concentration on within-community endpoints and a greater presence across community boundaries. Such a feature is somehow expected. It is well know in the literature that there in online social networks, few hubs are much more retweeted than they retweet~\cite{watts2007influentials,wu2011who,gonzalezbailon2013broadcasters}. Indeed, those are the nodes that are directly connected to the sink in the motif detection (see Section~\ref{sec:methods}). Since discursive communities are, by construction, pivoted on some particularly popular accounts, Motif~1 is simply detecting their presence. Furthermore, the preference towards motifs of short path length is due to the way data are collected. As mentioned elsewhere in this paper, the available data do not contain any information about the chain of retweets. Otherwise stated, Alice may have seen a retweet by Bob to the original message of Charlie, and then decided to retweet it, but the data accounts only for a retweet from Alice to Charlie. In this sense, all retweet chains are condensed in simple links from the retweeter to the author of the original message, thus reducing to a path of length 1. Since most standard users limit their activity to retweets, it is natural to observe many short paths from retweeters to the main leaders in each discursive communities.

Nonetheless, this shift is not uniform. Once first-order connectivity is taken into account, communities show markedly different behaviors: some retain more attention, while others lose it more rapidly. In particular, communities 7 and 10 (i.e., ``Civil Society \& Social Impact'' and ``Entertainment \& Light News'') are focused on Motif 1, while a slighter presence of Motif 3, is observed in the strictly political communities 2 and 8 (``Right-wing populists'' and ``Pro-EU Centre-left'') and in community 3 (``Sports \& COVID Bulletins''). The impact of longer-path motifs is more relevant for communities 1,4, 5, 6, and 9.

This heterogeneity becomes even clearer when looking at cross-community attention. Although the overall distribution remains broadly aligned with the first-order structure, higher-order pathways introduce systematic and non-trivial deviations. These deviations are not fully captured by community size or by direct retweet connectivity alone. Instead, longer chains selectively favor some communities while penalizing others. In particular, communities such as Digital \& Debunking, Technocratic Reformists, and Institutions \& Public Safety become increasingly prominent as endpoints of longer pathways, whereas communities such as Entertainment \& Light News, Civil Society, and Sports \& COVID Bulletins become progressively less central. These contrasts are substantively meaningful because they are broadly consistent with the semantic and functional profiles associated with the communities: some clusters appear better positioned to concentrate or attract structurally mediated attention under longer chains, while others remain more peripheral once attention is followed beyond direct links.

Before going on with the discussion of our results, let us remark that our results suggest that higher-order pathways should not be interpreted as a direct measure of observed diffusion. As already mentioned above, the non-availability of information about the chain of retweets, i.e., whether a retweet is induced or not by a retweet by some other followed user or from a genuine attention to the content created by the original author, does not permit to exactly trace the diffusion of attention. Nevertheless, it provides non-trivial information about the structure of the information flow.

A branch of the research in Computational Social Science is testing and adding observations about the presence of the two-step flow in Online Social Networks. The two-step flow was first proposed in the 1950's by ~\cite{katz1955personal} to describe the information flow in US politics. The main idea is that the information flows from mass media to citizens, mainly mediated by special figures, i.e., the \emph{opinion leaders}, who form their opinions based on news and re-elaborate the message by the media around their own narrative for other citizens. In the original proposal, the interaction between opinion leaders and other citizens was intended to be in person, for instance, during public social events. Besides the criticisms that it received, the two-step flow description of the information flow lasted until the popularization of TV, i.e., the possibility for individuals to autonomously look for the news sources they prefer, and the contemporary decline of the popularity of social events. Nevertheless, the two-step flow has recently revived in the study of online social platforms: some analyses show the presence of particularly active users who restructure the information from the main institutional news sources around their narrative and deliver their message to the public~\cite{wu2011who, hilbert2017onestep,bracciale2018fromsuper,dubois2020who, serafino2024analysis}. It is worth mentioning that not all scholars agree on this vision: for instance, instead of the two-step flow, \cite{bennett2006theonestep} proposed the one-step flow, in which the platform itself, through its recommendation systems, provides directly the users with the information that is most in line with their preferences.

Interpreting our results through the lens of the two-step flow theory, the presence of longer-path motifs can be seen as a deviation from a purely one-step flow structure, thus highlighting the role of content mediators. Indeed, Fig.~\ref{fig:motifs_by_start_ranked_stacked_brokeny} shows that communities such as Entertainment and Civil Society (7 and 10, respectively) are dominated by Motif~1, whereas more politically oriented communities — namely 2, 8, and 5 (Right-wing, Centre-left, and Five Stars Movement, respectively) — exhibit a relatively stronger contribution from longer-path motifs.

More specifically, Motif~1 indicates the presence of one-step flow, while Motifs~3 and~8 indicate the presence of two-step flows. %\edo{Qua forse mi sfugge qualcosa. Il two-step flow classico non dovrebbe corrispondere più naturalmente a Motif 3 (un intermediario), mentre Motif 8 sembrerebbe già un'estensione a tre step?} \fab{Il two-step flow prevede \emph{almeno} un intermediario, anzi, è facile che ce ne sia più di uno e che spesso si influenzino a vicenda, specie nella versione aggiornata dei social networks. Alcuni autori (Hilbert \& Co, per esempio) chiamano one, two-, three-, etc-step flow, per distinguere il numero di intermediari; non è una definizione condivisa, però. La differenza vera è se ci sono intermediari oppure no, indipendentemente da quanti sono.}
Here, we use two-step flow in the broader sense of mediated flow, independently of the exact number of intermediary nodes.
Both news sources and opinion leaders are structurally unbalanced nodes, in the sense that they are retweeted far more than they retweet others. However, this imbalance is considerably stronger for news sources, whose retweet activity is extremely limited, than for opinion leaders, who actively redistribute content both from news sources and from other like-minded opinion leaders. As a consequence, Motif~1 captures the presence of a direct link towards an unbalanced node, whereas Motifs~3 and ~8 capture paths of length 2 and 3, respectively, ending in such a node. In this interpretation, Motif~1 can be regarded as a signature of a one-step flow (i.e. direct access to news sources), while Motifs~3 and~8 reflect a two-step flow, in which political leaders act as intermediary nodes (i.e. opinion leaders) connecting sources and the broader public. Overall, the smaller frequency of Motifs~3 and~8 within political communities may therefore be interpreted as a proxy for the limited impact of the two-step information flows.

Other observations provide more details about the structure of the debate. For instance, the panel a of Fig.~\ref{fig:intercommunity_connectivity_combined} and the panel a of Fig.~\ref{fig:bar_sigma_links_within} show that the values on the diagonal are stronger than those off the diagonal, thus highlighting a particular polarized debate, intended as one in which the interaction between communities is particularly limited. Such a polarization signal is much stronger for Community 2, confirming the analysis of \cite{caldarelli2021flow}.

Furthermore, our analysis goes beyond such a distinction between one- and two-steps and tackles the role of different communities in the flow of information. In fact, while we still observe that mainly the various motifs start and end inside the community they originated in, longer-path motifs may connect different communities. As a matter of fact, the matrices in Fig.~\ref{fig:final_jump_heatmaps} are much less focused on the diagonal as the length of the path of the motif increases. Remarkably, this pattern is much more evident for communities from 1 to 6, i.e. those more related to official information and news sources (Communities 1, 3, 4, and 5) and to politics (Communities 2 and 6). In this case, the effect of the two-step flow seems to be more evident.

So far, the analysis was conducted following the flow of attention of the users, i.e., looking at who the various users are retweeting and interpreting the information flow from the bottom level. The analysis of the reversed orientation provides an additional perspective on this mechanism. Rather than serving as a robustness check, the reversed description permits studying how multi-step pathways characterize the flow of information. Indeed, the analysis of the frequency of directed motifs is strongly dependent on the directions of the link. The sink is introduced as a fictitious node intended to balance the outgoing with the ingoing flux for each node, see Section~\ref{sec:methods}. Therefore, the analysis of directed motifs in the retweet network using links from retweeters to retweeted and in the one considering links in the inverted direction are not equivalent and provide different insights.

Following this point of view, we observe that higher-order effects remain substantial but are not symmetric. Some communities gain prominence in longer pathways under both orientations, while others do so only in one direction. For example, community~4 (News Media) becomes more prominent in extended pathways under the attention orientation than under the reversed one, whereas community~8 (Pro-EU Centre-left) shows the opposite tendency, see Fig.~\ref{fig:figD1_scatter_delta8minus1_att_vs_rev}. This suggests that some communities are better understood as structurally effective attractors of attention, while others emerge more clearly as effective end-users of retweet chains when the same architecture is read in reverse. Such a discrepancy is due to the method used to account for the different frequency of motifs: in the attention representation, the focus is on the content creators, in the flow of information representation on the standard users. In this sense, the prominence of community 4, for instance, in the attention orientation is mainly due to the role of its main content creator in informing others, thus being the terminal of many random walkers. In the opposite representation, community 8, for instance, has a more pronounced tendency to display longer pathways as its retweeters are less active in producing new content, thus becoming the terminal of random walkers.

Beside their interpretation in terms of one- or two-step flow, these findings suggest that retweet networks are not only partitioned into communities, but also organized in ways that selectively reweight cross-community attention once multi-step pathways are taken into account. Communities may therefore differ not only in how much attention they retain internally, but also in how they participate in the wider structural circulation of attention across the network. In this sense, the analysis points to community-level roles that are partly directional and partly dependent on path length. These roles should not be treated as fixed or exhaustive properties of the communities themselves, but they remain informative because they align in meaningful ways with the broader semantic and functional character of the clusters identified in the network.

% ===================== CONCLUSIONS =====================
\section{Conclusions}

This work examined how the structure of retweet interactions changes when the network is observed beyond direct links alone. Combining a community-based reconstruction of online debate with a higher-order random-walk framework, we studied how short multi-step pathways reweight the distribution of attention across discursive communities in Italian COVID-19 Twitter. The aim was not to reconstruct observed cascades or the temporal propagation of individual tweets, but to describe how the aggregated retweet network organizes possible multi-step endpoints across communities.

Three main results emerge. First, direct retweet interactions already reveal a highly uneven and community-structured system, in which attention is concentrated on a limited number of users and communities. Second, when the analysis moves from direct links to longer motifs, within-community persistence declines and cross-community redistribution becomes more visible. Third, this redistribution is not uniform: some communities become progressively more prominent as endpoints of longer pathways, while others lose relative prominence. These differences are not fully captured by community size or by first-order connectivity alone, and they remain visible, with important asymmetries, when the network is analyzed under the reversed orientation.

Taken together, these findings suggest that retweet networks are not only modular, but also higher-order in a substantively relevant sense. Communities differ not only in who they connect to directly, but also in how they emerge as destinations of multi-step pathways. In this respect, looking beyond direct retweets changes the community-level representation of online debate. The resulting picture remains descriptive rather than causal, but it is nevertheless informative because it shows that some communities occupy structurally privileged positions once attention is followed beyond immediate interactions.

More broadly, the work shows that a higher-order perspective can add useful information to standard community-level analyses of social media networks. By combining validated community reconstruction with random-walk motifs, it becomes possible to characterize how attention is redistributed across communities under multi-step pathways, and to identify directional asymmetries that are not visible in first-order representations alone. Within the limits already discussed, the proposed framework therefore offers a structured and interpretable way to study how online debate is organized beyond direct retweet connectivity.

Our analysis naturally opens to future research. In the present paper, we display frequencies of patterns as they are measured on the empirical data. A natural extension concerns the use of null models for the statistical validation of the observed patterns. The first-order and size-based references adopted here are informative and easy to interpret, but they remain relatively simple comparison structures. Future work should compare motif and final-jump patterns against weighted null models that preserve more of the local structure of the network, including both degree and strength constraints. In this respect, maximum-entropy approaches such as the Enhanced Configuration Model \citep{mastrandrea2014enhanced} or other similar-in-spirit models \citep{gabrielli2019grand, parisi2020faster} appear especially promising, even if the dimension of the empirical data represents a strong limitation for the applicability of such methods using the existing tools~\cite{vallarano2021fast}.

Finally, the framework could be extended to other empirical settings. The retweet mechanism is unusually well suited to the present analysis, so direct portability to other platforms should not be taken for granted. Even so, similar questions could be explored in other contexts or in other issue domains, such as climate change, electoral campaigns, or war-related communication. Cross-country extensions may be especially informative when applied to relatively self-contained national debates, where the discursive structure remains more compact and comparable.

\bibliography{bibliography_article}

@article{gabrielli2019grand,
  title = {Grand canonical ensemble of weighted networks},
  author = {Gabrielli, Andrea and Mastrandrea, Rossana and Caldarelli, Guido and Cimini, Giulio},
  journal = {Phys. Rev. E},
  volume = {99},
  issue = {3},
  pages = {030301},
  numpages = {6},
  year = {2019},
  month = {Mar},
  publisher = {American Physical Society},
  doi = {10.1103/PhysRevE.99.030301},
  url = {https://link.aps.org/doi/10.1103/PhysRevE.99.030301}
}

@article{benjamini1995controlling,
author = {Benjamini, Yoav and Hochberg, Yosef},
title = {Controlling the False Discovery Rate: A Practical and Powerful Approach to Multiple Testing},
journal = {Journal of the Royal Statistical Society: Series B (Methodological)},
volume = {57},
number = {1},
pages = {289-300},
doi = {https://doi.org/10.1111/j.2517-6161.1995.tb02031.x},
url = {https://rss.onlinelibrary.wiley.com/doi/abs/10.1111/j.2517-6161.1995.tb02031.x},
eprint = {https://rss.onlinelibrary.wiley.com/doi/pdf/10.1111/j.2517-6161.1995.tb02031.x},
year = {1995}
}

@article{caldarelli2026thephysics,
  title = {The physics of news,  rumors,  and opinions},
  volume = {1186},
  ISSN = {0370-1573},
  url = {http://dx.doi.org/10.1016/j.physrep.2026.05.002},
  DOI = {10.1016/j.physrep.2026.05.002},
  journal = {Physics Reports},
  publisher = {Elsevier BV},
  author = {Caldarelli,  Guido and Artime,  Oriol and Fischetti,  Giulia and Guarino,  Stefano and Nowak,  Andrzej and Saracco,  Fabio and Holme,  Petter and De Domenico,  Manlio},
  year = {2026},
  month = Aug,
  pages = {1–75}
}

@article{gonzalezbailon2013broadcasters,
   author = {Sandra González-Bailón and Javier Borge-Holthoefer and Yamir Moreno},
   doi = {10.1177/0002764213479371},
   isbn = {0002-7642},
   issn = {00027642},
   journal = {American Behavioral Scientist},
   title = {Broadcasters and Hidden Influentials in Online Protest Diffusion},
   year = {2013},
}

@article{vallarano2021fast,
   author = {Nicolò Vallarano and Matteo Bruno and Emiliano Marchese and Giuseppe Trapani and Fabio Saracco and Giulio Cimini and Mario Zanon and Tiziano Squartini},
   doi = {10.1038/s41598-021-93830-4},
   issn = {2045-2322},
   issue = {1},
   journal = {Scientific Reports 2021 11:1},
   month = {7},
   pages = {1-33},
   publisher = {Nature Publishing Group},
   title = {Fast and scalable likelihood maximization for Exponential Random Graph Models with local constraints},
   volume = {11},
   url = {https://www.nature.com/articles/s41598-021-93830-4},
   year = {2021},
}

@article{bennett2006theonestep,
author = {W. Lance Bennett and Jarol B. Manheim},
title ={The One-Step Flow of Communication},
journal = {The ANNALS of the American Academy of Political and Social Science},
volume = {608},
number = {1},
pages = {213-232},
year = {2006},
doi = {10.1177/0002716206292266},
URL = {         https://doi.org/10.1177/0002716206292266    
},
eprint = {         https://doi.org/10.1177/0002716206292266
},
}

@article{serafino2024analysis,
  title = {Analysis of flows in social media uncovers a new multi-step model of information spread},
  volume = {2024},
  ISSN = {1742-5468},
  url = {http://dx.doi.org/10.1088/1742-5468/ad8748},
  DOI = {10.1088/1742-5468/ad8748},
  number = {11},
  journal = {Journal of Statistical Mechanics: Theory and Experiment},
  publisher = {IOP Publishing},
  author = {Serafino,  Matteo and Virginio Clemente,  Giulio and Flamino,  James and Szymanski,  Boleslaw K and Lizardo,  Omar and Makse,  Hernán A},
  year = {2024},
  month = Nov,
  pages = {113402}
}

@article{bracciale2018fromsuper,
   author = {Roberta Bracciale and Antonio Martella and Chiara Visentin},
   doi = {10.1285/I20356609V11I2P361},
   issn = {20356609},
   issue = {2},
   journal = {PaCO Vol. 11, No. 2 (2018). Special Issue: From Big Data in Politics to the Politics of Big Data},
   month = {1},
   pages = {361-393},
   publisher = {University of Salento},
   title = {From Super-Participants to Super-Echoed. Participation in the 2018 Italian Electoral Twittersphere},
   volume = {11},
   url = {https://www.academia.edu/40422799/From_Super_Participants_to_Super_Echoed_Participation_in_the_2018_Italian_Electoral_Twittersphere},
   year = {2018},
}

@article{dubois2020who,
author = {Elizabeth Dubois and Sara Minaeian and Ariane Paquet-Labelle and Simon Beaudry},
title ={Who to Trust on Social Media: How Opinion Leaders and Seekers Avoid Disinformation and Echo Chambers},
journal = {Social Media + Society},
volume = {6},
number = {2},
pages = {2056305120913993},
year = {2020},
doi = {10.1177/2056305120913993},
URL = {https://doi.org/10.1177/2056305120913993},
eprint = {https://doi.org/10.1177/2056305120913993}
}

@article{hilbert2017onestep,
author = {Martin Hilbert and Javier Vásquez and Daniel Halpern and Sebastián Valenzuela and Eduardo Arriagada},
title ={One Step, Two Step, Network Step? Complementary Perspectives on Communication Flows in Twittered Citizen Protests},
journal = {Social Science Computer Review},
volume = {35},
number = {4},
pages = {444-461},
year = {2017},
doi = {10.1177/0894439316639561},
URL = {https://doi.org/10.1177/0894439316639561},
eprint = {https://doi.org/10.1177/0894439316639561}
}

@book{katz1955personal,
   author = {Elihu Katz and Paul F Lazarsfeld},
   city = {New York,  NY,  US},
   journal = {Personal influence: the part played by people in the flow of mass communications.},
   pages = {400, xx, 400-xx},
   publisher = {Free Press},
   title = {Personal influence: the part played by people in the flow of mass communications.},
   year = {1955},
}

@article{garlaschelli2008maximum,
  title = {Maximum likelihood: Extracting unbiased information from complex networks},
  author = {Garlaschelli, Diego and Loffredo, Maria I.},
  journal = {Phys. Rev. E},
  volume = {78},
  issue = {1},
  pages = {015101},
  numpages = {4},
  year = {2008},
  month = {Jul},
  publisher = {American Physical Society},
  doi = {10.1103/PhysRevE.78.015101},
  url = {https://link.aps.org/doi/10.1103/PhysRevE.78.015101}
}

@article{fischer2015sampling,
  title = {Sampling Motif-Constrained Ensembles of Networks},
  author = {Fischer, Rico and Leit\~ao, Jorge C. and Peixoto, Tiago P. and Altmann, Eduardo G.},
  journal = {Phys. Rev. Lett.},
  volume = {115},
  issue = {18},
  pages = {188701},
  numpages = {5},
  year = {2015},
  month = {Oct},
  publisher = {American Physical Society},
  doi = {10.1103/PhysRevLett.115.188701},
  url = {https://link.aps.org/doi/10.1103/PhysRevLett.115.188701}
}

@ARTICLE{peixoto2026graphs,
       author = {{Peixoto}, Tiago P. and {Peel}, Leto and {Gross}, Thilo and {De Domenico}, Manlio},
        title = "{Graphs are maximally expressive for higher-order interactions}",
      journal = {arXiv e-prints},
         year = 2026,
        month = feb,
          eid = {arXiv:2602.16937},
        pages = {arXiv:2602.16937},
          doi = {10.48550/arXiv.2602.16937},
archivePrefix = {arXiv},
       eprint = {2602.16937},
 primaryClass = {physics.soc-ph},
       adsurl = {https://ui.adsabs.harvard.edu/abs/2026arXiv260216937P}
}

@article{chang2021analytical,
  title={Analytical solution of epidemic threshold for coupled information-epidemic dynamics on multiplex networks with alterable heterogeneity},
  author={Chang, Xin and Cai, Chao-Ran and Zhang, Ji-Qiang and Wang, Chong-Yang},
  journal={Physical Review E},
  volume={104},
  number={4},
  pages={044303},
  year={2021},
  publisher={APS}
}

@article{banos2013role,
    author = "Baños, Raquel A and Borge-Holthoefer, Javier and Moreno, Yamir",
    title = "The role of hidden influentials in the diffusion of online information cascades",
    volume = "2",
    ISSN = "2193-1127",
    url = {https://doi.org/10.1140/epjds18},
    DOI = "10.1140/epjds18",
    number = "1",
    journal = "EPJ Data Science",
    publisher = "Springer Science and Business Media LLC",
    year = "2013",
    month = "July"
}

@article{osullivan2015mathematical,
    author = "O'Sullivan, David J. P. and O'Keeffe, Gary J. and Fennell, Peter G. and Gleeson, James P.",
    title = "Mathematical modeling of complex contagion on clustered networks",
    journal = "Frontiers in Physics",
    volume = "Volume 3 - 2015",
    year = "2015",
    url = "https://www.frontiersin.org/journals/physics/articles/10.3389/fphy.2015.00071",
    doi = "10.3389/fphy.2015.00071",
    issn = "2296-424X"
}

@article{sobkowicz2018opinion,
    author = "Sobkowicz, Pawel",
    title = "Opinion Dynamics Model Based on Cognitive Biases of Complex Agents",
    journal = "Journal of Artificial Societies and Social Simulation (JASSS)",
    volume = "21",
    number = "4",
    pages = "8",
    year = "2018"
}

@inproceedings{wu2011who,
    author = "Wu, Shaomei and Hofman, Jake M. and Mason, Winter A. and Watts, Duncan J.",
    title = "Who Says What to Whom on Twitter",
    year = "2011",
    isbn = "9781450306324",
    publisher = "Association for Computing Machinery",
    address = "New York, NY, USA",
    url = "https://doi.org/10.1145/1963405.1963504",
    doi = "10.1145/1963405.1963504",
    booktitle = "Proceedings of the 20th International Conference on World Wide Web",
    pages = "705–714",
    numpages = "10",
    location = "Hyderabad, India",
    series = "WWW '11"
}

@article{goel2016structural,
  title={The structural virality of online diffusion},
  author={Goel, Sharad and Anderson, Ashton and Hofman, Jake and Watts, Duncan J},
  journal={Management science},
  volume={62},
  number={1},
  pages={180--196},
  year={2016},
  publisher={INFORMS}
}

@article{watts2007influentials,
    author = "Watts, Duncan J. and Dodds, Peter Sheridan",
    doi = "10.1086/518527",
    issn = "00935301",
    journal = "Journal of Consumer Research",
    month = "dec",
    number = "4",
    pages = "441--458",
    publisher = "Oxford Academic",
    title = "{Influentials, networks, and public opinion formation}",
    url = "https://academic.oup.com/jcr/article-lookup/doi/10.1086/518527",
    volume = "34",
    year = "2007"
}

@inproceedings{gomezrodriguez2010inferring,
author = {Gomez Rodriguez, Manuel and Leskovec, Jure and Krause, Andreas},
title = {Inferring networks of diffusion and influence},
year = {2010},
isbn = {9781450300551},
publisher = {Association for Computing Machinery},
address = {New York, NY, USA},
url = {https://doi.org/10.1145/1835804.1835933},
doi = {10.1145/1835804.1835933},
booktitle = {Proceedings of the 16th ACM SIGKDD International Conference on Knowledge Discovery and Data Mining},
pages = {1019–1028},
numpages = {10},
location = {Washington, DC, USA},
series = {KDD '10}
}

@article{flamino2023political,
    author = "Flamino, James and Galeazzi, Alessandro and Feldman, Stuart and Macy, Michael W. and Cross, Brendan and Zhou, Zhenkun and Serafino, Matteo and Bovet, Alexandre and Makse, Hern{\'{a}}n A. and Szymanski, Boleslaw K.",
    doi = "10.1038/s41562-023-01550-8",
    issn = "23973374",
    journal = "Nature Human Behaviour",
    month = "mar",
    number = "6",
    pages = "904--916",
    pmid = "36914806",
    publisher = "Nature Publishing Group",
    title = "{Political polarization of news media and influencers on Twitter in the 2016 and 2020 US presidential elections}",
    url = "https://www.nature.com/articles/s41562-023-01550-8",
    volume = "7",
    year = "2023"
}

@article{caldarelli2025physics,
  title={The physics of news, rumors, and opinions},
  author={Caldarelli, Guido and Artime, Oriol and Fischetti, Giulia and Guarino, Stefano and Nowak, Andrzej and Saracco, Fabio and Holme, Petter and de Domenico, Manlio},
  journal={arXiv preprint arXiv:2510.15053},
  year={2025}
}

@article{blondel2008fast,
  title={Fast unfolding of communities in large networks},
  author={Blondel, Vincent D and Guillaume, Jean-Loup and Lambiotte, Renaud and Lefebvre, Etienne},
  journal={Journal of statistical mechanics: theory and experiment},
  volume={2008},
  number={10},
  pages={P10008},
  year={2008},
  publisher={IOP Publishing}
}

@article{lancichinetti2009community,
  title={Community detection algorithms: a comparative analysis},
  author={Lancichinetti, Andrea and Fortunato, Santo},
  journal={Physical review E},
  volume={80},
  number={5},
  pages={056117},
  year={2009},
  publisher={APS}
}

@article{fortunato2010community,
  title={Community detection in graphs},
  author={Fortunato, Santo},
  journal={Physics reports},
  volume={486},
  number={3-5},
  pages={75--174},
  year={2010},
  publisher={Elsevier}
}

@article{picciolo2022weighted,
  title={Weighted network motifs as random walk patterns},
  author={Picciolo, Francesco and Ruzzenenti, Franco and Holme, Petter and Mastrandrea, Rossana},
  journal={New journal of physics},
  volume={24},
  number={5},
  pages={053056},
  year={2022},
  publisher={IoP Publishing}
}

@article{pratelli2024entropy,
  title={Entropy-based detection of Twitter echo chambers},
  author={Pratelli, Manuel and Saracco, Fabio and Petrocchi, Marinella},
  journal={PNAS nexus},
  volume={3},
  number={5},
  pages={pgae177},
  year={2024},
  publisher={Oxford University Press US}
}

@article{caldarelli2021flow,
  title={Flow of online misinformation during the peak of the COVID-19 pandemic in Italy},
  author={Caldarelli, Guido and De Nicola, Rocco and Petrocchi, Marinella and Pratelli, Manuel and Saracco, Fabio},
  journal={EPJ data science},
  volume={10},
  number={1},
  pages={34},
  year={2021},
  publisher={Springer Berlin Heidelberg}
}

@INPROCEEDINGS{8750923,
  author={Paul, Indraneil and Khattar, Abhinav and Kumaraguru, Ponnurangam and Gupta, Manish and Chopra, Shaan},
  booktitle={2019 IEEE 35th International Conference on Data Engineering Workshops (ICDEW)}, 
  title={Elites Tweet? Characterizing the Twitter Verified User Network}, 
  year={2019},
  volume={},
  number={},
  pages={278-285},
  doi={10.1109/ICDEW.2019.00006}}

@article{saracco2015randomizing,
  title={Randomizing bipartite networks: the case of the World Trade Web},
  author={Saracco, Fabio and Di Clemente, Riccardo and Gabrielli, Andrea and Squartini, Tiziano},
  journal={Scientific reports},
  volume={5},
  number={1},
  pages={10595},
  year={2015},
  publisher={Nature Publishing Group UK London}
}

@article{saracco2017inferring,
  title={Inferring monopartite projections of bipartite networks: an entropy-based approach},
  author={Saracco, Fabio and Straka, Mika J and Clemente, Riccardo Di and Gabrielli, Andrea and Caldarelli, Guido and Squartini, Tiziano},
  journal={New Journal of Physics},
  volume={19},
  number={5},
  pages={053022},
  year={2017},
  publisher={IOP Publishing}
}

@article{traag2019louvain,
  title={From Louvain to Leiden: guaranteeing well-connected communities},
  author={Traag, Vincent A and Waltman, Ludo and Van Eck, Nees Jan},
  journal={Scientific reports},
  volume={9},
  number={1},
  pages={5233},
  year={2019},
  publisher={Nature Publishing Group UK London}
}

@article{raghavan2007near,
  title={Near linear time algorithm to detect community structures in large-scale networks},
  author={Raghavan, Usha Nandini and Albert, R{\'e}ka and Kumara, Soundar},
  journal={Physical Review E—Statistical, Nonlinear, and Soft Matter Physics},
  volume={76},
  number={3},
  pages={036106},
  year={2007},
  publisher={APS}
}

@inproceedings{hanna2013partisan,
  title={Partisan alignments and political polarization online: A computational approach to understanding the French and US presidential elections},
  author={Hanna, Alexander and Wells, Chris and Maurer, Peter and Friedland, Lew and Shah, Dhavan and Matthes, J{\"o}rg},
  booktitle={Proceedings of the 2nd Workshop on Politics, Elections and Data},
  pages={15--22},
  year={2013}
}

@inproceedings{badawy2018analyzing,
  title={Analyzing the digital traces of political manipulation: The 2016 Russian interference Twitter campaign},
  author={Badawy, Adam and Ferrara, Emilio and Lerman, Kristina},
  booktitle={2018 IEEE/ACM international conference on advances in social networks analysis and mining (ASONAM)},
  pages={258--265},
  year={2018},
  organization={IEEE}
}

@inproceedings{conover2011political,
  title={Political polarization on twitter},
  author={Conover, Michael and Ratkiewicz, Jacob and Francisco, Matthew and Gon{\c{c}}alves, Bruno and Menczer, Filippo and Flammini, Alessandro},
  booktitle={Proceedings of the international aaai conference on web and social media},
  volume={5},
  number={1},
  pages={89--96},
  year={2011}
}

@article{conover2012partisan,
  title={Partisan asymmetries in online political activity},
  author={Conover, Michael D and Gon{\c{c}}alves, Bruno and Flammini, Alessandro and Menczer, Filippo},
  journal={EPJ Data science},
  volume={1},
  number={1},
  pages={6},
  year={2012},
  publisher={Springer}
}

@article{becatti2019extracting,
  title={Extracting significant signal of news consumption from social networks: the case of Twitter in Italian political elections},
  author={Becatti, Carolina and Caldarelli, Guido and Lambiotte, Renaud and Saracco, Fabio},
  journal={Palgrave Communications},
  volume={5},
  number={1},
  year={2019},
  publisher={Springer Science and Business Media LLC}
}

@misc{mattei2022bow,
  title={Bow-tie structures of Twitter discursive communities. Sci Rep 12 (1): 12944},
  author={Mattei, M and Pratelli, M and Caldarelli, G and Petrocchi, M and Saracco, F},
  year={2022}
}

@article{radicioni2021analysing,
  title={Analysing Twitter semantic networks: the case of 2018 Italian elections},
  author={Radicioni, Tommaso and Saracco, Fabio and Pavan, Elena and Squartini, Tiziano},
  journal={Scientific Reports},
  volume={11},
  number={1},
  pages={13207},
  year={2021},
  publisher={Nature Publishing Group UK London}
}

@article{cimini2019statistical,
  title={The statistical physics of real-world networks},
  author={Cimini, Giulio and Squartini, Tiziano and Saracco, Fabio and Garlaschelli, Diego and Gabrielli, Andrea and Caldarelli, Guido},
  journal={Nature Reviews Physics},
  volume={1},
  number={1},
  pages={58--71},
  year={2019},
  publisher={Nature Publishing Group UK London}
}

@article{guarino2021information,
  title={Information disorders during the COVID-19 infodemic: The case of Italian Facebook},
  author={Guarino, Stefano and Pierri, Francesco and Di Giovanni, Marco and Celestini, Alessandro},
  journal={Online Social Networks and Media},
  volume={22},
  pages={100124},
  year={2021},
  publisher={Elsevier}
}

@article{garrett2009echo,
  title={Echo chambers online?: Politically motivated selective exposure among Internet news users},
  author={Garrett, R Kelly},
  journal={Journal of computer-mediated communication},
  volume={14},
  number={2},
  pages={265--285},
  year={2009},
  publisher={Oxford University Press Oxford, UK}
}

@article{cinelli2021echo,
  title={The echo chamber effect on social media},
  author={Cinelli, Matteo and De Francisci Morales, Gianmarco and Galeazzi, Alessandro and Quattrociocchi, Walter and Starnini, Michele},
  journal={Proceedings of the national academy of sciences},
  volume={118},
  number={9},
  pages={e2023301118},
  year={2021},
  publisher={National Academy of Sciences}
}

@article{del2016spreading,
  title={The spreading of misinformation online},
  author={Del Vicario, Michela and Bessi, Alessandro and Zollo, Fabiana and Petroni, Fabio and Scala, Antonio and Caldarelli, Guido and Stanley, H Eugene and Quattrociocchi, Walter},
  journal={Proceedings of the national academy of Sciences},
  volume={113},
  number={3},
  pages={554--559},
  year={2016},
  publisher={National Academy of Sciences}
}

@article{guarino2024verified,
  title = {Leveraging content producer networks and user perception to detect online discursive communities},
  ISSN = {2045-2322},
  url = {https://doi.org/10.1038/s41598-026-39477-5},
  DOI = {10.1038/s41598-026-39477-5},
  journal = {Scientific Reports},
  publisher = {Springer Science and Business Media LLC},
  author = {Guarino,  Stefano and Mounim,  Ayoub and Caldarelli,  Guido and Saracco,  Fabio},
  year = {2026},
  month = mar 
}

@article{singh2025information,
  title={Information diffusion analysis: process, model, deployment, and application},
  author={Singh, Shashank Sheshar and Srivastava, Divya and Verma, Madhushi and Muhuri, Samya},
  journal={The Knowledge Engineering Review},
  volume={39},
  pages={e11},
  year={2025},
  publisher={Cambridge University Press}
}

@article{gong2023broadcast,
  title={Broadcast information diffusion processes on social media networks: Exogenous events lead to more integrated public discourse},
  author={Gong, Xuanjun and Huskey, Richard and Xue, Haoning and Shen, Cuihua and Frey, Seth},
  journal={Journal of Communication},
  volume={73},
  number={3},
  pages={247--259},
  year={2023},
  publisher={Oxford University Press}
}

@article{lambiotte2019networks,
  title={From networks to optimal higher-order models of complex systems},
  author={Lambiotte, Renaud and Rosvall, Martin and Scholtes, Ingo},
  journal={Nature physics},
  volume={15},
  number={4},
  pages={313--320},
  year={2019},
  publisher={Nature Publishing Group UK London}
}

@article{benson2016higher,
  title={Higher-order organization of complex networks},
  author={Benson, Austin R and Gleich, David F and Leskovec, Jure},
  journal={Science},
  volume={353},
  number={6295},
  pages={163--166},
  year={2016},
  publisher={American Association for the Advancement of Science}
}

@article{riascos2021random,
  title={Random walks on weighted networks: a survey of local and non-local dynamics},
  author={Riascos, Alejandro P and Mateos, Jos{\'e} L},
  journal={Journal of Complex Networks},
  volume={9},
  number={5},
  pages={cnab032},
  year={2021},
  publisher={Oxford University Press}
}

@article{larock2022sequential,
  title={Sequential motifs in observed walks},
  author={LaRock, Timothy and Scholtes, Ingo and Eliassi-Rad, Tina},
  journal={Journal of Complex Networks},
  volume={10},
  number={5},
  pages={cnac036},
  year={2022},
  publisher={Oxford University Press}
}

@inproceedings{scholtes2017network,
  title={When is a network a network? Multi-order graphical model selection in pathways and temporal networks},
  author={Scholtes, Ingo},
  booktitle={Proceedings of the 23rd ACM SIGKDD international conference on knowledge discovery and data mining},
  pages={1037--1046},
  year={2017}
}

@article{scholtes2016higher,
  title={Higher-order aggregate networks in the analysis of temporal networks: path structures and centralities},
  author={Scholtes, Ingo and Wider, Nicolas and Garas, Antonios},
  journal={The European Physical Journal B},
  volume={89},
  number={3},
  pages={61},
  year={2016},
  publisher={Springer}
}

@article{peixoto2017modelling,
  title={Modelling sequences and temporal networks with dynamic community structures},
  author={Peixoto, Tiago P and Rosvall, Martin},
  journal={Nature communications},
  volume={8},
  number={1},
  pages={582},
  year={2017},
  publisher={Nature Publishing Group UK London}
}

@article{vosoughi2018spread,
  title={The spread of true and false news online},
  author={Vosoughi, Soroush and Roy, Deb and Aral, Sinan},
  journal={science},
  volume={359},
  number={6380},
  pages={1146--1151},
  year={2018},
  publisher={American Association for the Advancement of Science}
}

@article{salnikov2016using,
  title={Using higher-order Markov models to reveal flow-based communities in networks},
  author={Salnikov, Vsevolod and Schaub, Michael T and Lambiotte, Renaud},
  journal={Scientific reports},
  volume={6},
  number={1},
  pages={23194},
  year={2016},
  publisher={Nature Publishing Group UK London}
}

@article{parisi2020faster,
  title={A faster horse on a safer trail: generalized inference for the efficient reconstruction of weighted networks},
  author={Parisi, Federica and Squartini, Tiziano and Garlaschelli, Diego},
  journal={New Journal of Physics},
  volume={22},
  number={5},
  pages={053053},
  year={2020},
  publisher={IOP Publishing}
}

@article{mastrandrea2014enhanced,
  title={Enhanced reconstruction of weighted networks from strengths and degrees},
  author={Mastrandrea, Rossana and Squartini, Tiziano and Fagiolo, Giorgio and Garlaschelli, Diego},
  journal={New Journal of Physics},
  volume={16},
  number={4},
  pages={043022},
  year={2014},
  publisher={IOP Publishing}
}

@article{lin1991divergence,
  title={Divergence measures based on the Shannon entropy},
  author={Lin, Jianhua},
  journal={IEEE Transactions on Information theory},
  volume={37},
  number={1},
  pages={145--151},
  year={1991},
  publisher={IEEE},
  doi={10.1109/18.61115}
}

@article{endres2003new,
  title={A new metric for probability distributions},
  author={Endres, Dominik Maria and Schindelin, Johannes E},
  journal={IEEE Transactions on Information theory},
  volume={49},
  number={7},
  pages={1858--1860},
  year={2003},
  publisher={IEEE},
  doi={10.1109/TIT.2003.813506}
}

@article{kullback1951information,
  title={On information and sufficiency},
  author={Kullback, Solomon and Leibler, Richard A},
  journal={The annals of mathematical statistics},
  volume={22},
  number={1},
  pages={79--86},
  year={1951},
  publisher={Institute of Mathematical Statistics},
  doi={10.1214/aoms/1177729694}
}

\cleardoublepage
% ===================== APPENDIX =====================
\appendix

\section{BiDC}~\label{appendix:biDC}
To derive the network of verified users by exploiting retweets, we will construct the bipartite binary network of retweets. In a bipartite network, nodes are organized into two distinct groups, called \emph{layers}, and links are allowed only between nodes belonging to different layers. Using this structure, we can place verified and unverified users on the respective layers and establish a link between two nodes whenever one has retweeted the other. 
In practice, following \cite{pratelli2024entropy}, we retain a verified--unverified link whenever at least one retweet interaction between the two accounts is observed, independently of its direction. Because verified users are the main producers of original content, the overwhelming majority of these interactions correspond to retweets from unverified users toward verified accounts.
%\fab{Since the network is undirected, we consider retweets in both directions [Avete fatto davvero così o avete preso solo i retweet fatti dagli unverified? La differenza è minima (gli unverified sono molto raramente popolari), ma di solito si prende solo i retweet fatti dai non verificati]}. 
Once completed, we would be able to derive the network of verified users by projecting the information from the bipartite network onto the layer of verified users.

\subsection{Validated projection of bipartite networks}

A bipartite network is defined as a graph $G = (U, V, E)$, where $U$ and $V$ are two disjoint and independent sets of nodes (i.e. layers) such that any edge $\in E$ has one endpoint in $U$ and one in $V$. We can represent the bipartite network using a biadjacency matrix $\mathbf{B}$ of size $N_\top\times N_\bot$, with $N_\top$ and $N_\bot$ being, the dimensions of the two layers, i.e., respectively, the total number of verified and unverified users. Then, $b_{\alpha i}$, i.e. the generic entry of matrix $\mathbf{B}$, is 1, if the unverified users $i$ has retweeted at least once the verified user $\alpha$. To quantify the similarity between two nodes belonging to the same layer, a first simple approach is to employ the total number of common neighbors. This corresponds, for example, to the case where the same unverified user has retweeted the tweets of two verified authors. Formally, given two verified users $\alpha$ and $\beta$, their observed similarity $V^*_{\alpha\beta}$ is given by

\begin{equation}
    V^*_{\alpha\beta} = \sum_{j=1}^{N_\bot} b_{\alpha j}b_{\beta j},
\end{equation}

\noindent where both $b_{\alpha i}$ and $b_{\beta i}$ are equal to 1 if verified users $\alpha$ and $\beta$ have been retweeted by the same unverified user $i$, 0 otherwise. However, if we were to straightforwardly apply this simple projection, we would get a densely connected structure with a very trivial topology and limited informative value.  
To obtain a more robust monopartite projection, we need a criterion that allows us to determine whether the set of retweeters shared by two verified users is indeed carrying a non-trivial genuine similarity. 
One possible approach to determine the statistical significance of the measured similarity is to compare it with a properly defined null model. 
In our case, we employed the Bipartite Configuration Model (BiCM) \cite{saracco2015randomizing, saracco2017inferring}, an entropy-based null model for binary bipartite networks.  In this type of model, the starting point is the definition of the set of all bipartite networks $\mathcal{B}$ with $N_\top$ and $N_\bot$ nodes in the two respective layers and with a number of links between 0 and the maximum $N_\top \times N_\bot$. From this set, it is possible to obtain a (fairly) unbiased probability distribution through the constrained maximization of Shannon entropy $S$. 

\begin{equation}
    S = - \sum_{\mathbf{B} \in \mathcal{B}} P(\mathbf{B})\ln{P(\mathbf{B})}.
\end{equation} \label{shannon}

\noindent Conceptually, the main idea is to derive a statistical benchmark capable of reproducing certain chosen quantities (i.e. constraints), while ensuring that everything else is kept as random as possible.
%\fab{. \st{By maximizing $S$, we therefore seek to obtain a probability distribution that minimizes the number of assumptions about the missing information}\cite{saracco2015randomizing, saracco2017inferring} [Il discorso precedente mi pareva facesse più confusione che altro.]}. 
\noindent In the case of BiCM, the constraints are represented by the degree of nodes belonging to both layers. Formally, the degrees of unverified users $k_i$ and verified users $h_\alpha$ are defined as follows %\fab{[Perché siamo passati da $b_{\alpha i}$ a $b_{ij}$?]} 

\begin{equation}
k_i =\sum_{\alpha=1}^{N_\top}b_{\alpha i},\quad i=1\dots N_\bot;
\end{equation}

\begin{equation}
h_\alpha =\sum_{i=1}^{N_\bot}b_{\alpha i},\quad \alpha=1\dots N_\top.
\end{equation}

\noindent The constrained maximization of the Shannon entropy, where $\boldsymbol{\theta}$ and $\boldsymbol{\eta}$ are the Lagrange multipliers introduced in order to constrain the degrees and $\psi$ to ensure the normalization of the probability, reads

\begin{equation} \label{maxShannon}S'=S-\psi\left[1-\sum_{\mathbf{B}\in\mathcal{B}}P(\mathbf{B})\right]-\sum_{i=1}^{N_\bot}\theta_i\left[k_i^*-\sum_{\mathbf{B}\in\mathcal{B}}P(\mathbf{B})k_i(\mathbf{B})\right]-\sum_{\alpha=1}^{N_\top}\eta_\alpha\left[h_\alpha^*-\sum_{\mathbf{B}\in\mathcal{B}}P(\mathbf{B})h_\alpha(\mathbf{B})\right].
\end{equation}
%\edo{Potreste ricontrollare anche voi questa formula? Mi pare che la somma sulle i vada fino $N_\bot$ e quella sulle $\alpha$ fino a $N_\top$, no? Anche quelle sotto in caso.}
% FS: vero! Sistemato

\noindent The constrained maximization of $S'$ provides the following functional form of the probability:

\begin{equation} \label{maxS_result}
P(\mathbf{B}|\boldsymbol{\theta},\boldsymbol{\eta})
=\prod_{i=1}^{N_\bot}\prod_{\alpha=1}^{N_\top}\frac{e^{-(\theta_i+\eta_\alpha)b_{\alpha i}}}{1+e^{-(\theta_i+\eta_\alpha)}}\nonumber =\prod_{i=1}^{N_\bot}\prod_{\alpha=1}^{N_\top}p_{\alpha i}^{b_{\alpha i}}(1-p_{\alpha i})^{1-b_{\alpha i}},
\end{equation}

i.e. the probability per graph can be factorized in terms of independent probability per link.\\

\noindent To estimate the numerical values of the parameters $\boldsymbol{\theta}$ and $\boldsymbol{\eta}$ we impose the constraints:

\begin{subequations}\label{eq:constraints}
\begin{align}
k_i&=\sum_{\alpha=1}^{N_\top}\frac{x_iy_\alpha}{1+x_iy_\alpha}=\langle k_i\rangle,\quad i=1\dots N_\bot;\\ 
h_\alpha&=\sum_{i=1}^{N_\bot}\frac{x_iy_\alpha} {1+x_iy_\alpha}=\langle h_\alpha\rangle,\quad \alpha=1\dots N_\top.
\end{align}
\end{subequations}

\noindent Remarkably, given the functional form of the probability in terms of the Lagrangian multipliers, maximizing the log-(likelihood) of the system returns conditions equivalent to Eq.s~\ref{eq:constraints}, and such identification is valid only for maximum entropy null models~\cite{garlaschelli2008maximum}.

Once the probability distribution has been obtained, we can proceed to quantify the statistical significance of node similarity. Given two verified users $\alpha$ and $\beta$, we compare the co-occurence $b_{\alpha j}b_{\beta j} = 1$ with the null model. In the BiCM, links are treated as independent random variables: therefore, the presence of a co-occurence can be described as the outcome of a Bernoulli trial. The similarity $V_{\alpha\beta}$ is then the sum of independent Bernoulli trials and its behavior can be described by the Poisson-Binomial (PB) distribution. Thus, we can compute the p-value for the observed similarity  $V^*_{\alpha\beta}$ as following

\begin{equation}
    \text{p-value}(V^*_{\alpha\beta}) = \sum_{V\geq V^*_{\alpha\beta}} f_{PB}(V).
\end{equation}

\noindent The procedure is repeated for each pair of nodes belonging to the verified layer, resulting in an $n\times n$ matrix of p-values, where $n = N_\top(N_\top -1)/2$~\cite{saracco2017inferring}.\\
\noindent Finally, to obtain the validated monopartite projection, the p-values are selected through a validation procedure based on the False Discovery Rate (FDR, \cite{benjamini1995controlling}). The pairs of nodes corresponding to the p-values selected by the FDR are then connected by an edge in the final monopartite projection. In our case, we selected pairs of nodes applying the FDR with a single-test significance level $t = 0.05$.

\subsection{Community Detection}
%\noindent An interesting property of many real-world networks, including online social networks \citep{zhao2025modularity}, is their tendency to exhibit a modular structure, i.e., to form dense subgroups of nodes, known as clusters or communities \citep{fortunato2010community}. The few connections between different communities contrast with the high number of links within each cluster, which may reflect, for example, similarities or shared characteristics among the nodes that comprise it.\\ 
%\noindent Identifying these communities is no trivial task, and several methods have been proposed over the years, each of which often reflects different definitions and operationalization of communities \citep{newman2018networks}.
Besides the latest developments in the literature of community detection, in~\cite{guarino2024verified} it was shown that communities found with inferential and descriptive methods are almost equivalent, once applied to the validated projection of the verified-unverified users bipartite network. In our analyses, we used the Leiden algorithm~\cite{traag2019louvain}, i.e. a revised version of the celebrated Louvain algorithm~\cite{blondel2008fast}, employing the modularity. The modularity is one of the most widely used measures to assess the quality of a given partition, and it is defined as

\begin{equation}
\label{eq:modularity}
    Q=\frac{1}{2L}\sum_{i=1}^{N_\top}\sum_{j=1}^{N_\top}\left[a_{ij}-\frac{k_ik_j}{2L}\right]\delta_{c_i,c_j}
\end{equation}
where $a_{ij}$ represents the generic entry of the adjacency matrix $\mathbf{A}$, $k_i=\sum_j a_{ij}$ is the degree of node $i$, $L=\sum_i k_i$/2 is the total number of links, and $c_i$ represent the community to which the node $i$ belongs to. The term  $k_ik_j/2L$ is the probability that nodes $i$ and $j$ establish a connection (according to the Chung-Lu model) and $\delta_{c_i,c_j}$ is the Kronecker delta which equals 1 if nodes $i$ and $j$ belong to the same cluster (i.e. $c_i=c_j$) and 0 otherwise. In other words, modularity is a score function measuring the optimality of a given partition by comparing the empirical pattern of interconnections with the one predicted by a properly-defined benchmark model.\\

%In this work, to identify communities in the network of verified users, we employ a method based on modularity maximization, specifically the Leiden algorithm \citep{traag2019louvain}.
%This algorithm is actually an extension of the well-known Louvain algorithm \citep{blondel2008fast}. 
The Louvain algorithm starts by assigning each node to a single cluster. Then, each vertex is moved to a different cluster until the maximum increase in modularity in the system is achieved. The process is performed for each node in the network and then repeated iteratively until it is no longer possible to increase modularity. Once this phase is complete, the procedure is applied again in the same way, but this time the algorithm moves clusters, which are treated as supernodes. The process continues until no further increases in modularity can be achieved.\\
\noindent The Louvain algorithm is faster and more accurate than other methods \citep{lancichinetti2009community}, however, it suffers from some limitations. The partitions obtained depend on the order in which the nodes are selected \citep{fortunato2010community} and can create internally disconnected communities \citep{traag2019louvain}. The latter issue arises because, in its first phase, the algorithm evaluates only the movement of one node at a time, with the possibility of removing nodes that act as “bridges” within a community, fragmenting it. The subsequent irreversible aggregation of the community into a single node then makes it impossible to correct this error, leading to the final partition including communities that may be internally disconnected. 
To address this limitation, the Leiden algorithm begins by generating an initial partition, such as the Louvain algorithm, followed by a refinement phase. 

% In the refinement phase, Leiden improves upon Louvain by checking whether the communities obtained in the first optimization step are internally well connected. Rather than immediately aggregating each detected community into a single supernode, the algorithm first revisits the nodes within each community and may split them into smaller subcommunities when this improves the quality function while preserving internal connectivity. These refined subcommunities are then used in the aggregation step of the next iteration. This additional refinement is the key difference with respect to Louvain: it reduces the risk of obtaining communities that are only weakly connected internally or even disconnected, while retaining the same general logic of iterative quality-function optimization. The procedure is repeated until no further improvement of the chosen quality function is possible.

\noindent Leiden algorithm is equipped with a parameter $\gamma$ controlling the resolution of the procedure~\cite{traag2019louvain}. In our case, we applied the Leiden algorithm with resolution parameter $\gamma=2$, employing a multi-start relabeling strategy. The partition maximizing modularity was selected at the same $\gamma$. The resulting structure comprises 77 communities of verified users.

%\noindent FS: dopo un paragrafo così, uno sis aspetterebbe di vedere delle analisi che mostrano che effettivamente non ci siano grosse differenze. Se non c'è, sembra sparata lì a caso. Provo a raccontarla in modo da evitare altri test inutili da aggiungere nelle appendici.
%More broadly, we explored different community-detection settings, including alternative algorithms, different values of the resolution parameter, and multiple realizations at the same resolution, in order to assess the robustness and interpretability of the resulting partitions. This exploration showed that finer resolutions tended to produce increasingly fragmented structures, while lower resolutions yielded coarser partitions that were less informative for the purposes of this chapter. Although the exact partition varied slightly across realizations, especially after label propagation to the full retweet network, the largest communities and their most central accounts remained broadly stable. For this reason, we retain one representative Leiden partition at $\gamma=2$, which provided a reasonable balance between granularity, coherence, and interpretability for the downstream analysis. 
%\fab{[Di solito qui si conclude dicendo che alcune analisi usando valori di $\gamma$ diversi sono presenti nelle appendici e mostrano che qualitativamente i risultati sono molto simili. È così?].} \edo{In realtà no, ho fatto solo delle prove per vedre come cambiavano le comunità discorsive (quindi senza calcolare i motifs), ma non ho salvato i risultati.}

\noindent More broadly, we explored different community-detection settings, including different values of the resolution parameter and multiple realizations at the same resolution to assess the robustness and interpretability of the resulting partitions. This exploration showed that finer resolutions tended to produce increasingly fragmented structures, while lower resolutions yielded coarser partitions that were less informative for the purposes of our analyses. Although the exact partition varied slightly across realizations, the largest communities and their most central accounts remained broadly stable. For this reason, we retain one representative Leiden partition at $\gamma=2$, which provided a reasonable balance between granularity, coherence, and interpretability for the downstream analysis. 
%\fab{[Di solito qui si conclude dicendo che alcune analisi usando valori di $\gamma$ diversi sono presenti nelle appendici e mostrano che qualitativamente i risultati sono molto simili. È così?].} \edo{In realtà no, ho fatto solo delle prove per vedre come cambiavano le comunità discorsive (quindi senza calcolare i motifs), ma non ho salvato i risultati.}

\subsection{Label Propagation}
The communities obtained through the community detection process described in the previous paragraph are limited to verified users. To identify communities across the full network, including unverified users, it is necessary to assign community labels to the latter as well. To achieve this, we employed the label propagation algorithm developed by \cite{raghavan2007near}. The algorithm initially assigns a unique label to each vertex in the network. Subsequently, for each iteration, every node updates its label by adopting the dominant local label, i.e. the most frequent label in its neighborhood; in the event of a tie, the choice is made randomly among the dominant labels only. The process continues until consensus is reached, stopping when each node has the majority label of its neighborhood. This criterion can lead to different final configurations starting from the same initial conditions. In our case, we applied the label propagation algorithm with fixed seeds, assigning the initial labels obtained through community detection to verified users. These labels remained fixed throughout the entire label propagation process. The presence of a fixed initial set of labels also reduces the final number of possible configurations.
After propagation, communities are reindexed by descending size within the largest weakly connected component (LWCC). This canonical labeling is used throughout the paper.
The resulting partition should therefore be interpreted as a practical extension of the validated verified-user structure to the broader retweet network, adopted to obtain a coherent community-level representation for the subsequent analyses.

\section{Additional global motifs analysis}
\label{appendix:global_motifs}

\begin{table}[!htbp]
    \centering
    \scriptsize
    \renewcommand{\arraystretch}{1.1}
    \begin{tabular}{l|r|rrrrrrrr}
    \toprule
    Starting community & Share & Motif 1 & Motif 2 & Motif 3 & Motif 4 & Motif 5 & Motif 6 & Motif 7 & Motif 8 \\
    \midrule
    Global (All communities) & 1 & 0.911339 & 0.000085 & 0.074391 & 0.000155 &  0.000015 & 0.000141 & 0.000579 & 0.013295 \\
    1 (Technocratic Reformists) & 0.267282 & 0.887209 & 0.000077 & 0.093629 & 0.000136 & 0.000013 & 0.000173 & 0.000565 & 0.018197 \\
    2 (Right-wing Populists) & 0.214314 & 0.920030 & 0.000040 & 0.067486 & 0.000090 & 0.000009 & 0.000083 & 0.000352 & 0.011911 \\
    3 (Sports \& COVID Bulletins) & 0.192471 & 0.936638 & 0.000035 & 0.055540 & 0.000040 & 0.000003 & 0.000056 & 0.000206 & 0.007482 \\
    4 (News Media) & 0.145791 & 0.905224 & 0.000132 & 0.078519 & 0.000259 & 0.000021 & 0.000174 & 0.000880 & 0.014790 \\
    5 (Institutions \& Public Safety) & 0.053234 & 0.892162 & 0.000181 & 0.090782 & 0.000348 & 0.000056 & 0.000362 & 0.001422 & 0.014687 \\
    6 (Five Star Movement) & 0.033092 & 0.904086 & 0.000131 & 0.078218 & 0.000174 & 0.000013 & 0.000087 & 0.000395 & 0.016896 \\
    7 (Civil Society) & 0.027479 & 0.959314 & 0.000121 & 0.034928 & 0.000290 & 0.000024 & 0.000105 & 0.000407 & 0.004811 \\
    8 (Pro-EU Centre-left) & 0.015448 & 0.916666 & 0.000165 & 0.069749 & 0.000423 & 0.000007 & 0.000265 & 0.001133 & 0.011591 \\
    9 (Digital \& Debunking) & 0.009311 & 0.866333 & 0.000417 & 0.102118 & 0.000881 & 0.000131 & 0.000464 & 0.002463 & 0.027193 \\
    10 (Entertainment \& Light News) & 0.009254 & 0.956369 & 0.000036 & 0.038519 & 0.000048 & 0.000000 & 0.000024 & 0.000132 & 0.004873 \\
    0 \ \ (Other) & 0.032325 & 0.925933 & 0.000182 & 0.061878 & 0.000192 & 0.000021 & 0.000213 & 0.001327 & 0.010256 \\
    \bottomrule
    \end{tabular}
    \caption{Motif distribution by starting community in the attention network (top-10 communities plus Other). Rows are normalized within each starting community; \emph{Share} indicates the fraction of realized paths originating in that community.}
    \label{tab:motifs_by_start_comm_attention_top10}
\end{table}

% Figure 1: Ranked log-profile lines
\begin{figure}[!htbp]
    \centering
    \includegraphics[width=0.75\linewidth]{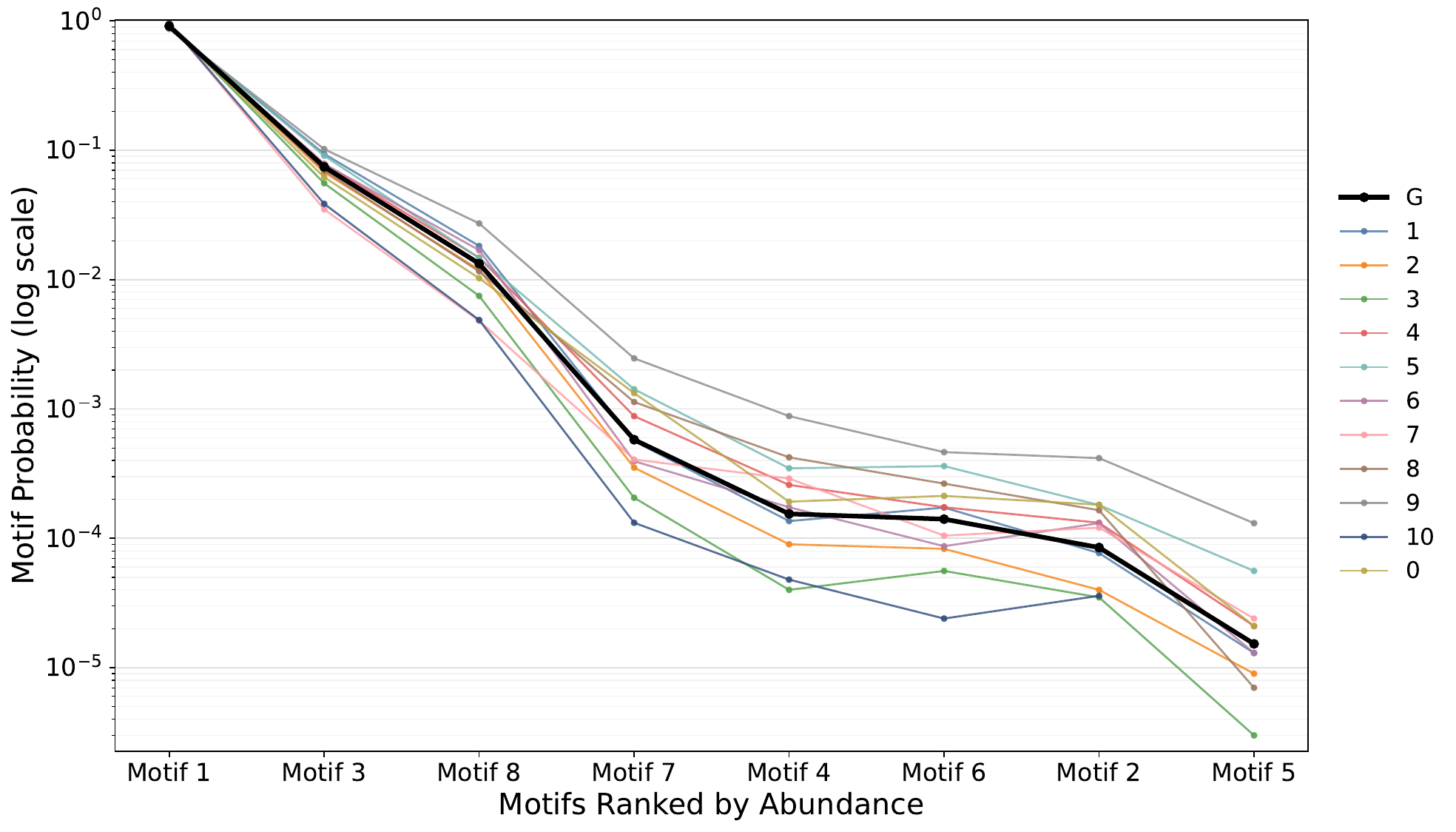}
    \caption{Motif probabilities by starting community, with motifs ordered by global abundance and y-axis on log scale. Each line corresponds to one starting community (G, 1--10, 0).}
    \label{fig:motifs_by_start_ranked_log_profiles}
\end{figure}

% Figure 2: Stacked bars (linear y)
\begin{figure}[!htbp]
    \centering
    \includegraphics[width=0.75\linewidth]{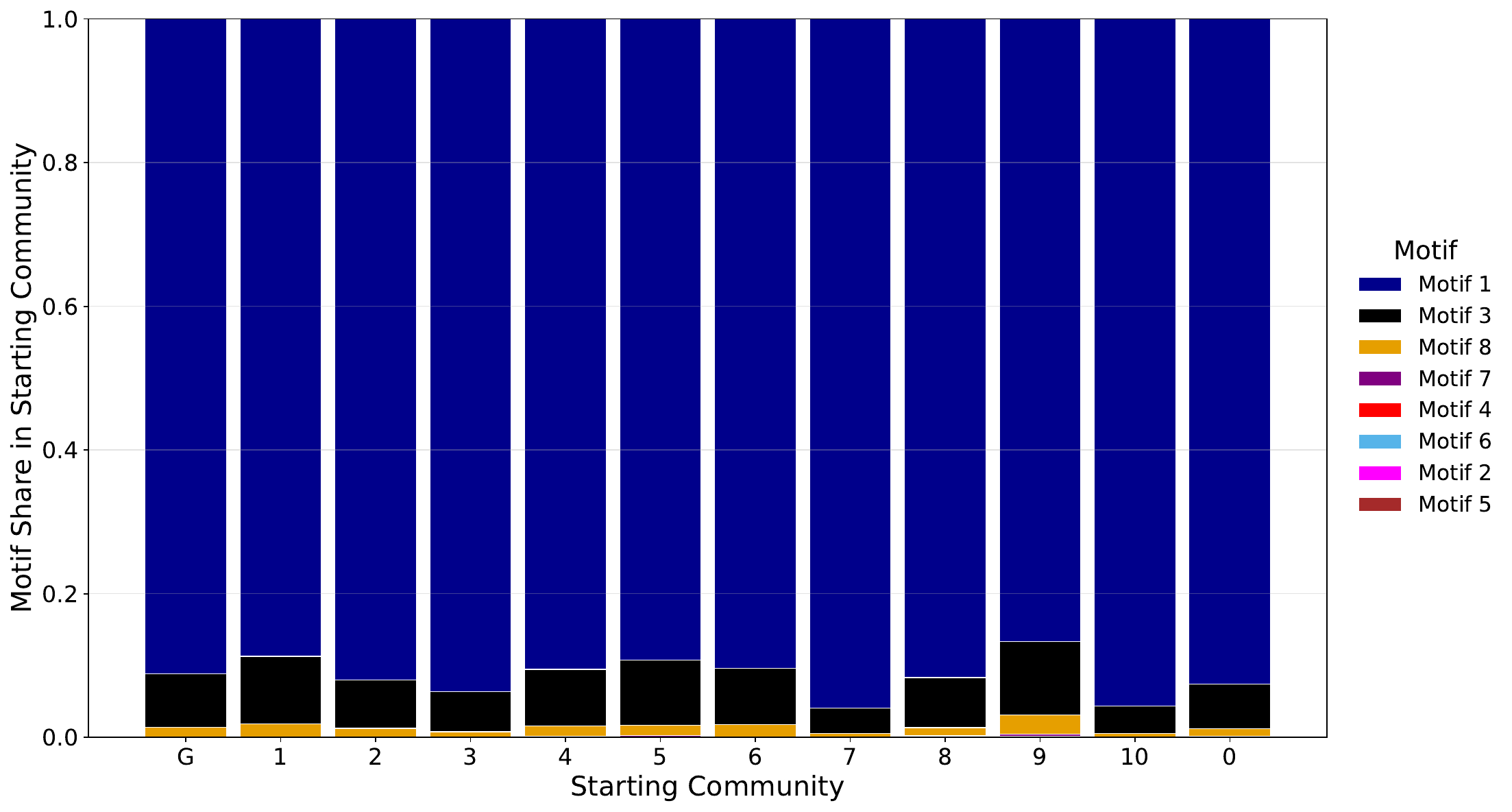}
    \caption{Motif composition by starting community as stacked bars (linear scale), with motifs ordered by global abundance.}
    \label{fig:motifs_by_start_ranked_stacked_linear}
\end{figure}

\section{Additional community-level diagnostics}
\label{appendix:community_diagnostics}

The figures collected in this appendix provide a more granular view of the community-level distributions discussed in the main text. In particular, they confirm that both within-community persistence and incoming attention are highly heterogeneous across communities, and that the associated baseline-adjusted deviations are not uniform across motif lengths.

\begin{figure}[h!]
    \centering
    \includegraphics[width=0.75\linewidth]{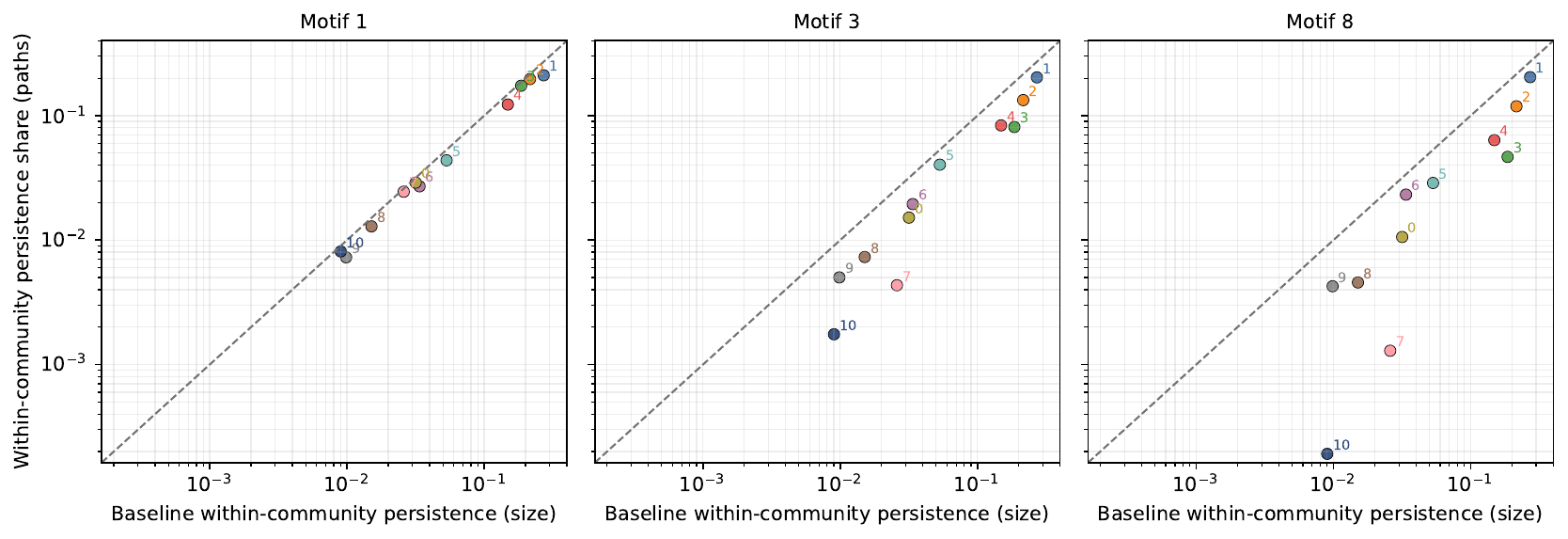}
    \caption{Observed versus size-based baseline within-community persistence. Each point represents a community. The x-axis reports the size-based reference $\sigma^{\mathrm{size}}_{\mathrm{within},c}$, while the y-axis reports the corresponding persistence share derived from paths $\sigma^{\mathrm{paths},(m)}_{\mathrm{within},c}$. Both axes are shown on logarithmic scales. The dashed line indicates equality between observed and baseline values.}
    \label{fig:persistence_vs_size}
\end{figure}

\begin{figure}[h!]
    \centering
    \includegraphics[width=0.75\linewidth]{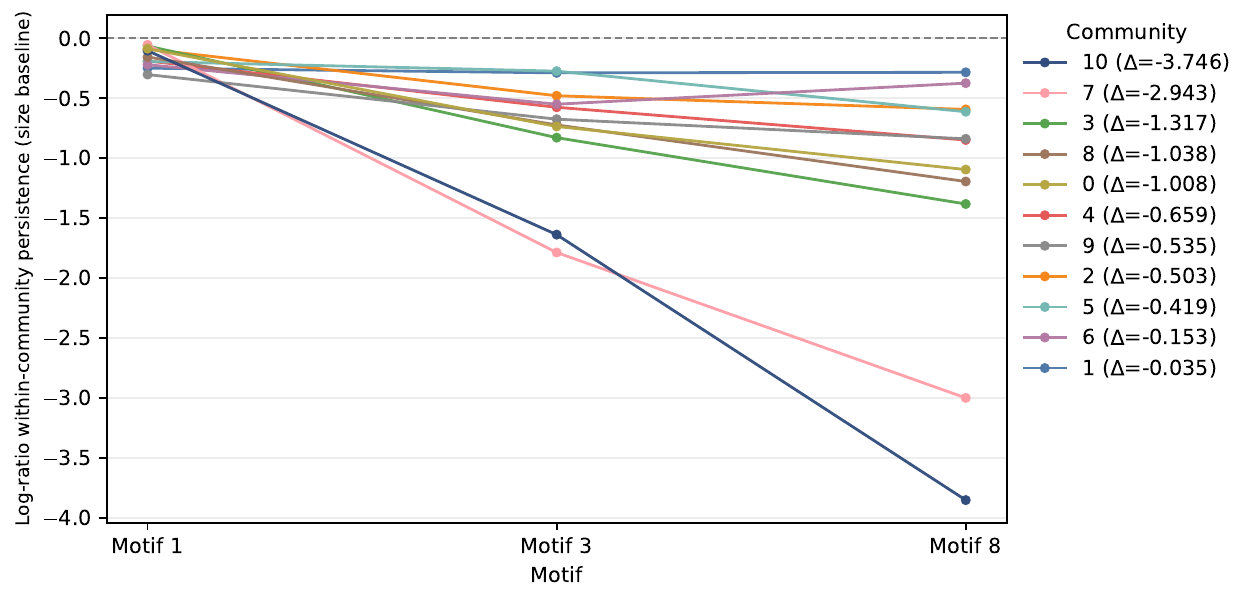}
    \caption{Benchmark-adjusted within-community persistence across motif length under the size-only baseline. Lines report $\mathrm{log\_ratio}^{(m)}_{\mathrm{within},c,\mathrm{size}}$ for Motifs~1, 3, and 8. Negative values indicate lower persistence relative to the size-based reference.}
    \label{fig:persistence_by_length_size}
\end{figure}

\begin{figure}[h!]
    \centering
    \includegraphics[width=0.8\linewidth]{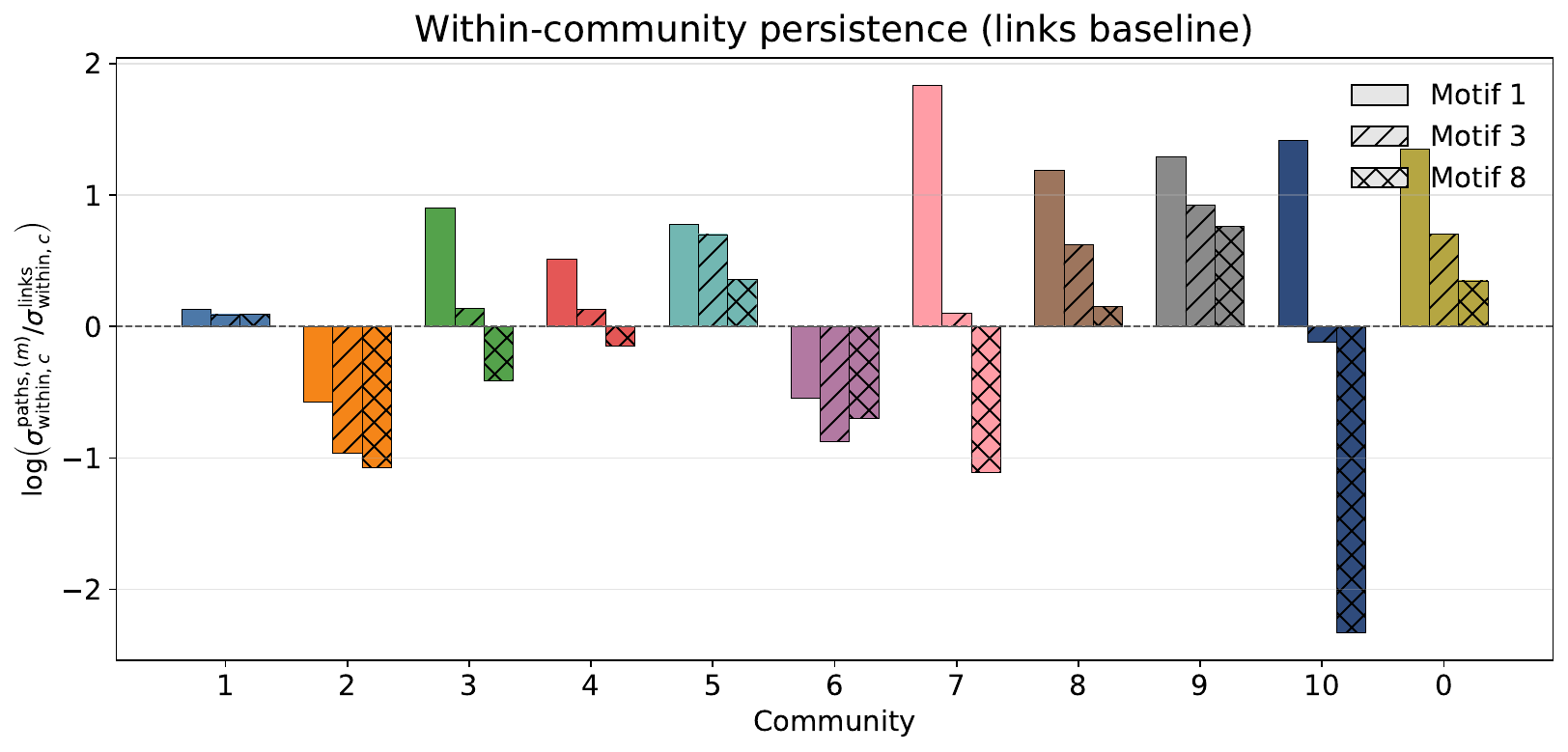}
    \caption{Community-level log-ratio of within-community persistence under the $W^{\mathrm{links}}$ baseline. Bars report $\log\left(\sigma^{\mathrm{paths},(m)}_{\mathrm{within},c} / \sigma^{\mathrm{links}}_{\mathrm{within},c}\right)$ for Motifs~1, 3, and 8.}
    \label{fig:persistence_logratio_links_by_comm}
\end{figure}

\begin{figure}[h!]
    \centering
    \includegraphics[width=0.8\linewidth]{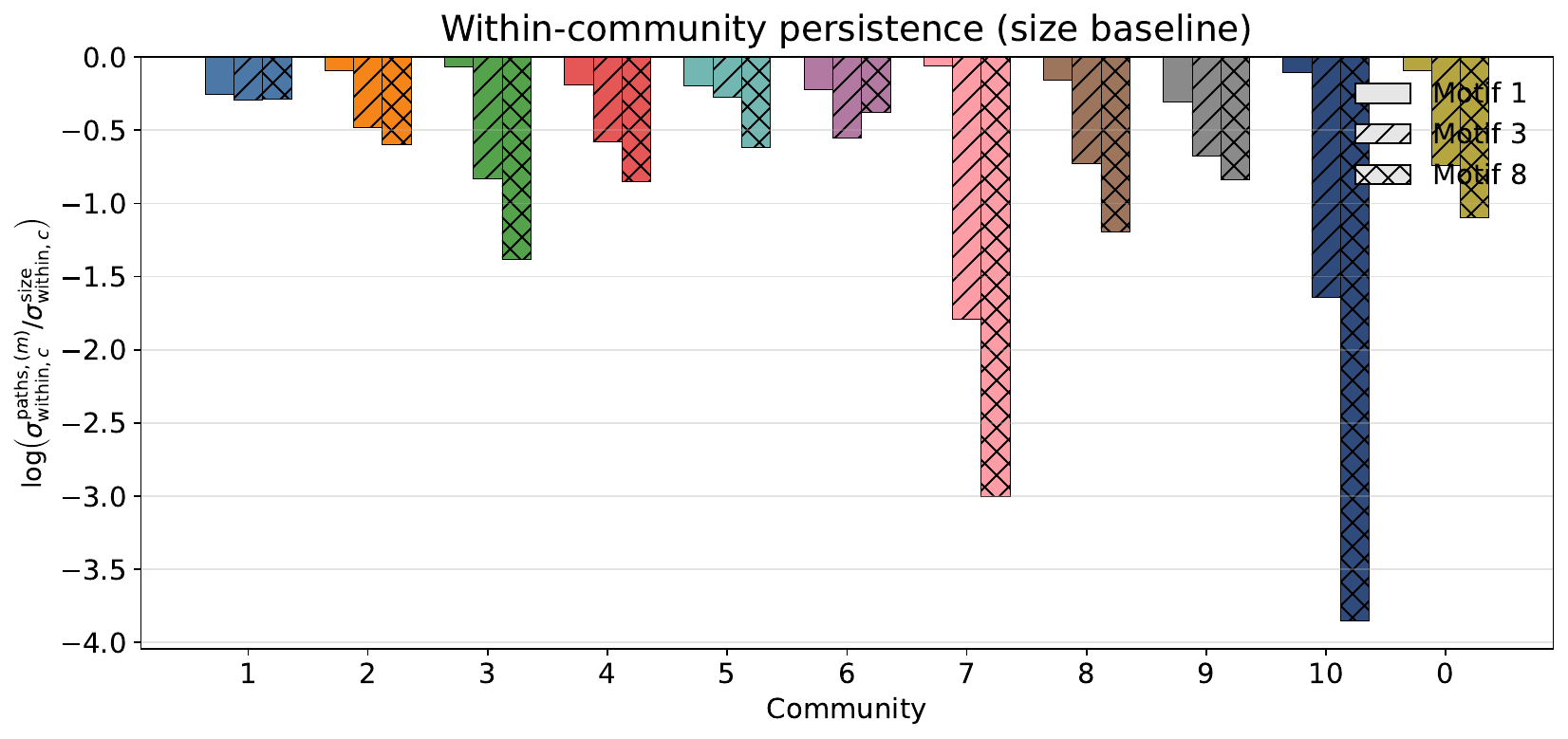}
    \caption{Community-level log-ratio of within-community persistence under the size-only baseline. Bars report $\log\left(\sigma^{\mathrm{paths},(m)}_{\mathrm{within},c} / \sigma^{\mathrm{size}}_{\mathrm{within},c}\right)$ for Motifs~1, 3, and 8.}
    \label{fig:persistence_logratio_size_by_comm}
\end{figure}

\begin{figure}[h!]
    \centering
    \includegraphics[width=0.8\linewidth]{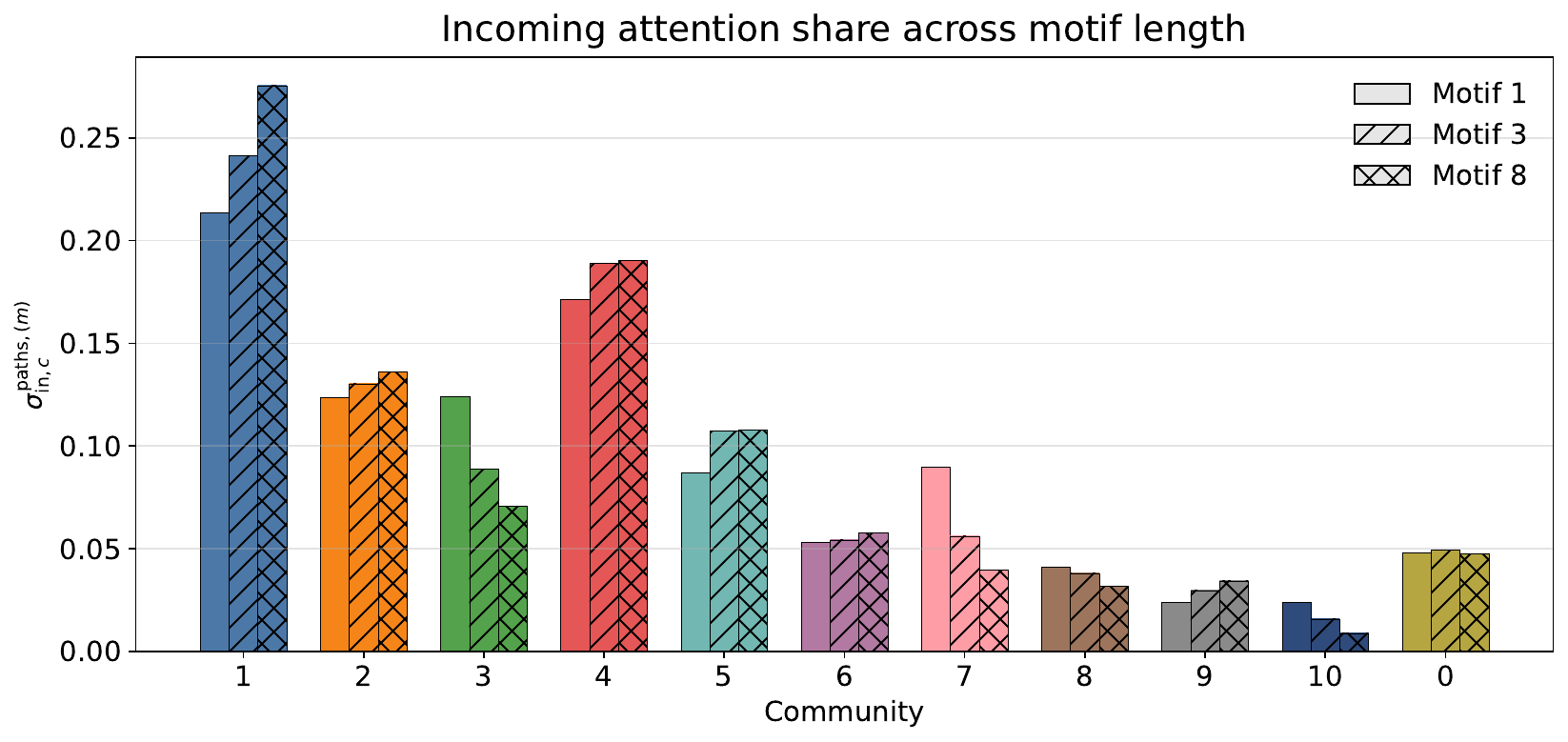}
    \caption{Distribution of incoming attention across communities. Bars report the share of paths $\sigma^{\mathrm{paths},(m)}_{\mathrm{in},c}$ for Motifs~1, 3, and 8.}
    \label{fig:incoming_spillover_raw_by_comm}
\end{figure}

\begin{figure}[h!]
    \centering
    \includegraphics[width=0.8\linewidth]{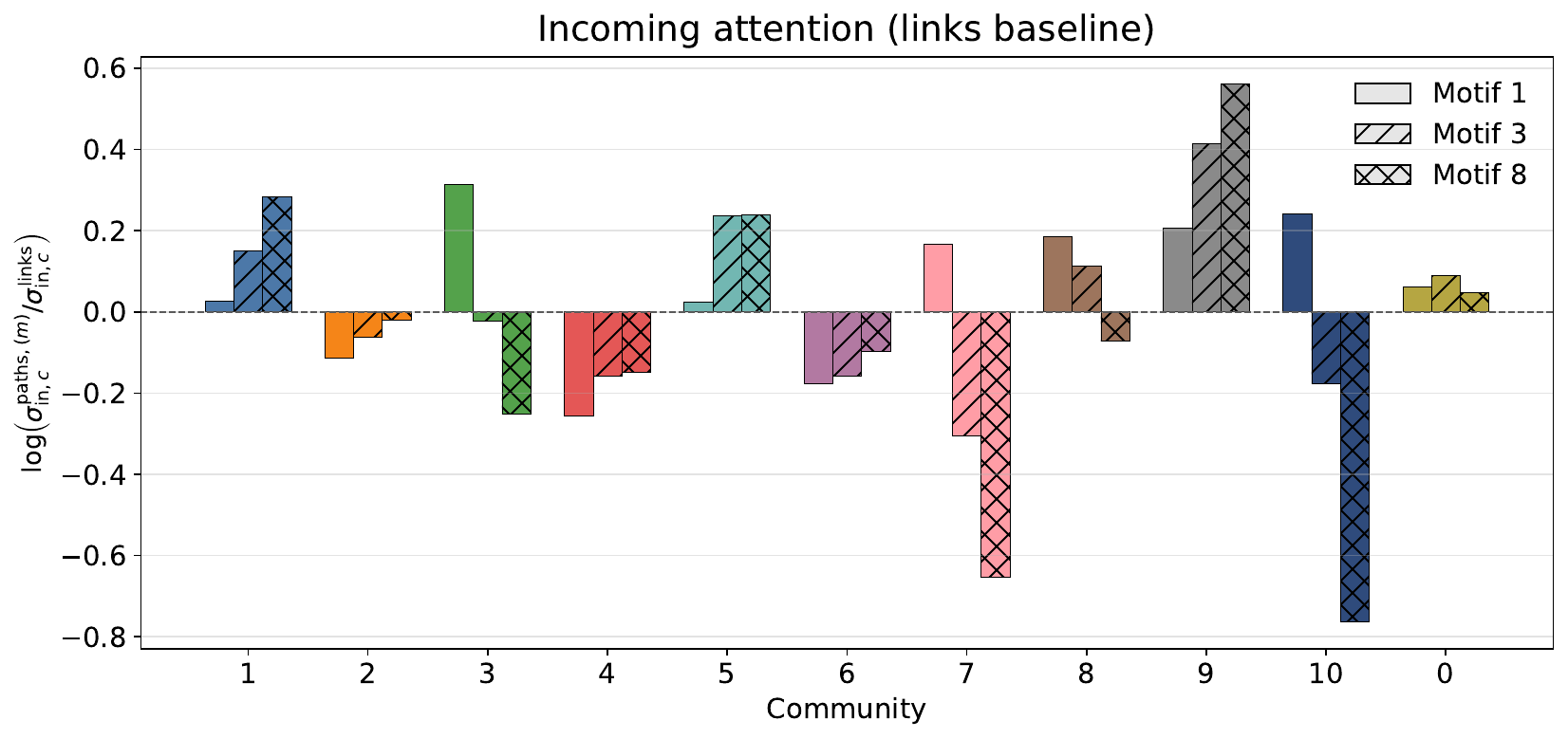}
    \caption{Community-level log-ratio of incoming attention under the $W^{\mathrm{links}}$ baseline. Bars report $\log\left(\sigma^{\mathrm{paths},(m)}_{\mathrm{in},c} / \sigma^{\mathrm{links}}_{\mathrm{in},c}\right)$ for Motifs~1, 3, and 8.}
    \label{fig:incoming_spillover_logratio_links_by_comm}
\end{figure}

\begin{figure}[h!]
    \centering
    \includegraphics[width=0.8\linewidth]{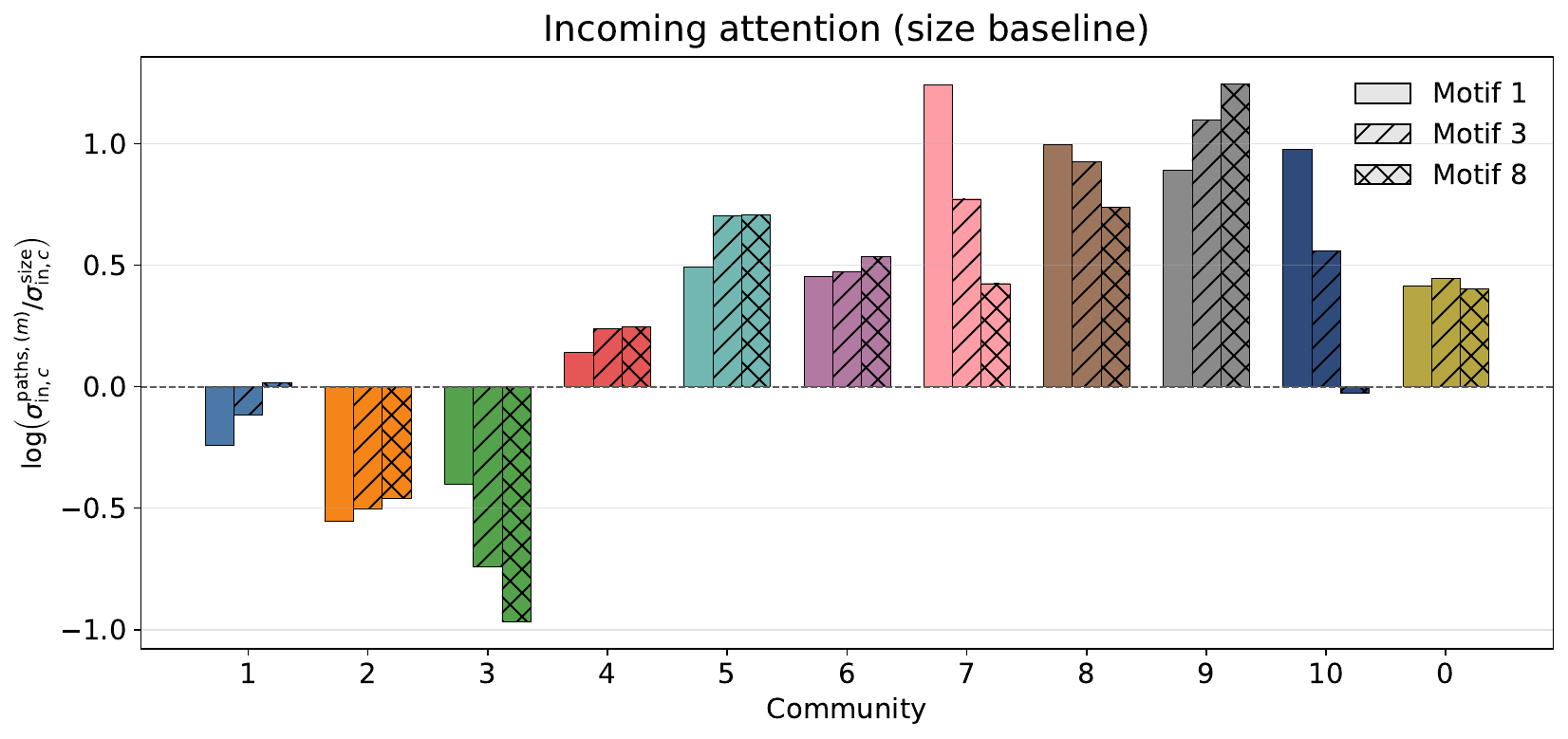}
    \caption{Community-level log-ratio of incoming attention under the size-only baseline. Bars report $\log\left(\sigma^{\mathrm{paths},(m)}_{\mathrm{in},c} / \sigma^{\mathrm{size}}_{\mathrm{in},c}\right)$ for Motifs~1, 3, and 8.}
    \label{fig:incoming_spillover_logratio_size_by_comm}
\end{figure}

\begin{figure}[h!]
    \centering
    \includegraphics[width=0.75\linewidth]{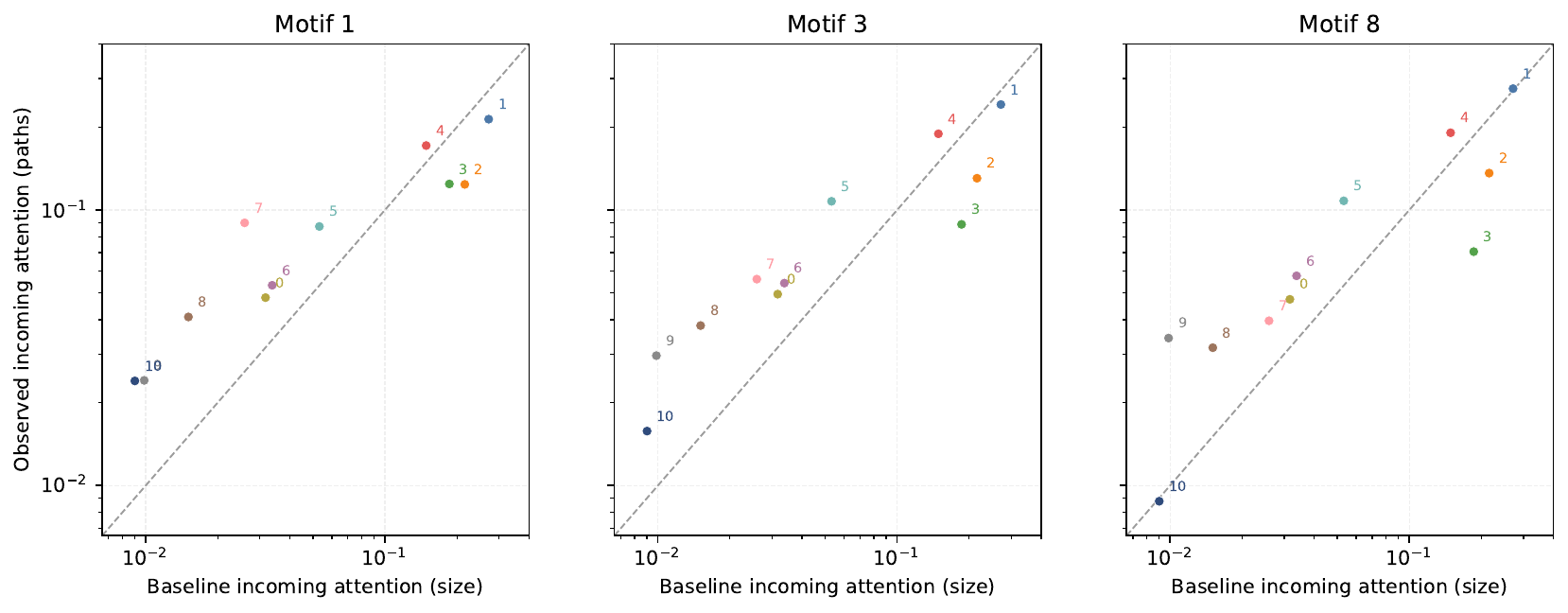}
    \caption{Observed versus baseline share of received cross-community attention under the $W^{\mathrm{size}}$ benchmark (final-jump, attention orientation). Each point is a community; panels correspond to Motif~1, Motif~3, and Motif~8. The dashed line indicates equality between observed and baseline shares.}
    \label{fig:obs_vs_size}
\end{figure}

\begin{figure}[h!]
    \centering
    \includegraphics[width=0.75\linewidth]{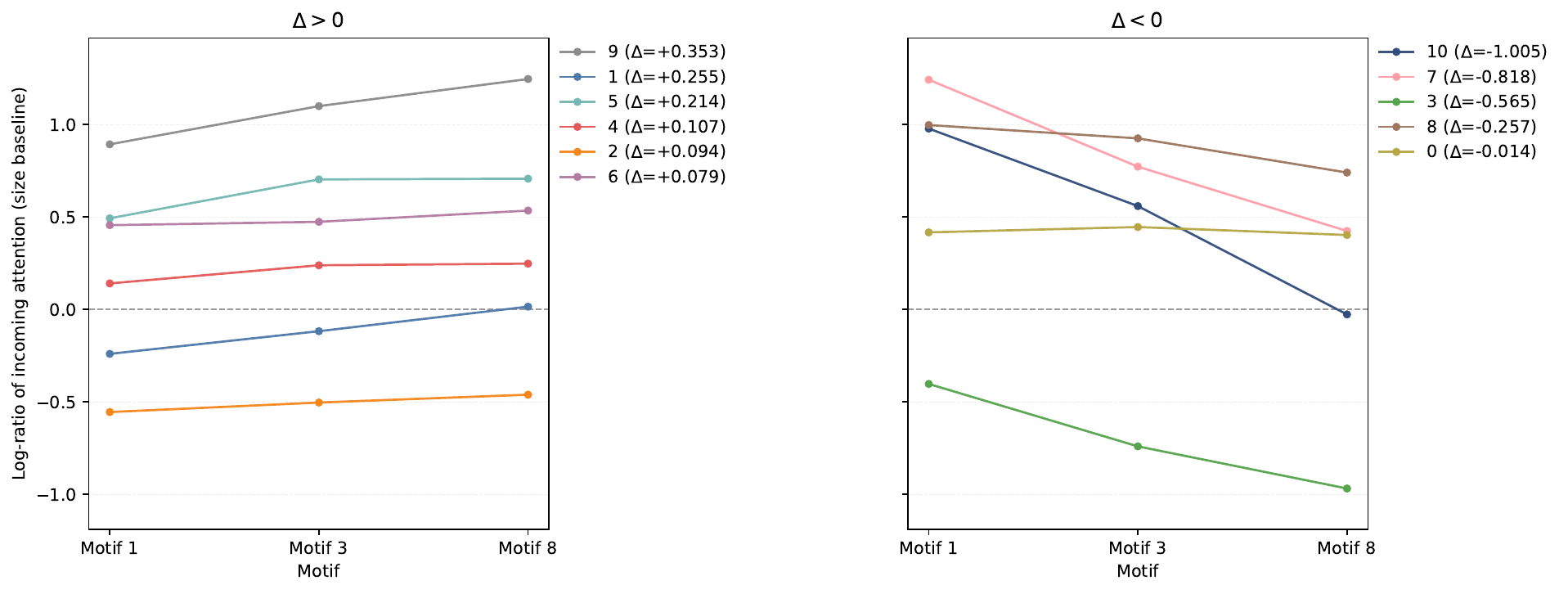}
    \caption{Baseline deviations across motif length under the $W^{\mathrm{size}}$ benchmark (final-jump, attention orientation). Lines show $\mathrm{log\_ratio}^{\mathrm{size},(m)}_{c}$ from Motif~1 to Motif~8. Left panel: communities with $\Delta^{\mathrm{size},(8-1)}_{c}>0$ (increasing deviation). Right panel: communities with $\Delta^{\mathrm{size},(8-1)}_{c}<0$ (decreasing deviation). The dashed horizontal line marks $0$ (no deviation from the baseline).}
    \label{fig:split_delta_size}
\end{figure}

\newpage
\clearpage
\section{Node-level absorption and hubness}
\label{appendix:absorption_hubness}

To further characterize the structural mechanisms underlying random-walk pathways, we examine node-level absorption and hubness patterns. As discussed in the Methods, the quantity $p_{\mathrm{sink}}$ captures the local imbalance between incoming and outgoing strength and therefore how likely a node is to act as a structural endpoint of a random-walk path.

Table~\ref{tab:psink_hubness_attention_terminal} summarizes the terminal position of the three main motifs in the attention orientation. Across Motifs~1, 3, and 8, terminal nodes are very frequently characterized by high $p_{\mathrm{sink}}$, and a non-negligible fraction are perfectly absorbing ($p_{\mathrm{sink}}=1$). Moreover, among nodes with high $p_{\mathrm{sink}}$, a large share also belongs to the top $1\%$ of the in-degree distribution, indicating that absorption and hubness substantially overlap in practice.

These results support the interpretation used in the main text: path termination and attention concentration are not driven by uniform exploration of the network, but by structurally prominent nodes that combine strong incoming concentration with limited continuation capacity.

\begin{table}[h!]
    \centering
    \small
    \begin{tabular}{lccc}
        \hline
        Motif (terminal position) & $p_{\mathrm{sink}}>0.90$ & $p_{\mathrm{sink}}=1$ & top-$1\%$ in-degree among $p_{\mathrm{sink}}>0.90$ \\
        \hline
        Motif~1 (position 2) & 0.920 & 0.302 & 0.866 \\
        Motif~3 (position 3) & 0.856 & 0.212 & 0.842 \\
        Motif~8 (position 4) & 0.691 & 0.162 & 0.835 \\
        \hline
    \end{tabular}
    \caption{Node-level absorption and hubness at the terminal position of the three main motifs in the attention orientation.}
    \label{tab:psink_hubness_attention_terminal}
\end{table}

\newpage
\section{Reversed orientation}
\label{appendix:reversed_orientation}

In this appendix we report the full set of figures for the reversed orientation, mirroring the analysis presented in the main text for the attention orientation. These figures provide a complete view of the structural patterns under direction inversion and serve as a diagnostic complement to the main results.

\begin{figure}[h!]
    \centering
    \includegraphics[width=0.9\linewidth]{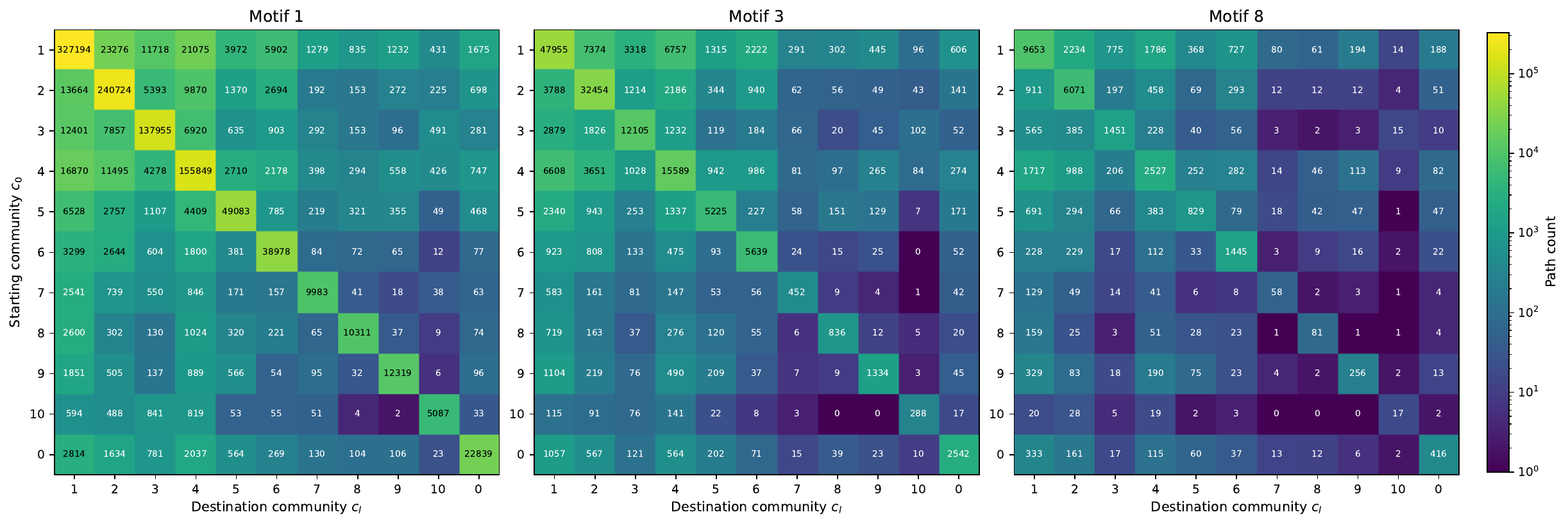}
    \caption{Community-level final-jump matrices (raw counts) for Motifs~1, 3, and 8 under the reversed orientation.}
    \label{fig:final_jump_heatmaps_reversed}
\end{figure}

\begin{figure}[h!]
    \centering
    \includegraphics[width=0.85\linewidth]{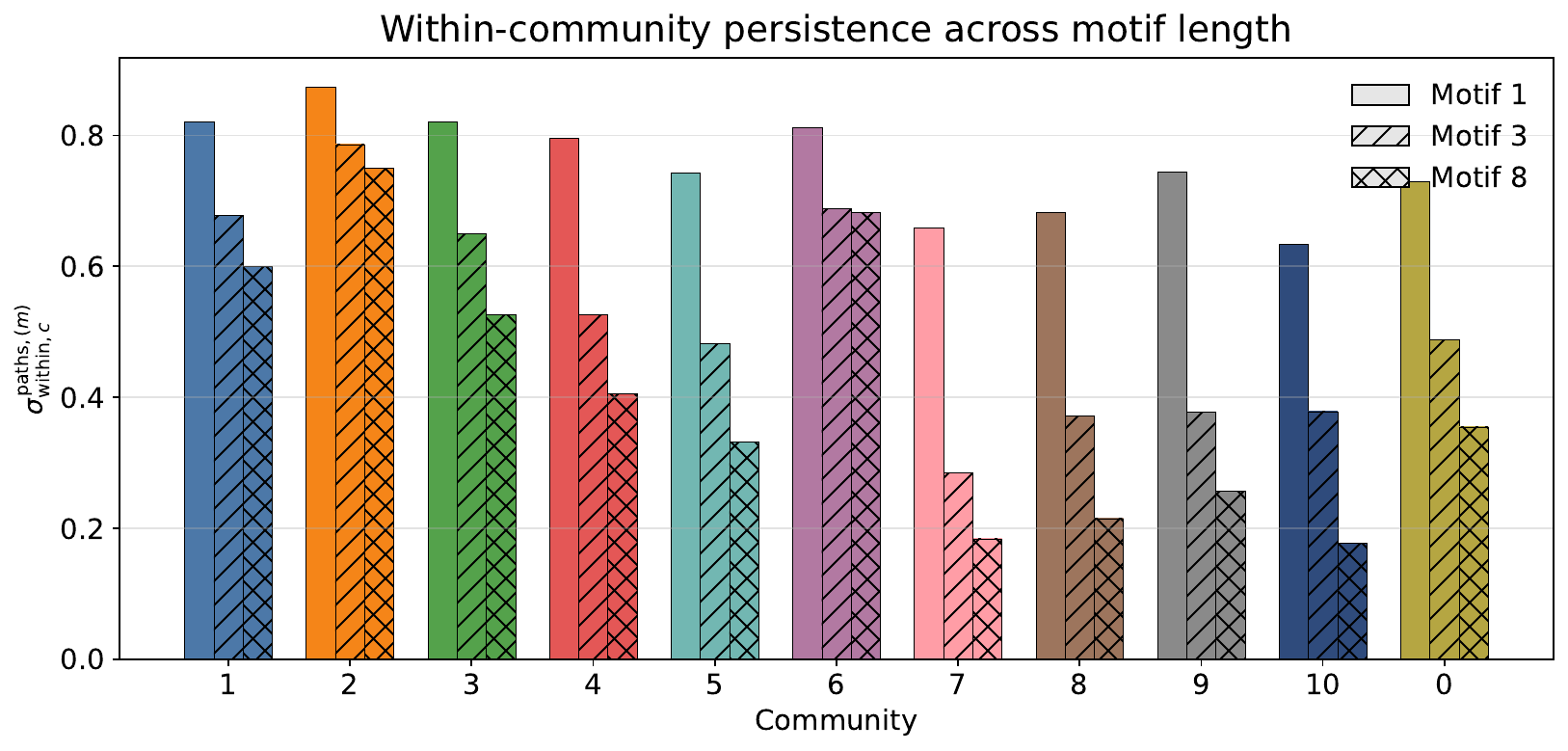}
    \caption{Within-community persistence across motif length under the reversed orientation.}
    \label{fig:persistence_raw_reversed}
\end{figure}

\begin{figure}[h!]
    \centering
    \includegraphics[width=0.85\linewidth]{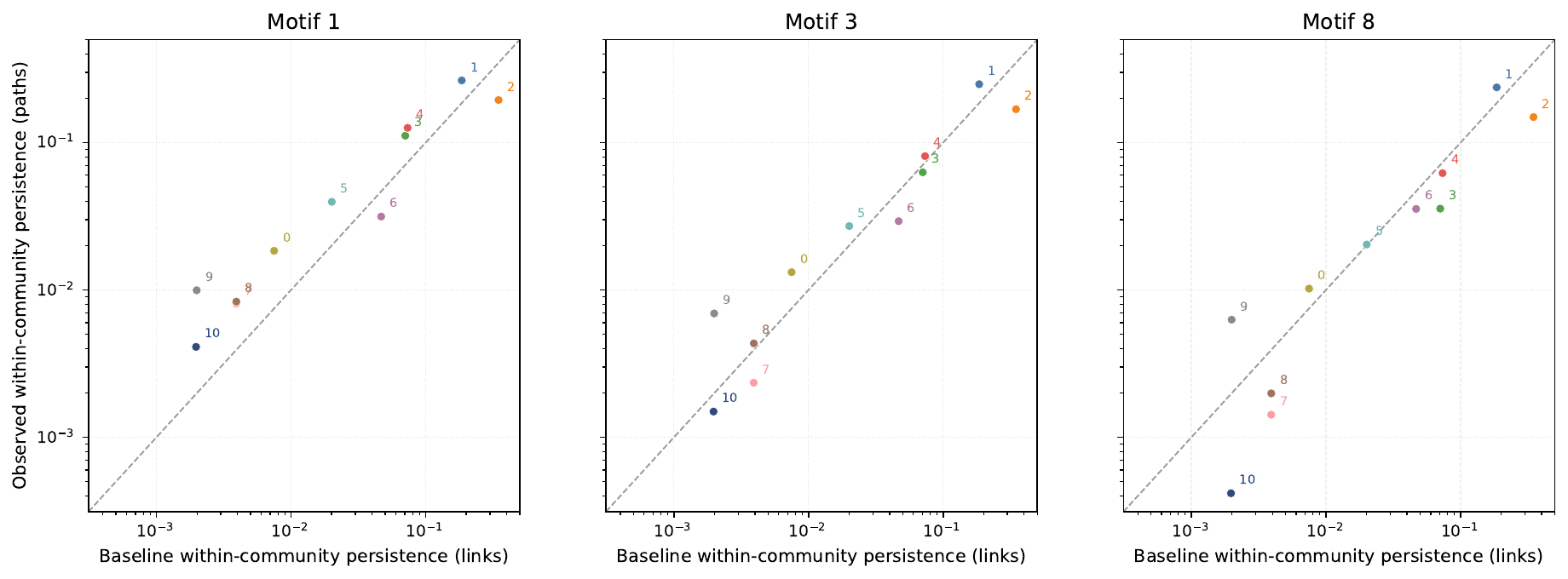}
    \caption{Observed versus baseline within-community persistence under the $W^{\mathrm{links}}$ benchmark (reversed orientation).}
    \label{fig:persistence_vs_links_reversed}
\end{figure}

\begin{figure}[h!]
    \centering
    \includegraphics[width=0.85\linewidth]{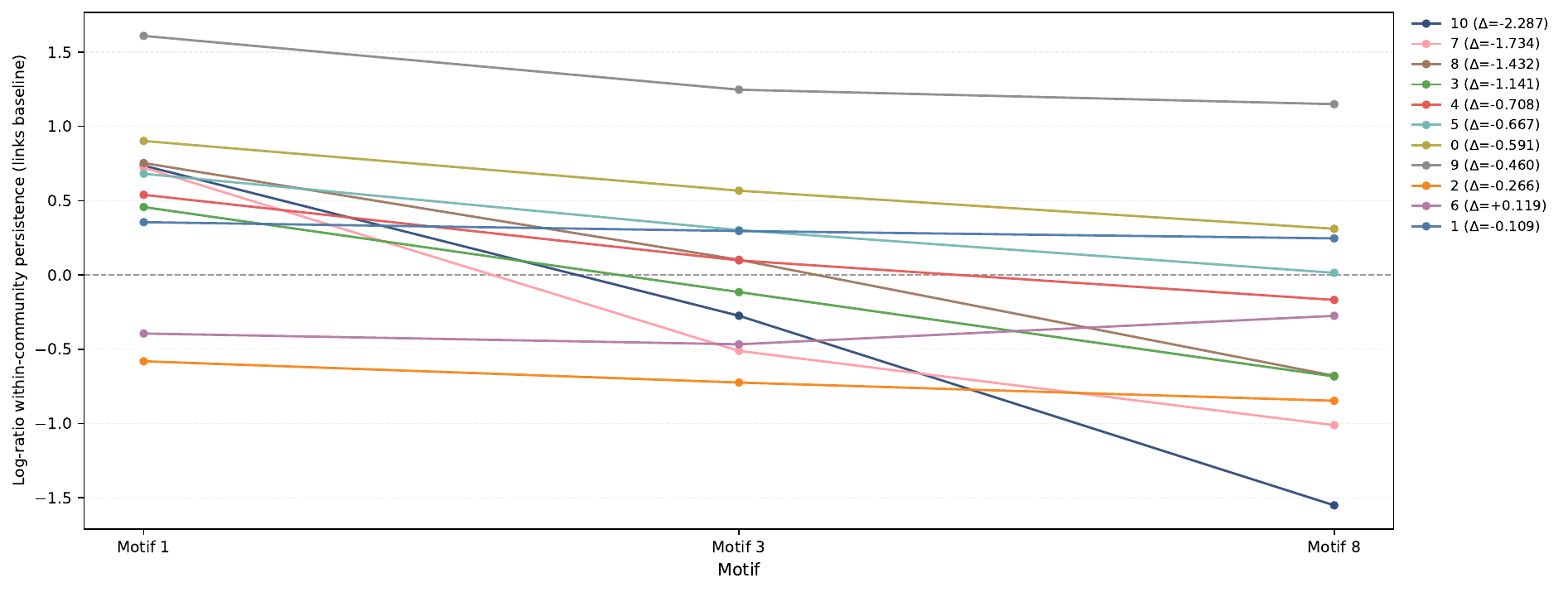}
    \caption{Benchmark-adjusted within-community persistence across motif length under the reversed orientation.}
    \label{fig:persistence_by_length_reversed}
\end{figure}

\begin{figure}[h!]
    \centering
    \includegraphics[width=0.9\linewidth]{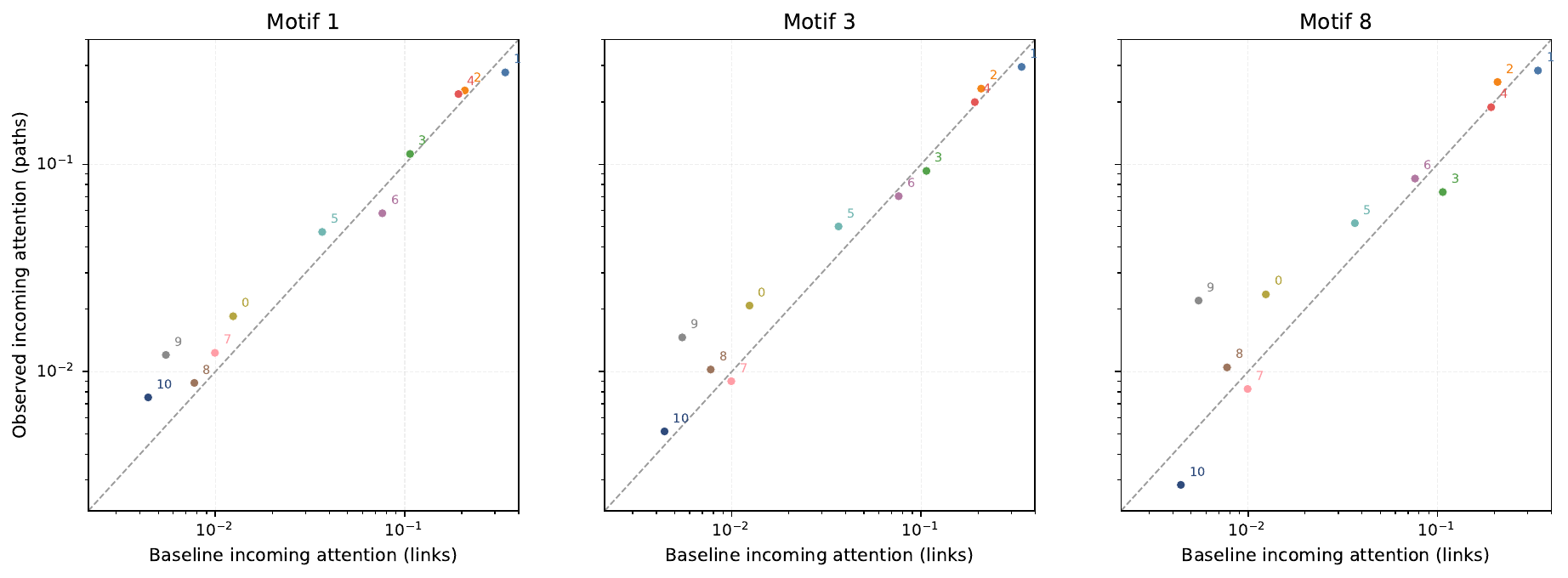}
    \caption{Observed versus baseline share of received cross-community attention under the $W^{\mathrm{links}}$ benchmark (reversed orientation).}
    \label{fig:obs_vs_links_reversed}
\end{figure}

\section{Distributions}
\begin{figure}[h]
    \centering
    \begin{subfigure}[b]{0.48\textwidth}
        \centering
        \includegraphics[width=\textwidth]{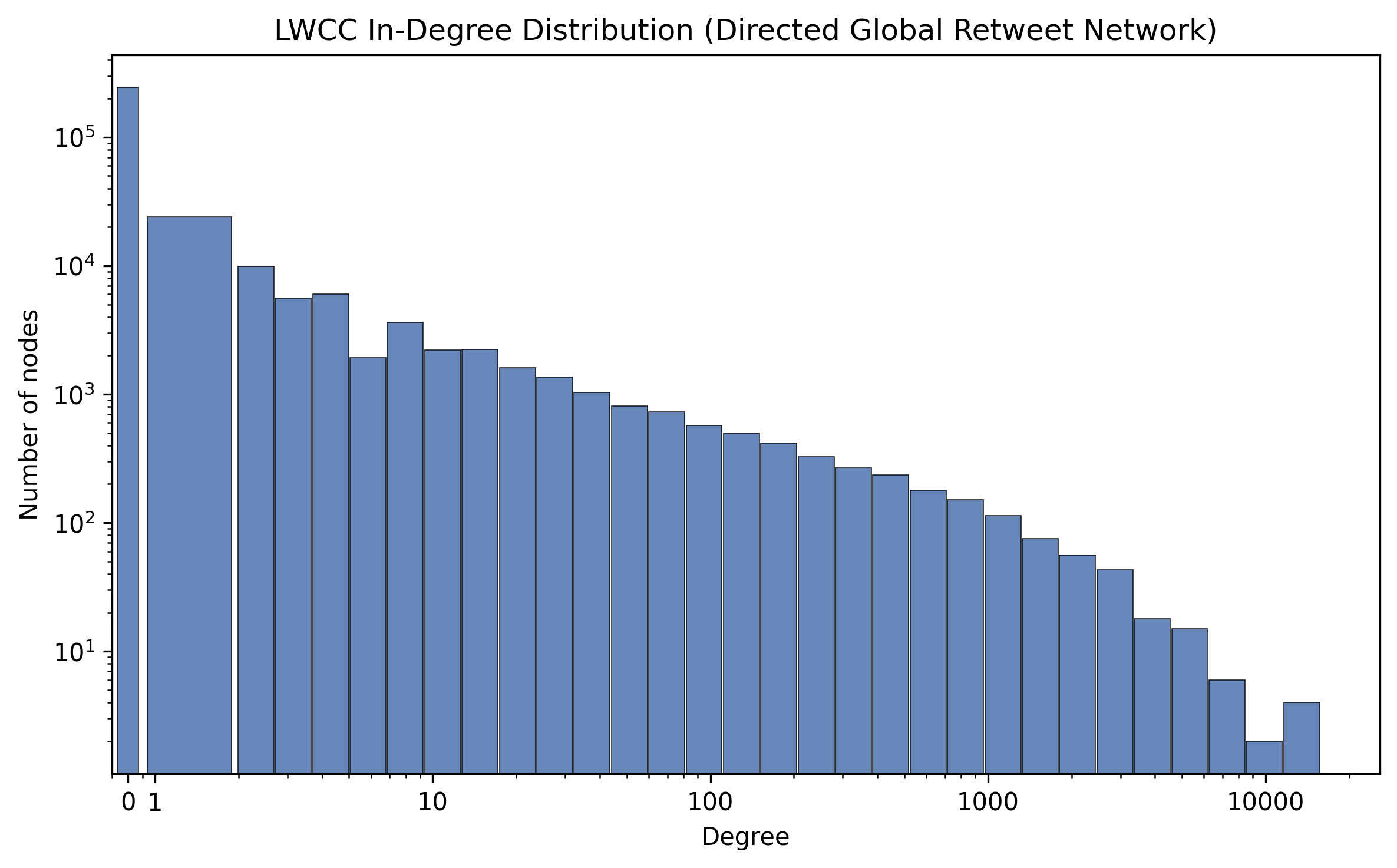}
        \caption{In-degree distribution.}
        \label{fig:lwcc_in_degree_distribution}
    \end{subfigure}
    \hfill
    \begin{subfigure}[b]{0.48\textwidth}
        \centering
        \includegraphics[width=\textwidth]{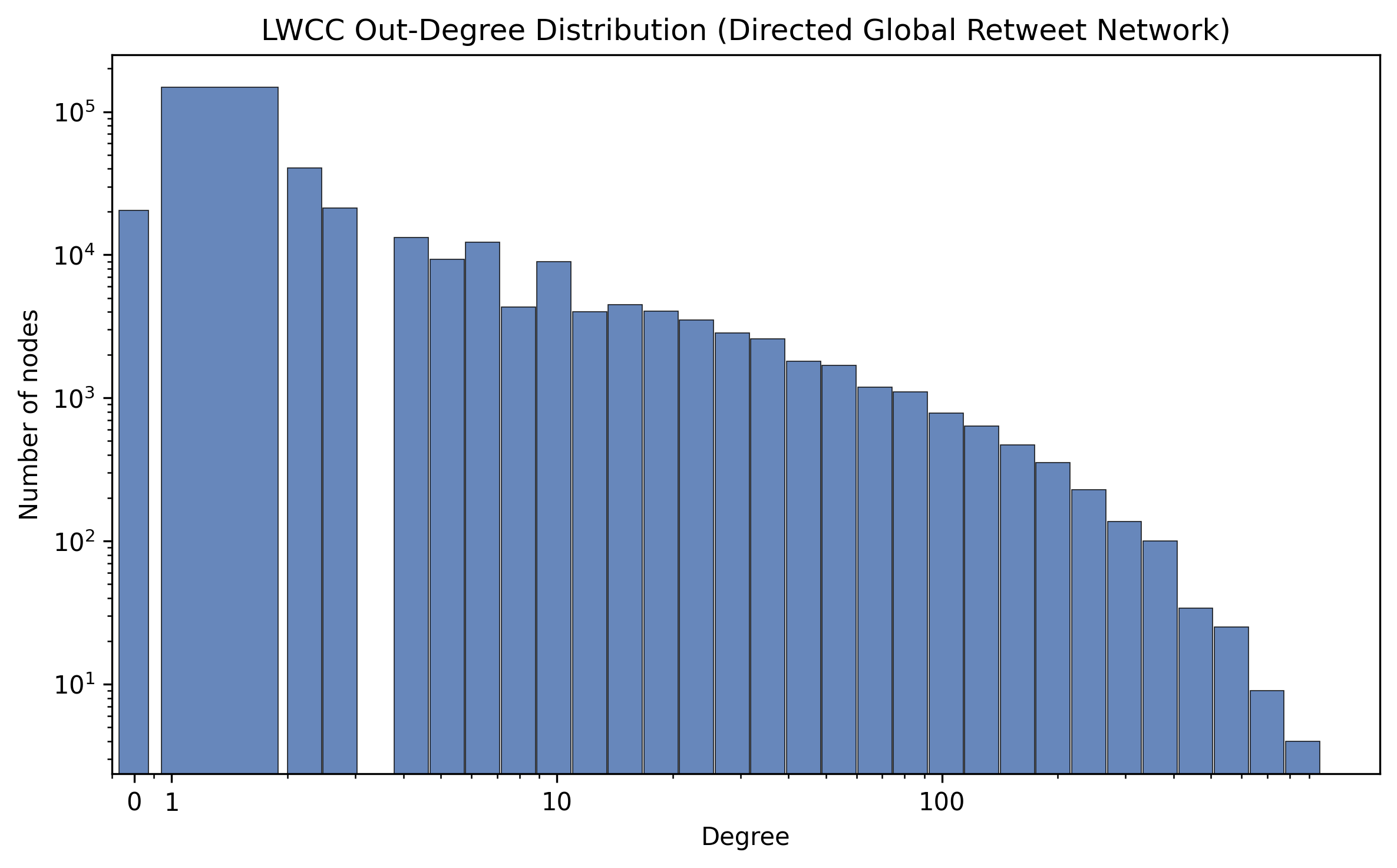}
        \caption{Out-degree distribution.}
        \label{fig:lwcc_out_degree_distribution}
    \end{subfigure}
    \caption{In- and out-degree distributions in the largest weakly connected component (LWCC) of the global retweet network. Both distributions are heavy-tailed, with a small fraction of users having very high degree.}
    \label{fig:lwcc_degree_distribution}
\end{figure}

\begin{figure}[h]
    \centering
    \begin{subfigure}[b]{0.48\textwidth}
        \centering
        \includegraphics[width=\textwidth]{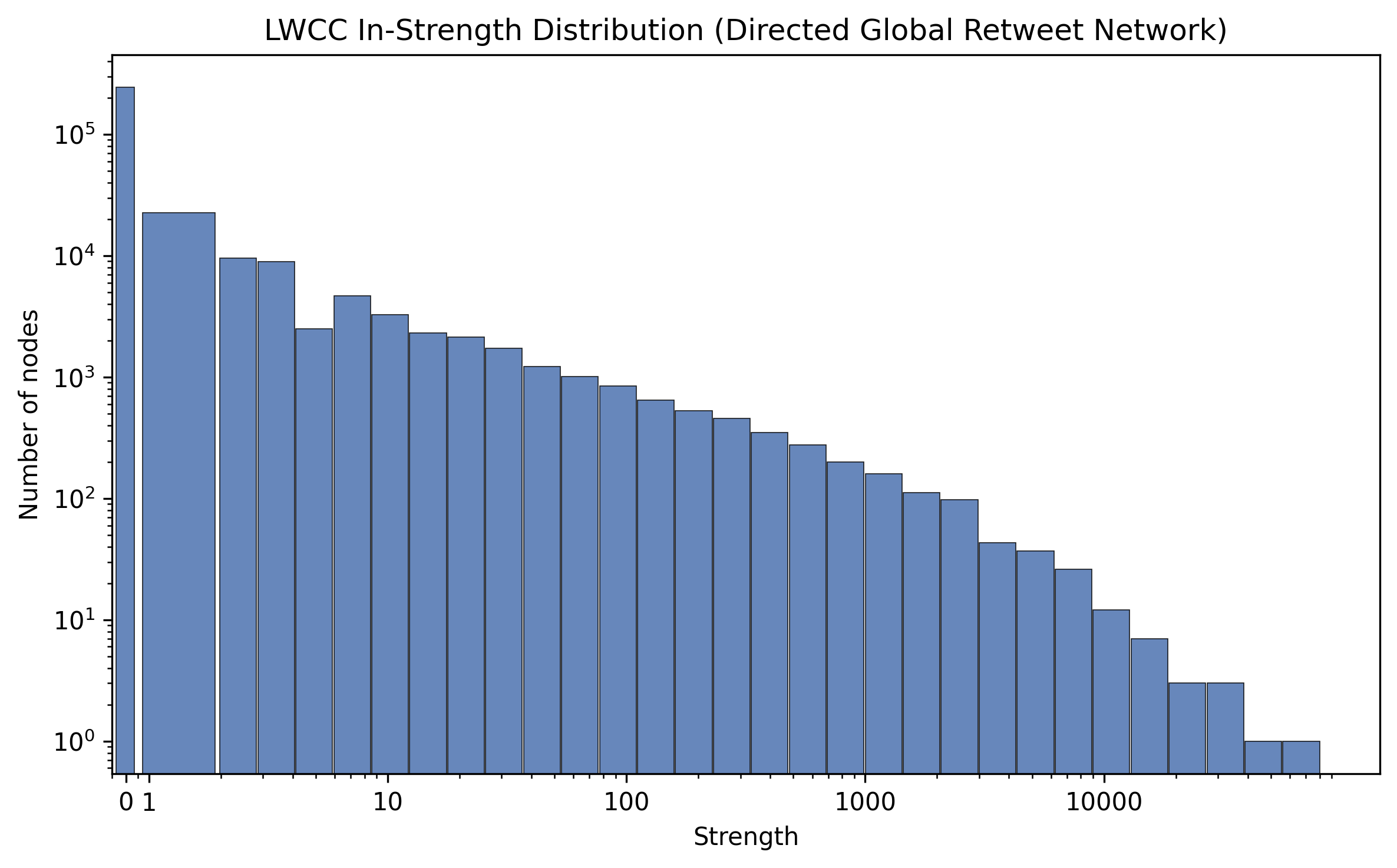}
        \caption{In-strength distribution.}
        \label{fig:lwcc_in_strength_distribution}
    \end{subfigure}
    \hfill
    \begin{subfigure}[b]{0.48\textwidth}
        \centering
        \includegraphics[width=\textwidth]{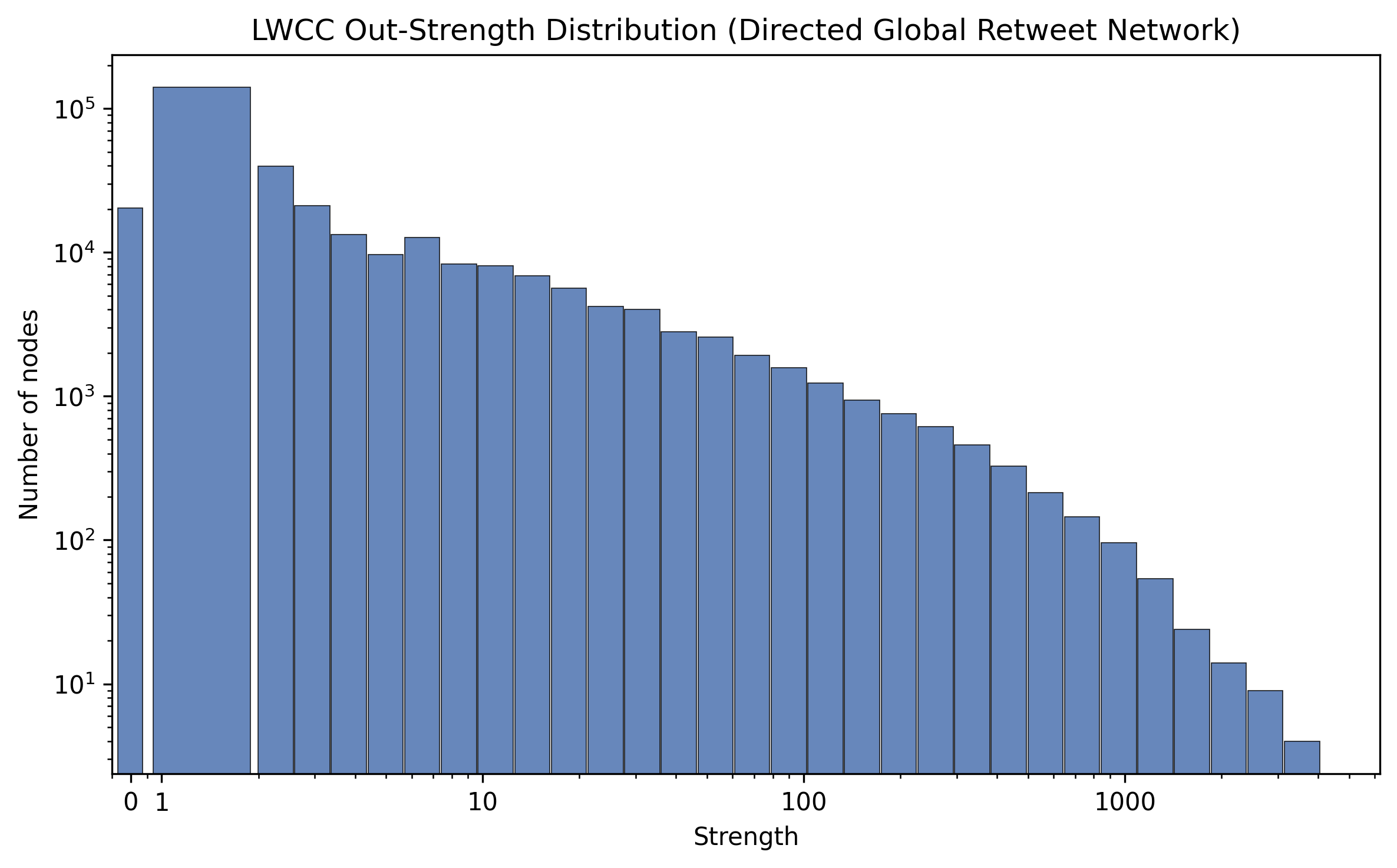}
        \caption{Out-strength distribution.}
        \label{fig:lwcc_out_strength_distribution}
    \end{subfigure}
    \caption{In- and out-strength distributions in the largest weakly connected component (LWCC) of the global retweet network. Both distributions are heavy-tailed, with a small fraction of users having very high degree.}
    \label{fig:lwcc_strength_distribution}
\end{figure}

\end{document}